\UseRawInputEncoding
\documentclass[11pt]{article}
\pdfoutput=1
\usepackage{jheppub}
\usepackage{amsmath}
\usepackage{bm}
\usepackage{setspace}
\usepackage{amsthm}

\usepackage{hyperref}

\usepackage{subcaption}

\newcommand{\xnew}{x_{\text{new}}}
\newcommand{\xold}{x_{\text{old}}}
\newcommand{\dwddw}[2]{\frac{\partial W}{\partial x^{#2}} \frac{\partial^2 \bar W}{\partial \bar x^{#2} \partial \bar x^{#1}}}
\newcommand{\twk}{$\bm w_k$}

\usepackage[font=scriptsize]{caption}
\def\R{{\mathbb R}}
\def\S{{\mathbb S}}
\def\Z{{\mathbb Z}}

\def\ZN{{\mathbb Z}_N}

\def\beq{\begin{equation}}
\def\eeq{\end{equation}}

\def\R{{\mathbb R}}
\def\S{{\mathbb S}}
\def\Z{{\mathbb Z}}

\def\ZN{{\mathbb Z}_N}

\def\beq{\begin{equation}}
\def\eeq{\end{equation}}

\title{Confinement on $\R^3 \times \S^1$ and Double-String Collapse}
\author[a]{Mathew W. Bub,}\author[a]{Erich Poppitz,} \author[a,b]{Samuel S. Y. Wong}
\affiliation[a]{Department of Physics,   University of Toronto, Toronto, ON M5S 1A7, Canada}
\affiliation[b]{Department of Physics, Stanford University, Stanford, CA 94305, USA}
\emailAdd{mathew.bub@mail.utoronto.ca}\emailAdd{poppitz@physics.utoronto.ca}  \emailAdd{samswong@stanford.edu}    
 \abstract{

 {\flushleft{We}} study confining strings in ${\cal{N}}=1$ supersymmetric $SU(N_c)$ Yang-Mills theory in the semiclassical regime on $\R^{1,2} \times \S^1$. 
Static quarks are expected to be confined by double strings composed of two domain walls---which are lines in $\R^2$---rather than by a single flux tube.  Each domain wall carries part of the quarks' chromoelectric flux. We numerically study this  mechanism and find that  double-string confinement holds for strings of all $N$-alities, except for those between fundamental quarks. We show that, for $N_c \ge 5$, the two domain walls confining unit $N$-ality quarks attract and form non-BPS bound states, collapsing to a single flux line.  We determine the $N$-ality dependence of the string tensions for $2 \le N_c\le 10$. Compared to known scaling laws, we find a  weaker, almost flat  $N$-ality dependence, which is qualitatively explained by the properties of  BPS domain walls. We also quantitatively study the behavior of confining strings upon increasing the $\S^1$ size by including the effect of virtual ``$W$-bosons" and show  that the qualitative features of double-string  confinement persist.}

\begin{document}

\maketitle

\section{Introduction, Summary, and Outlook}
\label{sec:1}

Quark confinement  is an old and difficult problem in quantum field theory.\footnote{The confinement problem and many approaches to it are comprehensively reviewed in \cite{Greensite:2011zz}.}  A direct analytical attack in non-supersymmetric gauge theories on $\R^{1,3}$ is beyond current capabilities. The study of ``toy" theories approximating different aspects of the physical world  is thus a worthwhile endeavour.
Supersymmetric Yang-Mills (SYM) theory stands out in this regard. It has been the subject of many studies   due to its tractability and  its similarity  to non-supersymmetric pure Yang-Mills (YM) theory, to which it is connected by decoupling the gaugino. 

Of special interest to us in this paper is the theory formulated on  $\R^{3} \times \S^1$, with supersymmetric boundary conditions on $\S^1$ \cite{Seiberg:1996nz,Aharony:1997bx}. Here, the dynamics is drastically simplified when $L N \Lambda \ll 1$; $L$ is the $\S^1$ circumference, $N$ is the number of colors, and $\Lambda$ is the strong coupling scale of the $SU(N)$ gauge group \cite{Davies:1999uw,Davies:2000nw}. Remarkably, in this regime, center stability, confinement, and discrete chiral symmetry breaking  become intertwined and are all due to the proliferation of various
topological molecules in the Yang-Mills vacuum: magnetic \cite{Unsal:2007jx,Unsal:2007vu} and neutral ``bions" \cite{Poppitz:2011wy,Poppitz:2012sw,Argyres:2012ka,Argyres:2012vv,Poppitz:2012nz}, composite objects made of various monopole-instantons \cite{Lee:1997vp,Kraan:1998pm}. The magnetic bions, in particular, provide a fascinating and non-trivial locally four-dimensional  generalization  \cite{Unsal:2007jx,Unsal:2007vu} of the Polyakov mechanism \cite{Polyakov:1976fu} of confinement. Despite looking three-dimensional (3D) at long distances, the theory remembers much about its four-dimensional (4D) origin. In many cases, it is known or believed that the small-$L$ theory is connected to the $\R^4$ theory ``adiabatically," i.e. without a phase transition. Lattice studies \cite{Bergner:2015cqa,Bergner:2018unx,Bergner:2019dim} have already provided evidence for the adiabaticity. 

The $\R^{3} \times \S^1$ setup described above provides a rare example of  analytically tractable nonperturbative phenomena in four dimensional gauge theories and many of its aspects have been the subject of previous investigations, see \cite{Dunne:2016nmc} for a review.
Here, we continue the study  of the properties of confining strings in this calculable framework. 
Our goal is to study the detailed structure of confining strings in SYM in the semiclassical  regime: we want to determine the string tensions, their $N$-ality dependence, and get at least a glimpse at how the confining string configurations evolve upon increase of $L$ towards the $\R^4$ limit. 

 In Ref.~\cite{Anber:2015kea}, it was shown that heavy fundamental quarks are confined by particular ``double-string" configurations:  a quark-antiquark pair has its chromoelectric flux split into two---rather than a single---flux tube. These flux tubes are actually domain walls (DW) connecting the $N$ different vacua of SYM. 
Static DWs  are lines in $\R^2$ of finite energy per unit length. Along their length, they carry a  fraction of the chromoelectric flux of quarks. These fractional fluxes are precisely such  that a pair of DWs form a ``double string"  stretched between a quark and antiquark,   leading to quark confinement with a linearly rising potential.\footnote{A typical example of a double string is shown on Fig.~\ref{fig:DSexample} in the bulk of this paper.}

The double-string nature of confining strings is closely tied to the recently discovered  mixed 0-form chiral/1-form center anomaly  \cite{Gaiotto:2014kfa,Gaiotto:2017yup,Gaiotto:2017tne,Sulejmanpasic:2016uwq,Komargodski:2017smk,Hsin:2018vcg}  in SYM.\footnote{Historically, within quantum field theory, the 
 double-string   picture was the first used to argue  that quarks become deconfined on DWs---the quark's chromoelectric flux is absorbed by the DWs, of equal tension,   to the left and the right. The equal tensions of different walls in SYM is due to supersymmetry, as many of the walls are BPS objects (in non-supersymmetric generalizations the unbroken 0-form center symmetry also plays a crucial role). The anomaly inflow connection, understood later, gives an additional general argument valid beyond semiclassics \cite{Gaiotto:2014kfa,Gaiotto:2017yup,Gaiotto:2017tne,Sulejmanpasic:2016uwq,Komargodski:2017smk,Hsin:2018vcg}. Earlier arguments within string theory originated in \cite{Witten:1997ep}.
}  The anomaly-inflow aspect and the quark deconfinement on DWs  was studied in detail in ref.
\cite{Cox:2019aji}. There, the chromoelectric fluxes carried by Bogomolnyi-Prasad-Sommerfield (BPS) DWs connecting the $N$ vacua of SYM were determined.  
 In our study of confining strings, we shall make extensive use of these results.

\subsection{Summary of Results and Outline}
\label{sec:1.1}

In this paper, we study in detail the double-string confinement mechanism, in the framework of the abelian EFT valid at $L N \Lambda \ll 1$, also accounting for the leading effect of virtual $W$-boson loops. This EFT is described in some detail in Section~\ref{sec:2}.

For the reader familiar with the $\R^3$ Polyakov model, but less familiar with confinement in circle-compactified theories at small $L$, we review, in Section \ref{sec:polyakovsection}, the main differences between Polyakov confinement and confinement in SYM on $\R^3 \times \S^1$, for gauge group $SU(2)$. This discussion does not suffer from too much Lie-algebraic notation needed to study $SU(N)$, but is sufficient to stress the main peculiarities of small-$L$ confinement in SYM theory, which persist to arbitrary values of $N$.

For general $SU(N)$ gauge groups, the double-string confinement mechanism is explained 
in Sections~\ref{sec:3.1}, \ref{sec:3.2}.
The taxonomy of the double-string configurations of Sections~\ref{sec:3.2.1}-\ref{sec:3.2.4} captures most qualitative aspects of the physics for any number of colors $N$ and for quark sources of any $N$-ality $k$. It is found, via numerical simulations, to fail only for $N\ge5$, $k=1$, when DWs exhibit attraction and the double string collapses.

Our numerical simulations were performed for $SU(N)$ gauge theories with $N \le 10$.
The numerical computations of the two-dimensional classical equations of motion with quark sources (see Section~\ref{sec:4}) can become quite demanding, especially when sources of all $N$-alities for up to $N=10$  are considered. Including the effect of virtual $W$-bosons demands even more resources due to the broadening of the double strings. Our computations for the confining string configurations were primarily performed on the Niagara cluster at the SciNet HPC Consortium \cite{niagara, scinet}. Each quark source configuration was run on a single hyperthreaded processor core, allowing us to run 80 quark sources simultaneously on a single compute node. In total, our simulations utilized approximately 15 core-years of compute time on Niagara.

\begin{figure}[t]
\centering
  \includegraphics[width= 1 \textwidth]{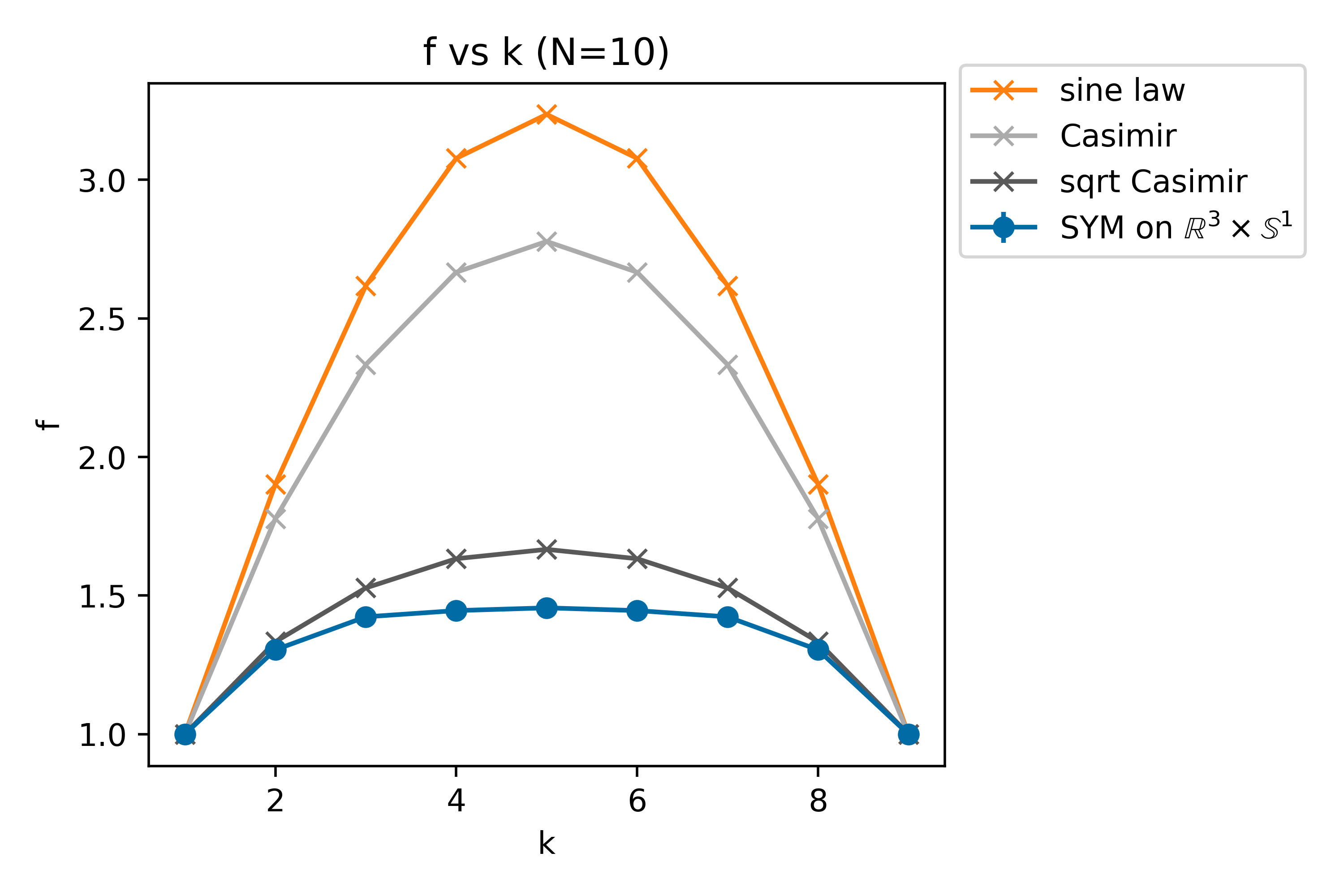}
  \caption{The dependence of the $k$-string tension ratio $f = {T(k,N) \over T(1,N)}$ on $k$ for $SU(10)$, compared with other known scaling laws. Here, $T(k,N)$ denotes the tension of the stable $k$-string for $SU(N)$. Recall that the sine law approximately holds  in softly-broken Seiberg-Witten theory and the MQCD embedding of SYM. The square-root law was found in the MIT Bag Model, as well as, approximately, in deformed Yang-Mills theory on $\R^3 \times \S^1$. Our results show that the $N$-ality dependence of $f$ in SYM on $\R^3 \times \S^1$ at small $L$ is   flatter than any of the other known scaling laws. A qualitative explanation follows from the BPS DW properties and is given in the text. The flat $N$-ality dependence shows that in the abelian large-$N$ limit, confining strings remain interacting.  }
  \label{fig:f(10,k)}
\end{figure}

{\flushleft{Our main results are as follows:}}

\begin{enumerate}
\item We first argue, in Section~\ref{sec:3.2}, that for static quark sources of $N$-ality $k$, the lowest tension confining strings are double-string configurations composed of two of the BPS 1-walls of SYM theory. These BPS 1-walls carry electric fluxes that add up to match the flux of the quark source $\vec w_k$, the highest weight of the $k$-index antisymmetric representation.  

We also show that quarks of $N$-ality $k$ with charges  different from $\vec w_k$ are confined by   double-string configurations made of  two BPS $p$-walls, with\footnote{$p=0$ denotes non-BPS walls.} $0 \le p \le k$. These metastable configurations are expected to decay to the lowest tension ones by the pair production of $W$-bosons, an effect that cannot be seen by the small-$L$ abelian EFT. 

\item
Numerically (see Section~\ref{sec:5}) the above qualitative expectations are found to hold for $N \le 4$ for all $k$. However,  for $N \ge 5$ double strings form only for $k>1$, i.e. for all but the fundamental representation quarks. Fundamental quark sources for $N\ge 5$ were found to be confined by a single-string configuration, formed upon a collapse of the two BPS 1-walls into a non-BPS bound state. See Section~\ref{sec:5.1}.

The collapse of the double string for $k=1$  is the most unexpected result of our findings and, at present, we have no simple qualitative explanation for it.  To make sure it is not a numerical artifact, we perform many checks on this collapse. One includes simulating the largest possible distance between the sources, making sure that the collapse is not a finite separation effect (see Fig.~\ref{fig:SU5k1energy}). Most importantly, we show that the fact that this collapse occurs only for $k=1$, $N \ge 5$ can already be seen by studying the interactions of parallel BPS DWs with appropriate fluxes. These one-dimensional configurations are studied in great detail in Section~\ref{sec:6}, confirming that the collapse occurs only for BPS 1-walls whose fluxes add to that of a fundamental quark.  

\item The results of the two items above imply that the $N$-ality dependence of the string tensions is rather weak. Qualitatively, this is because the $k$-strings of lowest tension with $k>1$ are all double strings made of BPS $1$-walls carrying appropriate fluxes. Hence, one expects that the  $k$-string tension is approximately twice the tension of the BPS $1$-wall, for all $k>1$. This expectation is borne out in our numerical simulations, see  Fig.~\ref{fig:f(10,k)} for $SU(10)$, as well as Section~\ref{sec:5.2}. The observed slight increase of the string tension with $k$ seen in the data is consistent with the increase of the length of the strings due to the larger transverse separation for larger $N$-ality.  On the other hand, the fact that for $k=1$ the tension is significantly lower is due to the double-string collapse and reflects the binding energy of the non-BPS bound state.

The $N$-ality dependence found here does not conform to the usual large-$N$ argument that $T(k,N)$ should approach $k T(1,N)$. The difference between the abelian and nonabelian large-$N$ has been emphasized many times \cite{Douglas:1995nw,Poppitz:2011wy,Cherman:2016jtu,Poppitz:2017ivi}: the confining strings discussed here do not become free in the $N\rightarrow \infty$, $LN$-fixed abelian large-$N$ limit. 

\item
We find that, whenever confinement is due to a double string, the separation between the strings increases as a logarithm of the distance between the sources. This qualitatively matches  with a simple model of exponential DW repulsion proposed earlier for an $SU(2)$ theory \cite{Anber:2015kea}. The fit of the parameters to the logarithmic growth is in  Sec.~\ref{sec:5.3}.

\item
We probe larger values of $L$ by including the effect of virtual $W$-boson (and superpartner) loops on double-string confinement for $N \le 9$ in Sec.~\ref{sec:5.4}. These effects become more important as the $\S^1$ size increases. 
The main lesson is that, while still in the semiclassical regime, these effects do not change the qualitative picture of double-string confinement for all $k$ and $N \le 9$: no separation of collapsed $k=1$ strings for $N >4$ occurs, nor do the $N\le 4$ $k=1$ strings collapse. The quantitative changes due to increasing $L$ are the larger transverse separation  of the double string and the change of  the $k$-string tension ratio, see Appendix \ref{appx:C.0.3} and \ref{appx:C.0.4}. Upon increasing $L$ within the abelianized regime, the range of variation of $f(N,k)$ is seen to be even smaller than that shown in Fig.~\ref{fig:f(10,k)} (see Fig.~\ref{fig:kratioL}).

The expectation for the string-tension ratio on $\R^4$, based on large-$N$ arguments and the M-theory embedding of SYM, is that $f(N,k)$ is smilar to the sine-law and Casimir-scaling curves on Fig.~\ref{fig:f(10,k)}. The behavior we find  indicates that the evolution of the string-tension ratio towards $\R^4$ may not be monotonic. It is also possible that at the finite values of the coupling (see eqns.~(\ref{epsilondef}, \ref{twolengths})) we chose to study, the unknown higher-order corrections to the K\" ahler potential qualitatively affect the string-tension ratios.
\end{enumerate}

\subsection{Outlook}
\label{sec:1.2}

In this paper, we studied the details of double-string confinement in the semiclassical regime on $\R^3 \times \S^1$, in the framework of SYM. While originally found independently, double-string confinement in semiclassical circle-compactified theories is, in fact, required by anomaly inflow of the mixed anomalies involving the spontaneously broken discrete 0-form (e.g. chiral or CP) symmetry and the 1-form center symmetry \cite{Cox:2019aji}. This confinement mechanism transcends SYM and also holds in other theories that have similar higher-form anomalies. Two examples are deformed Yang-Mills theory and QCD with adjoint fermions, and more can be found in \cite{Anber:2017pak,Anber:2019nfu}. It would be of interest to study double-string confinement in more detail in those theories as well. We note that QCD with adjoint fermions stands out among the many theories, since it has gapless modes in the bulk.

The question of the behavior of confining strings as $L$ is increased beyond the $\Lambda N L < 1$ regime is, of course, of great interest. Lattice studies of adiabatic continuity in SYM, as in \cite{Bergner:2018unx}, and in other theories should be able to shed light on the transition between the abelian confinement studied in this paper and the full-fledged non-abelian confinement in the $\R^4$ theory. 

There are several other features of circle-compactified theories, their domain walls, and their relation to confinement that would be nice to understand better. All our studies here, and in virtually all papers on the subject, are done in the framework of the 3D EFT of the lightest fields, the Kaluza-Klein zero modes of the 4D fields. The small-$L$ theory, however, is weakly coupled at all scales and it should be possible---although it may be complicated, see \cite{Anber:2013xfa} for related work and the recent remarks in  \cite{Unsal:2020yeh}---to study more interesting dynamical questions that necessitate the incorportation of the Kaluza-Klein modes in the description. One such question would be to understand dynamically how quarks ``become anyons" on the domain walls. This  aspect of anomaly inflow on domain walls, see e.g. \cite{Hsin:2018vcg}, is not possible to see in our 3D EFT. Further, the pair production of ``$W$-bosons,"  which we allude to many times and which makes the string tensions  $N$-ality-only dependent, is also outside of the realm of validity of the 3D EFT. Naturally, incorporating the Kaluza-Klein modes in the dynamics  might shed further light on the large-$L$ behavior.

We hope to return to these interesting questions in the future.

 \section{SYM on $\R^3 \times \S^1$: Effective Field Theory, Scales, and Symmetries}
 \label{sec:2}
 
 We shall not describe in detail  the  microscopic dynamics leading to the long-distance theory\footnote{In historical order, see \cite{Seiberg:1996nz, Aharony:1997bx} and the instanton calculation of \cite{Davies:2000nw,Davies:1999uw}, completed recently in \cite{Poppitz:2012sw,Anber:2014lba} by the calculations of the non-cancelling one-loop determinants in monopole-instanton backgrounds.} described below. Our starting point here is the result that 
the infrared dynamics of SYM on $\R^3 \times \S^1$ is described by an effective field theory (EFT)  of chiral superfields. These superfields are the duals of the Cartan subalgebra photons and their superpartners. The simplicity of the EFT is due to the fact that the theory abelianizes, breaking spontaneously the gauge group $SU(N) \rightarrow U(1)^{N-1}$ at a scale of order the lightest $W$-boson mass, $m_W = {2 \pi \over LN}$. Weak coupling is then owed to the small-$L$ condition $m_W \gg \Lambda$.

The  lowest components of the chiral superfields of the EFT are 
the Cartan-subalgebra valued  bosonic fields, the dual photons and holonomy scalars combined into dimensionless complex scalar fields:
\begin{eqnarray}
\label{fields1}
x^a = \phi^a + i \sigma^a~, ~ a= 1, \ldots N-1, ~~ {\rm with } ~\; \sigma^a  \simeq  \sigma^a + 2 \pi w_k^a~, ~ k = 1, \ldots N-1~.  
\end{eqnarray}
Here $\phi^a$ are real scalar fields describing the deviation of the $\S^1$ holonomy eigenvalues from its center-symmetric value\footnote{Explicitly, the $N$ eigenvalues of the $\S^1$-holonomy in the fundamental representation, labeled by $A$$=$$1$,...,$N$, are $e^{i \bm \nu_A \cdot( {2 \pi \over N} \bm\rho - {g^2 \over 4 \pi} \bm \phi)} \equiv e^{ i {2 \pi   \over N}({N+1 \over 2} - A)} \;e^{-i {g^2 \over 4 \pi} \bm \nu_A \cdot \bm \phi} $, where $\bm \rho$ and $\bm \nu_A$ denote the Weyl vector and the weights of the fundamental representation (see  our notations and conventions below). We give this explicit formula in order to stress that  ${\cal{O}}(1)$ excursions of $\bm \phi$ away from the origin do not lead to a breakdown of semiclassics. Such a breakdown requires that two eigenvalues of the holonomy coincide, causing the appearance of massless $W$-bosons, but this requires ${\cal{O}}(1/g^2)$ deviations of $\bm \phi$ away from the origin. However,  all our numerical results for DWs and strings do not involve such large excursions---instead, the range of variation of $\bm \phi$ in the classical solutions of our EFT is ${\cal{O}}(1)$.} and $\sigma^a$ are the duals of the Cartan-subalgebra photons.  
The more precise relation to 4D fields is as follows:\footnote{The spacetime metric is $(+,-,-)$ and $\epsilon^{012} = \epsilon_{012} = 1$.}
\begin{eqnarray} \label{fieldrelation}
{g^2  \over 4\pi L}\; \partial_\mu \phi^a &=& F_{\mu 3}^a,~~  \mu,\nu = 0,1,2 ,\nonumber \\
{g^2 \over 4 \pi L}\; \epsilon_{\mu\nu\lambda} \partial^\lambda \sigma^a &=&  F_{\mu\nu}^a.
\end{eqnarray}
Here $F_{\mu 3}^a$  denotes the mixed $\R^3$-$\S^1$ component of the Cartan field strength tensor of the 4D theory, $F_{\mu\nu}^a$ is the field strength along $\R^3$, all taken independent of the $\S^1$-coordinate $x^3$, and $g^2$ is the 4D SYM gauge coupling at a scale of order\footnote{See the discussion around Eqn.~(\ref{lambdascale}) for more detail on the renormalization scale.} $1\over L$.
Physically, Eqn.~(\ref{fieldrelation}) implies that spatial derivatives of $\phi^a$ are 2D duals of the 4D magnetic field's components along  $\R^2$;  likewise, spatial derivatives of $\sigma^a$ are 2D duals of the 4D electric field's  components along  $\R^2$.\footnote{\label{gradient1}For example, labeling the spacetime coordinates as ($0,1,2$)$=$($t, y,z$), as we do later in this paper (reserving $x$ for the complex field (\ref{fields1})),  the spatial gradients  ${g^2 \over 4 \pi L} \partial_z \sigma^a = - F_{0 y}^a =  - E_y^a$ and $ {g^2 \over 4 \pi L} \partial_y \sigma^a = F_{0z}^a =  E_z^a$ are 2D duals to (or a $90^o$ rotation of) the  electric field.}

As indicated in (\ref{fields1}), the target space of the  dual photons fields is the unit cell of the $SU(N)$ weight lattice spanned by $\bm w_k = (w_k^1, ..., w_k^{N-1})$, $k=1\ldots N-1$, the fundamental weights.\footnote{Vectors in the Cartan subalgebra will be sometimes denoted by bold face: $\bm \sigma = (\sigma^1, ..., \sigma^{N-1})$, $\bm \phi = (\phi^1, ..., \phi^{N-1})$ and the complex field $\bm x = (x^1, ..., x^{N-1})$ and similar for their complex conjugates $\bar{\bm x}$. The dot product used throughout is the usual Euclidean one.  We shall interchangeably also use a notation $\vec\sigma$ instead of $\bm \sigma$ (and likewise for group-lattice vectors)  to denote these Cartan subalgebra vectors.} A simple way to understand the $\sigma$-field periodicity is that it  allows for non vanishing monodromies corresponding to the insertion of probe electric charges (quarks) of any  $N$-ality, as appropriate in an $SU(N)$ theory.\footnote{\label{dualfootnote}Explicitly, from (\ref{fieldrelation}), the monodromy of the $\bm \sigma$ field around a spatial loop $C \in \R^2$ (the $y$-$z$ plane), taken in the counterclockwise direction, is, within our conventions, $\oint d \sigma^a = -{4 \pi L \over g^2 } \oint \vec E^a \cdot d \vec n$, where $\vec n$ is an outward normal to $C$ and the arrows denote vectors in $\R^2$. Thus, the monodromy $\oint d \bm \sigma$ measures the total electric charge bounded by $C$. It does, in fact, equal $2 \pi \bm \lambda$, where $\vec\lambda$ is the vector of $U(1)^{N-1}$ charges inside $C$. See Eqn.~(\ref{eq:eom}) and Figure \ref{fig:01} below.}

\subsection{EFT}
\label{sec:2.1}

At small $L N \Lambda \ll 1$, the bosonic part of the long distance theory is described by the weakly-coupled $\R^{1,2}$ Lagrangian:
\begin{equation}
\label{lagrangian1}
L_{\rm EFT} = M \; \partial_\mu { x}^a K_{ab} \; \partial^\mu { \bar{x}}^b - M\; {  m^2  \over 4}\; {\partial W({\bm x}) \over \partial x^a} \; K^{ab}\; {\partial  \bar W(\bar{\bm x}))  \over \partial \bar{x}^{b}}  ~.
\end{equation}
Here, $W(\bm{x})$ is the holomorphic superpotential:
\begin{equation}
\label{lagrangian2}
W({\bm x}) = \sum_{A=1}^N e^{\bm\alpha_A \cdot {\bm x}}~,
\end{equation}
 $\bm\alpha_{1}, \ldots, \bm\alpha_{N-1}$ are the simple roots, and $\bm\alpha_N = - \sum\limits_{A=1}^{N-1} \bm\alpha_A$ is the affine (or lowest) root of the $SU(N)$ algebra.\footnote{Roots are normalized to have square length $2$; roots and coroots for $SU(N)$ are identified, and $\bm \alpha_A \cdot \bm{w}_B = \delta_{AB}$, $A,B=1, \ldots N-1$.}  The K\" ahler metric and its inverse are denoted by 
$K_{ab}$ and $K^{ab}$ in (\ref{lagrangian1}).\footnote{Because our K\" ahler metric (\ref{oneloopkahler},\ref{corrntokahler}) is real, symmetric, and constant, we do not distinguish between holomorphic and antiholomoprhic indices.}

 As opposed to the superpotential $W$, whose form can be determined by holomorphy and symmetries, the K\" ahler potential is under control only at small $L$. In fact, only one term in its weak-coupling expansion at $\Lambda N L \ll 1$ has been computed. The expression for the inverse K\" ahler metric is\begin{eqnarray}
\label{kahler}\label{oneloopkahler}
K^{ab} &=& \delta^{ab} + {3 g^2 N\over 16 \pi^2} \; k^{ab} ~.
\end{eqnarray}
where the one-loop correction is $k^{ab}$.
Physically, the matrix $k^{ab}$ determining the inverse K\" ahler metric represents the mixing of the electric Cartan photons due to the non-Cartan ``$W$-boson" and superpartner loops. The  form of the inverse K\" ahler metric $K^{ab}$ (\ref{kahler}) is justified in the semiclassical $L N \Lambda \ll 1$ limit and the leading correction $k^{ab}$  was computed in various ways in \cite{Poppitz:2012sw,Anber:2014sda,Anber:2014lba}.
The Lagrangian (\ref{lagrangian1}) in terms of  the dual photon fields is obtained after a linear-chiral superfield duality transformation, which takes into account the 
one-loop mixing. The duality is performed in an expansion in powers of $g^2$ and is described in  detail in \cite{Anber:2014lba}.\footnote{For use below, we only note that when the Cartan-photon mixing is accounted for, the photon-dual photon duality relation (\ref{fieldrelation}) is replaced by
\begin{eqnarray}\label{dualitymixed}
K^{ab} F^b_{\mu\nu} = {g^2 \over 4 \pi L}\epsilon_{\mu\nu\lambda} \partial^\lambda \sigma^a.
\end{eqnarray}} 
The one-loop correction to the (inverse) K\" ahler metric is currently known at the center symmetric point:
\begin{eqnarray}\label{corrntokahler}
k^{ab} &=&{ 1\over N} \sum\limits_{1 \le A < B \le N} (\vec\beta^{AB})^a (\vec\beta^{AB})^b \left(\psi({B-A \over N}) + \psi(1 - { B-A \over N}) \right).
\end{eqnarray}
Here  $\psi$ is the logarithmic derivative of the gamma function and $\vec\beta^{AB}$ denote the positive roots of $SU(N)$. These are $N-1$ dimensional vectors in the root lattice (whose components are denoted $(\vec\beta^{AB})^a$, $a=1,...,N-1$); recall that the simple roots $\vec\alpha_A= \vec\beta^{A, A+1}$, $A = 1, ...N-1$ are a subset of the positive roots. We note that (\ref{corrntokahler}) is a purely group theoretic factor; its implications will be discussed after we introduce the relevant scales.

\subsection{Scales}
\label{sec:2.2}

 The scales appearing in the long-distance theory (\ref{lagrangian1}) are determined by the microscopic dynamics of the underlying 4D SYM theory on $\S^1$ of size $L$. We begin with the strong coupling scale $\Lambda$ of the 
$SU(N)$ theory
\begin{eqnarray}\label{lambdascale}
\left({\Lambda L \over 4 \pi}\right)^3 = {16 \pi^2 \over 3 N g^2} \;e^{- {8 \pi^2 \over N g^2}}~.
\end{eqnarray}
Here, and everywhere in this paper, $g$ denotes the 4D $SU(N)$ gauge coupling taken at the scale $4\pi \over L$ (of order the mass of the heaviest $W$-boson). The mass of the lightest $W$-boson is $m_W = {2 \pi \over LN}$ and the condition for validity of our EFT, usually stated as $\Lambda L N \ll 1$, can more precisely be phrased as  ${\Lambda } \ll m_W = {2 \pi \over LN}$.

All scales appearing in the long-distance theory (\ref{lagrangian1})  can be expressed in terms of $\Lambda$ and $L$ (or $\Lambda$ and $m_W$). The scale $M$ appearing in the kinetic term is
\begin{eqnarray}\label{scaleM}
M = {g^2 \over 16 \pi^2 L}\; \sim  {m_W \over | 3 \log {\Lambda L \over 4 \pi}|}(1 + \ldots)~,
\end{eqnarray}
while the nonperturbative scale $m \ll M$, which determines the strength of the 
instanton effects generating the superpotential $W$ (see \cite{Poppitz:2012nz} for the $SU(N)$ normalizations) is:
\begin{eqnarray}\label{scalemsquared}
 m &=& 6 \sqrt{2} \; \left( {4 \pi \over LN}\right) \left( {16 \pi^2 \over g^2 N}\right) \left({\Lambda L N \over 4 \pi}\right)^3\; \sim m_W \left({\Lambda L N \over 4 \pi}\right)^3 \left( |3 \log {\Lambda L  \over 4 \pi}| + \ldots \right)~.
\end{eqnarray}
In both (\ref{scaleM}) and (\ref{scalemsquared}), we used (\ref{lambdascale}) to express $g^2N$ through the strong coupling scale $\Lambda$ and $L$, to leading accuracy at $g^2 N \ll 1$ (as indicated by the dots).

The one-loop mixing matrix $k^{ab}$ can be diagonalized using a discrete Fourier transform (as done in unpublished work with A. Cherman), but  for our purposes a numerical evaluation will suffice. As already noted, $k^{ab}$ from (\ref{corrntokahler}) is purely group-theoretic in nature and it is easy to see that it has only  negative eigenvalues, the smallest of which is about $- 6$, for the range of $N$ we study. This means that, upon increasing $g^2 N$, the eigenvalues of $K^{ab}$ decrease away from unity (their vanishing would  signal a strong coupling singularity where our EFT breaks down).  As (\ref{lambdascale}) shows, at fixed strong coupling scale $\Lambda$, an increase of $g^2 N$ is tantamount to increasing $L$. Thus, the decrease of the $K^{ab}$ eigenvalues shows that a strong coupling regime is approached, as expected, in the large-$L$ limit. Conversely, a tiny value of $g^2 N$ corresponds to small $\Lambda LN$ and implies that the corrections $k^{ab}$ are negligible.

We shall begin  our study of confining strings with values of $g^2N$ such that the effect of $k^{ab}$ is negligible, i.e. we shall first take $K^{ab} = \delta^{ab}$. Later on, we  shall also use the one-loop correction (\ref{corrntokahler}) as a probe of the $L$-dependence of the confining strings, while keeping the theory still well in (or at least on the ``safe" side of) the calculable regime.

Following the usual terminology,  we often call the scale $m$ the ``dual photon mass." We should, however, keep in mind that $m$ is really the mass of the heaviest of the $N-1$ dual photons, whose mass spectrum is given by $m_k \sim m \sin^2 {\pi k \over N}$, $k=1,...,N-1$.\footnote{We cannot  help but mention that the EFT (\ref{lagrangian1}), in addition to providing a semiclassically calculable framework to study confinement, exhibits many other, perhaps more unusual, phenomena.  For example, in the abelian large-$N$ limit \cite{Cherman:2016jtu}, the 3D theory  (\ref{lagrangian1}) becomes effectively four dimensional by generating  an emergent latticized dimension, with many features similar to $T$-duality of string theory. Further, as shown in  \cite{Aitken:2017ayq,Anber:2017ezt}, the theory generates doubly exponentially small nonperturbative scales, $\sim e^{- e^{1/g^2}}$, which are manifested in the existence of exponentially weakly bound states of the dual photons of (\ref{lagrangian1})---the 3D remnant of 4D glueball supermultiplets.} The widths of DWs are generally determined by the lightest dual photon mass.

\subsection{Symmetries and Vacua}
\label{sec:2.3}

Of special interest to us here are the discrete chiral and center global symmetries of SYM:
\begin{enumerate}
\item A 0-form chiral symmetry, $\Z_{2 N}^{(0)}$. This is the usual discrete $R$-symmetry of SYM that  acts on the fermionic superpartners of (\ref{fields1}) (and on the superpotential),  by a phase rotation. However, of most relevance to us is that  it also shifts the dual photons:\footnote{Notice that $\Z_{2N}$ acts as $\Z_{N}$ on the dual photon and the superpotential.}
\begin{eqnarray}\label{chiral}
\Z_{2 N}^{(0)}: ~\bm \sigma &\rightarrow& \bm \sigma + {2 \pi \bm \rho \over N}~,  \\
e^{\bm\alpha_a \cdot \bm{x}} &\rightarrow& e^{i {2\pi \over N}} e^{\bm\alpha_a \cdot \bm{x}} ~,~ a = 1, \ldots, N,\nonumber\end{eqnarray}
where we also show the action of  $\Z_{2 N}^{(0)}$ on the terms appearing in the superpotential (\ref{lagrangian2}). Here, $\bm \rho$ denotes the Weyl vector, $\bm\rho = \bm{w}_1 + \ldots +\bm{w}_{N-1}$. As we shall see, this symmetry is spontaneously broken to $\Z_2$ (fermion number), leading to the existence of $N$ ground states and DWs between them.
\item A ``0-form" center symmetry
 $\Z_N^{(1),\S^1_L}$. The notation is chosen to emphasize that this is the dimensional reduction of the    $\S^1_L$-component of the 4D 1-form center symmetry. As explained in detail in 
 \cite{Anber:2015wha,Cherman:2016jtu,Aitken:2017ayq}, the  0-form center symmetry  acts on the dual photons and their superpartners:
 \begin{eqnarray} 
 \Z_N^{(1),\S^1_L}: ~ \bm{x} &\rightarrow& {\cal P} \bm{x},\label{zeroform0} \\
 e^{\bm\alpha_a \cdot \bm{x}} &\rightarrow&  e^{\bm\alpha_{a+1 ({\rm mod} N)} \cdot \bm{x}}~ \label{zeroform1},
 \end{eqnarray}
 and is an exact unbroken global symmetry of the theory.
 
In a basis independent way, the operation denoted by $\cal{P}$  in (\ref{zeroform0})  is the product of Weyl reflections with respect to all simple roots \cite{Anber:2015wha},   while in the standard $N$-dimensional basis for the weight vectors it is a $\Z_N$ cyclic permutation of the vector's components \cite{Cherman:2016jtu,Aitken:2017ayq}. This clockwise  action is evident from the way  $\Z_N^{(1),\S^1_L}$ acts on the  $e^{\bm\alpha_a \cdot \bm{x}}$ terms in the superpotential, eqn. (\ref{zeroform1}). \item A 1-form center symmetry $\Z_N^{(1),\R^3}$, acting on Wilson  line operators in $\R^3$  by multiplication by appropriate $\Z_N$ phases. 
 \end{enumerate}

 The 0-form and 1-form center symmetries discussed above are parts of the  $\Z_{N}^{(1), \R^4}$ 1-form center of the $\R^4$ theory. 
As recently realized \cite{Gaiotto:2014kfa}, SYM on $\R^4$ has a mixed 't Hooft anomaly between the 0-form  $\Z_{2N}^{(0)}$ chiral symmetry and the $\Z_{N}^{(1), \R^4}$ 1-form center symmetry. This 't Hooft anomaly persists upon an $\R^3 \times \S^1$ compactification and anomaly matching is saturated by the spontaneous breaking of the discrete chiral symmetry,  $\Z_{2N}^{(0)} \rightarrow \Z_2^{(0)}$, as in (\ref{vacua}) below. Anomaly inflow on the resulting DWs implies that the DW worldvolume is nontrivial, supporting the phenomena of quark deconfinement (and braiding of Wilson lines, for DWs with three-dimensional worldvolume). Aspects of this inflow has been  studied in many works \cite{
Gaiotto:2017yup,Gaiotto:2017tne,Argurio:2018uup,Bashmakov:2018ghn,Anber:2018jdf,Anber:2018xek}. In the semiclassical small-$L$ setup of this paper, anomaly inflow and the deconfinement of quarks on DWs were studied in \cite{Cox:2019aji}. For a recent discussion of these symmetries see also \cite{Poppitz:2020tto}.

As already stated, our SYM  EFT (\ref{lagrangian1},\ref{lagrangian2}) has $N$ vacua, determined by solving for the stationary points of the superpotential $W(\bm x)$. These are labelled by $k= 0, \ldots, N-1$. In the $k$-th vacuum, the dual photon field $\bm\sigma$ has a nonzero expectation value $\langle \bm{\sigma} \rangle_k$, while $\bm \phi$ is fixed at the center symmetric point:
\begin{eqnarray}
\label{vacua} 
\langle \bm{\sigma} \rangle_k &=& { 2 \pi k \over N} \bm\rho,\; \langle\bm\phi\rangle_k = 0,\; {\rm such \; that} \; \\
 \langle e^{\bm\alpha_a \cdot \bm{x}} \rangle_k  &=&  e^{i {2 \pi k \over N}}, a = 1,\ldots N,\; {\rm and } ~
\langle W \rangle_k \equiv W_k  =  N e^{i {2 \pi k \over N}}~. \nonumber
\end{eqnarray}
 We also showed the expectation value $W_k$ of the superpotential in the $k$-th ground state. The $N$ vacua (\ref{vacua}) are interchanged by the action of the spontaneously broken $\Z_{2 N}^{(0)} \rightarrow \Z_2^{(0)}$ symmetry (\ref{chiral}), while the $\Z_N^{(1),\S^1_L}$ symmetry is unbroken.\footnote{This follows from ${\cal{P}} {2 \pi \bm\rho \over N} = {2\pi \bm\rho \over N} - 2 \pi \bm{w}_1$. In words,  the action of the 0-form center $\Z_N^{(1),\S^1_L}$ on the vacua (\ref{vacua}) is a weight-lattice shift of $\langle\bm\sigma\rangle_k$, which is an identification, as per (\ref{fields1}).} The 1-form $\Z_N^{(1),\R^3}$ symmetry is also unbroken in the bulk of SYM, corresponding to the confinement of quarks.
 
 There are DWs between the various $N$ vacua. A $k$-wall is one that connects vacua $k$ units apart, i.e. $\bm\sigma$ changes from $\langle \bm\sigma\rangle_0$ to $\langle\bm\sigma\rangle_k$. The BPS $k$-walls are the ones of lowest tension among all DW connecting the same two vacua. 
  The tension of a BPS $k$-wall depends only on the asymptotics of the superpotential on the two sides of the DW, $W_q$ and $W_{q+k ({\rm mod} N)}$, and is independent on the details of the K\" ahler metric, see ref.~\cite{Hori:2000ck}. The  BPS DW tensions in SYM are
  \begin{eqnarray}\label{dwtension}
 T_{k-{\rm wall}} = M m \; 2 N \sin {\pi  k \over N}~, ~ k=1, \ldots, N-1~. 
\end{eqnarray}
 These BPS $k$-walls and their tensions
 play an important role in the constructions of candidate minimum-tension double-string confining configurations.

\section{Relation To (and Differences From) Confinement in the $\R^3$ Polyakov Model}
\label{sec:polyakovsection}

Before we continue our detailed study of confinement in $SU(N)$ SYM, we pause to discuss the similarities and differences between confinement in SYM on $\R^3 \times \S^1$ and the well-known 3D Polyakov model. This discussion is intended for the reader who is familiar with the Polyakov model, but has not delved into the intricacies of confinement in abelianized gauge theories on $\R^3 \times \S^1$.

In this Section, we outline the difference between confinement in the $SU(2)$ Polyakov model on $\R^3$ and $SU(2)$ SYM on $\R^3 \times S^1$. The discussion of $SU(2)$ suffices to show the necessity to consider double-string configurations even for fundamental quarks in the simplest setting, without much of the Lie-theoretic technology necessary to study $SU(N)$ SYM.

{\flushleft{\bf $SU(2)$ Polyakov model on $\R^3$:} }The 3D Polyakov model is a nonsupersymmetric $SU(2)$ gauge theory with an adjoint Higgs field, breaking the gauge group to $U(1)$. Like the SYM theory discussed earlier, the long-distance physics is described by a single dual photon (in SYM, there are additional fields, the dual photon's superpartners). The dual photon is a dimensionless field, whose periodicity is in the weight lattice, $\sigma \equiv \sigma + 2 \pi w_1$, where $w_1 = 1/\sqrt{2}$ is the fundamental weight of $SU(2)$ and the simple root of $SU(2)$ is $\alpha_1 = \sqrt{2}$ (in order to have a unified notation with the rest this paper, we keep these unusual normalizations for $SU(2)$). 
Thus the $U(1)$ charge of a fundamental quark is $w_1 = \nu_1 = - \nu_2 = 1/\sqrt{2}$, while that of a $W$-boson (heavy gluon) is $\alpha_1 = \sqrt{2}$.\footnote{Usually one takes the charge of  W-bosons to be $\pm 1$ and the charges of fundamental quarks to be $\pm 1/2$; clearly the different normalization does not affect the physics. We stick with using roots and weights for $SU(2)$ in order to emphasize the fact that our  comparison of the $SU(2)$ Polyakov model and $SU(2)$ SYM  generalizes to $SU(N)$. } Thus, in our normalization of the dual photons from Footnote \ref{dualfootnote}, the monodromy (denoted by $\Delta\sigma_{n F}$) of the dual photon $\sigma$ around a charge equal to $n$ times the fundamental charge $(n= \pm1, \pm 2,...)$ is:
\begin{equation} \label{monodromysu2}
\Delta \sigma_{n F} =2 \pi n w_1 =  {2 \pi n \over \sqrt{2}}  , n \in \Z~.
\end{equation}
With this preamble and normalization, 
the long-distance effective Lagrangian, first written by Polyakov \cite{Polyakov:1976fu}, is:\footnote{We have shifted the vacuum energy of  $L_{Polyakov}$ to zero and have written  the lagrangian up to a normalization constant in the kinetic term (a number of order unity, not essential for our discussion); further, a power-law dependence on $v/g_3^2$ of the potential term is absorbed into the coefficient $c$. Here $g_3$ is the  3D gauge coupling of mass dimension $1/2$ and $v \gg g_3^2$ is the scale of breaking $SU(2) \rightarrow U(1)$. 
The potential terms are due to a dilute gas of  instantons ($\sim e^{i \alpha_1 \sigma}$)
 and anti-instantons ($\sim e^{-i \alpha_1 \sigma}$) of action $\sim v/g_3^2 \gg 1$, as shown on the first line in (\ref{polyakov1}).}
\begin{eqnarray}\label{polyakov1}
L_{Polyakov} &=& g_3^2 \; \partial_\mu \sigma \partial^\mu \sigma - c v^3 e^{- {{\cal{O}}(1) v \over g_3^2} }(2 - e^{i \alpha_1 \sigma} - e^{- i \alpha_1 \sigma})~, \\
&=& g_3^2 \; \partial_\mu \sigma \partial^\mu \sigma - 2 c v^3 e^{- {{\cal{O}}(1) v \over g_3^2} }  (1 - \cos \alpha_1 \sigma)~, ~ {\rm where}~ \sigma \equiv \sigma + 2 \pi w_1,   ~w_1 \alpha_1 = 1 .\nonumber 
\end{eqnarray}
The Polyakov model (\ref{polyakov1}) has a single vacuum at $\alpha_1 \sigma = 0$. The periodic nature of the dual photon, $\sigma \equiv \sigma + 2 \pi w_1$, and the relation $w_1 \alpha_1 = 1$ imply that the vacuum at $\alpha_1 \sigma = 2 \pi$ is equivalent to the one at $\sigma = 0$ (as are all periodic repetitions thereof). Domain wall-like configurations interpolating between these two values of $\sigma$ have a jump  $\Delta \sigma =  {2\pi \over \sqrt{2}}$ upon crossing the wall. Thus, from (\ref{monodromysu2}), we conclude that the jump of $\sigma$ across the ``domain wall" means that the ``DW" configuration carries flux appropriate to a flux tube confining a single fundamental quark/anti-quark pair. 

If one considers, within the semiclassical regime, higher charges in the Polyakov model, obtaining confining flux configurations with $n>1$, as per (\ref{monodromysu2}), clearly requires double- as well as multiple-string configurations. For example, confining a massive $W$-boson would require two $n=1$ ``DWs" (as in the ``gluon chain" model \cite{Greensite:2001nx}); naturally, if stretched too long these adjoint  strings will necessarily decay via $W$-boson pair production as discussed already in \cite{Ambjorn:1999ym}, see also our related discussion of the breakdown of the abelianized EFT in Section \ref{sec:6.5}. 

Our main point, as we discuss below for $SU(2)$, and in the rest of the paper for $N>2$, is that in SYM on $\R^3 \times \S^1$, one has to consider double-string configurations even for fundamental quarks. This is, in fact, necessitated by the new mixed discrete chiral/center symmetry anomaly, as discussed already in \cite{Cox:2019aji}.

 {\flushleft{\bf $SU(2)$ SYM on $\R^3 \times \S^1$:} } Let us now contrast Polyakov's effective Lagrangian (\ref{polyakov1}) with the long distance Lagrangian of SYM on $\R^3\times \S^1$.
 The latter, setting the $\phi$ field to its vanishing vev (recall (\ref{vacua})), has the form
 \begin{eqnarray}\label{sym1}
L_{SYM} &=& {g_4^2 \over L}\; \partial_\mu \sigma \partial^\mu \sigma - {c' \over L^3} e^{- {{\cal{O}}(1)  \over g_4^2} }(1 - \cos 2 \alpha_1 \sigma)~,~ {\rm where}~ \sigma \equiv \sigma + 2 \pi w_1,   w_1 \alpha_1 = 1.
\end{eqnarray}
Eq. (\ref{sym1}) is obtained from our general formula for $SU(N)$ SYM (\ref{lagrangian1}), taken for $N=2$ and with $W$ from (\ref{lagrangian2}), after setting $\phi = 0$. Apart from the appearance of the 4D gauge coupling and the compactification radius $L$, the lagrangians (\ref{sym1}) and (\ref{polyakov1}) are very similar. There is, however, one striking difference between (\ref{sym1}) and (\ref{polyakov1}): the factor of 2 inside the cosine potential. This innocent-looking factor is due to the nature of the confining objects---the ``magnetic bions", composite instantons made of BPS monopole-instantons and KK antimonopole-instantons \cite{Unsal:2007jx}, which have magnetic charge $2 \alpha_1$ (rather than $\alpha_1$, as the monopole-instantons in the Polyakov model). 

Before we continue with the description of confining strings via (\ref{sym1}), let  us note that if we take $L \rightarrow 0$, we arrive to the 3D version of SYM, rather than to the Polyakov model (\ref{polyakov1}). The difference between the two is elucidated in, e.g. \cite{Aharony:1997bx}: the SYM on $\R^3$ has a ``runaway vacuum", a peculiarity occuring in supersymmetric theories. To see this, one has to study the behaviour of the $\phi$-field, the bosonic superpartner of $\sigma$, which has been set to its vev in (\ref{sym1}). The expectation value of $\phi$, which is not a compact field in the $\R^3$ limit, as it is not related to the $\S^1$ holonomy, is then seen to run off to infinity.

Continuing with the confining strings, we note that the Lagrangian (\ref{sym1}) implies that SYM has two vacua in the fundamental domain for $\sigma$ ($\sigma \equiv \sigma + 2 \pi w_1$): at $\alpha_1 \sigma=0$ and $\alpha_1 \sigma = \pi$. Notice that these two vacua are not related by a $2 \pi w_1$ shift of $\sigma$, thus they are not equivalent (however, the vacuum at $\alpha_1 \sigma = 2 \pi$ is equivalent to the one at $\sigma =0$ by the weight-lattice periodicity of $\sigma$). Rather, these two vacua correspond to the spontaneous breaking of the $\Z_4 \rightarrow \Z_2$ chiral symmetry of SYM (\ref{chiral}). The domain wall  between the vacua at $\sigma = 0$ and $\sigma = \pi/\sqrt{2}$ thus has a jump   $\Delta \sigma = \pi/\sqrt{2}$. On the other hand, the domain wall between $\sigma =2  \pi/\sqrt{2}$ (equivalent to the $\sigma=0$ vacuum) and  $\sigma = \pi/\sqrt{2}$ also carries   flux $|\Delta\sigma| = \pi/\sqrt{2}$. (Notice that these are distinct DWs between the same vacua and the fluxes they carry are in accordance of our general results for BPS $k$-wall fluxes given in Eq.~(\ref{fluxes1}) of the next Section.)

Thus, the domain walls in the $SU(2)$ SYM theory carry electric flux $\Delta\sigma = \pi/\sqrt{2}$---exactly half the electric flux, eq.~(\ref{monodromysu2}) with $n=1$, needed to confine fundamental quarks. A single DW does not have sufficient flux  to confine a single fundamental quark and we arrive at the necessity to study ``double string" configurations made of two domain walls, even for the confinement of fundamental quarks, in contrast with the Polyakov model.\footnote{We stress that this difference holds for general $SU(N)$ SYM theories, as shown in detail in the rest of the paper.}

It was numerically shown, already in \cite{Anber:2015kea}, that it is double string configurations that confine fundamental quarks in $SU(2)$ SYM on $\R^3 \times \S^1$ (i.e. they do not collapse to a single flux tube). The novelty  of this paper is to make use of the analytical results of \cite{Cox:2019aji} classifying the BPS DW fluxes (and the further insight into their properties) in $SU(N>2)$ SYM to study the double-string confinement mechanism for $SU(N)$ SYM. 

Apart from finding the weakest  $N$-ality dependence, among any known confining theory, the unexpected feature we found is the collapse of the double string confining fundamental quarks for $N\ge 5$, but not for $N=2,3,4$. This has to do with the interactions of the appropriate DWs, which changes from repulsion to attraction upon increase of $N$. As we discuss in Section \ref{sec:5.1}, this is not analytically understood, but has been subjected to many tests (see Section \ref{sec:6}).

We now continue with describing the details of our study for general $SU(N)$ SYM thoeries.

 \section{Taxonomy of DW Fluxes and Double-String Confinement \label{taxonomy}}
 \label{sec:3}
 
 \subsection{Setup}
 \label{sec:3.1}
 
To study confinement, we add a static source term to the action for quarks of weights\footnote{We stress that since the long-distance dynamics is abelian, Wilson loops can be classified by the weights $\vec\lambda$. These are $SU(N)$ weight lattice elements, representing the charges  of the quarks under the unbroken $U(1)^{N-1}$. These abelian sources can be given a fully gauge invariant description using   Wilson loops representing static sources in $SU(N)$ representations $R$,  extending in $\R^{1,2}$, but ``decorated" by the insertion of appropriate powers of the $\S^1$ holonomies. In the abelianized phase where the holonomy has a $\Z_{N}^{(1), \S^1_L}$-preserving expectation value, individual weights can be selected by a discrete Fourier transform, see  \cite{Poppitz:2017ivi} for details.} $\pm \vec\lambda$ at fixed positions $\vec{r}_{1,2} \in \R^2$:
 \begin{eqnarray}
S_{{\rm source}}= \int dt \; \vec{\lambda} \cdot (\vec A_{0}(\vec{r}_1) - \vec A_0(\vec{r}_2))~.
\end{eqnarray}
We denote, recalling footnote \ref{gradient1}, the $\R^2$ coordinates as $(z,y)$, and take $\vec{r}_{2}$$=$$(z$$=$$0$, $y$$=$$0)$ and $\vec{r}_1$$=$$(z$$=$$0$, $y$ $=$ $R)$. Then,  using the duality relation (\ref{dualitymixed}), we find
\begin{eqnarray}
S_{\rm source}&=& -  \int dt \int\limits_{0}^R dy \;\vec\lambda \cdot \vec F_{0y}(0,y) = {g^2 \over 4 \pi L} \; (\vec\lambda)^a K_{ab} \; \int dt   \int\limits_0^R dy\; \partial_z \sigma^b(z, y)\vert_{z=0} \nonumber \\
&=& - {g^2 \over 4 \pi L} (\vec\lambda)^a K_{ab}  \int dt dz dy \; \partial_z \delta(z)  \int\limits_0^R d y' \delta(y-y') \sigma^b(z,y)~.
\end{eqnarray}
We then add the action of the EFT (\ref{lagrangian1}), $S_{\rm EFT} = \int dt dz dy L_{\rm EFT}$, to the source action $S_{\rm source}$ and introduce dimensionless coordinates $\tilde t, \tilde z, \tilde y$, via $t = {\tilde t \over m}, z = {\tilde z \over m}, y= {\tilde y\over m}$, to obtain, after dropping the tilde from the dimensionless coordinates\footnote{Here, $\vec\nabla$ denotes the gradient operator in $\R^2$ and not a vector in the group lattice.}
\begin{eqnarray}\label{fullaction}
S_{\rm EFT+source}&=&{M \over m}\int dt dz dy \left[ K_{ab} (\partial_t x^a \partial_t \bar x^b - \vec\nabla x^a \vec\nabla \bar x^b) - {1 \over 4} {\partial W \over \partial x^a} K^{ab} {\partial \bar W \over \partial \bar x^b} \right. \nonumber \\
&& \left. \qquad \qquad\qquad - 4 \pi  (\vec\lambda)^a K_{ab}  \; \partial_z \delta(z)  \int\limits_0^R d y' \delta(y-y') {\rm Im}[ x^b(z,y)] \right]~.
\end{eqnarray}
We numerically solve the static equation of motion\footnote{This equation implies that $\bm\sigma$ exhibits a $2 \pi \bm \lambda$  jump upon crossing the horizontal ($y$) axis in the positive $z$ direction, for $y$ between $0$ and $R$. See also the original derivation in  \cite{Polyakov:1976fu,Polyakov:1987ez}, also given in \cite{Poppitz:2017ivi}.} following from (\ref{fullaction}) 
\begin{equation} \label{eq:eom}
    \nabla^2 x^a = \frac{1}{4} K^{ab} K^{cd}\; \frac{\partial W}{\partial x^c}   \frac{\partial^2 \bar W}{\partial \bar x^{b} \partial \bar x^{ d}} + 2 \pi i   (\vec{\lambda})^a \partial_z \delta(z) \int_{0}^{ R} dy' \delta(y-y')~.
\end{equation}
Distance scales in the equations above are all dimensionless numbers, which are converted to physical units upon  multiplication by $1/m$, i.e. the (heaviest, at $K_{ab}=\delta_{ab}$) dual photon Compton wavelength, with $m$  defined in (\ref{scalemsquared}). 
Likewise, the dimensionless static energy, 
\begin{eqnarray}\label{dimlessenergy}
\tilde E = \int dz dy \left[ K_{ab}  \vec\nabla x^a \vec\nabla \bar x^b + {1 \over 4} {\partial W \over \partial x^a} K^{ab} {\partial \bar W \over \partial \bar x^b}\right]~
\end{eqnarray} is converted to the physical energy $E = M \tilde E$. A dimensionless string tension (or a DW tension (\ref{dwtension}))  $\tilde T$ computed from the simulation,  is converted to physical units as  $T = M m \; \tilde T$. When we study the effect of changing $L$ at fixed $\Lambda$ (i.e. changing $g^2 N$ in (\ref{oneloopkahler}) according to (\ref{lambdascale})), being mindful of these scalings and the definitions of $M$ and $m$ in terms of $L$ and $\Lambda$  will be important.

 \begin{figure}[h]
  \includegraphics[width= 1 \textwidth]{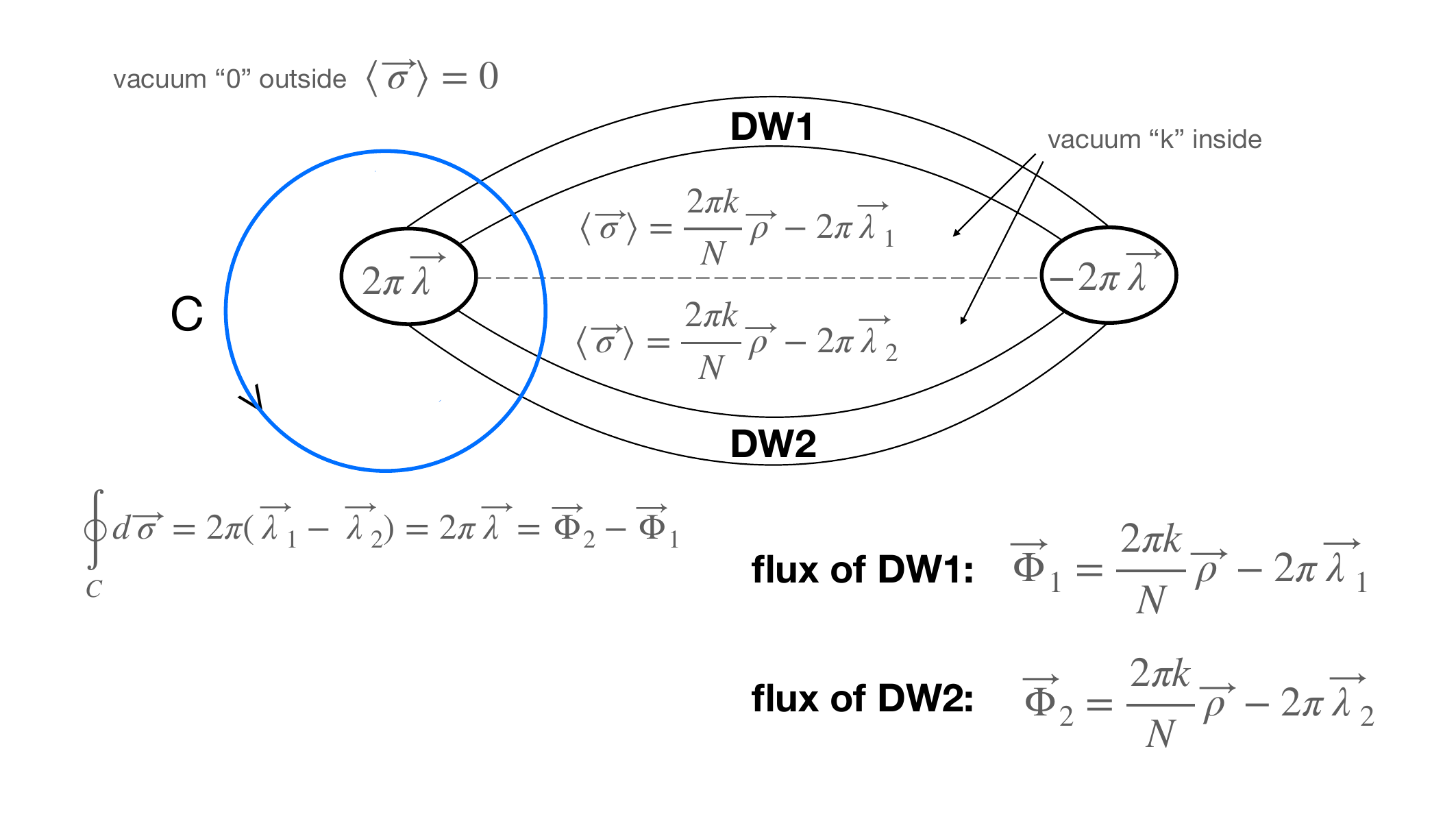}
  \caption{{\bf The double-string confinement mechanism:} The static quark/antiquark sources impose monodromies of  $\vec\sigma$ around contours $C$ surrounding each quark. These monodromies are   weight-lattice elements  $\pm 2 \pi \vec\lambda = 2 \pi n^i \vec{w}_i$, $n_i \in \Z$. We consider, without  loss of generality, strings in the vacuum $\langle\vec\sigma\rangle = 0$. Due to screening by magnetic bions, chromoelectric flux can only penetrate the vacuum in thin flux tubes  instead of spreading into all of $\R^2$. The natural candidates for such flux lines are the DWs of SYM theory, connecting vacua $k$ units apart, $\langle\vec\sigma\rangle = 0$ to $\langle\vec\sigma\rangle = {2 \pi k \over N} \vec\rho$. 
  The quarks' flux, $2 \pi \vec\lambda = 2\pi (\vec\lambda_1 - \vec\lambda_2)$, is then  split between the flux carried by DW$1$, $\vec \Phi_1 = 2 \pi {k \vec\rho\over N} - 2\pi \vec\lambda_1$, and that of DW$2$, $\vec\Phi_2 = 2 \pi {k \vec\rho\over N} - 2 \pi \vec\lambda_2$, where $\vec\lambda_1$ and $\vec\lambda_2$ are weight-lattice vectors.  Assuming that the DWs repel, a double-string configuration like the one shown will then be created.
  Among all DWs connecting vacua $k$ units apart, the ones of least tension are the BPS ones. Ignoring the possibility of   attractive interactions   between the DWs, in the text we describe the candidate lowest tension double-string configurations confining quarks in  representations of any $N$-ality. We argue that quarks with weights in the $k$-index $N$-ality $k$ representation, with $k \le  \lfloor {N\over 2} \rfloor$, are confined by double strings made of BPS $p$-walls with $0 \le p \le k$ (we use $p=0$ to denote appropriate non-BPS walls). Most of these $N^k$ double strings are metastable and are expected to decay into double strings of lowest tension, made of BPS $1$-walls.
 \smallskip
  \newline
  The dotted line inside the double string denotes a weight-lattice discontinuity of the dual photon field, which is a gauge identification. The relation of this picture to deconfinement on DWs can be seen by noting that when the two $k$-walls are BPS, they have equal tensions. Thus, upon ``opening up" the picture, we find that quarks can move along the DWs with no energy cost, giving a microscopic realization of deconfinement of quarks on DWs between the $0$-th and $k$-th vacua, as argued for by anomaly inflow.
  }
  \label{fig:01}
\end{figure}

\subsection{BPS DW Fluxes and Double-String Configurations}
\label{sec:3.2}

The double-string confinement mechanism is illustrated on Fig.~\ref{fig:01} and is explained in more detail in the caption. 
As already alluded to, the confining string consists of two DWs between the $N$ vacua of SYM. Its essence is that due to confinement, the chromoelectric flux of color sources can only propagate along one dimensional lines, the DWs of SYM theory. As explained at length below, the fluxes carried by the DWs are such that one needs two DWs in order to absorb the quark fluxes.

We shall  consider quark sources in different representations of $SU(N)$ and discuss what kind of double-string configurations are expected. But first, we remind the reader of the results of \cite{Cox:2019aji}, which are essential for our arguments. In that paper, the electric fluxes carried by BPS $k$-walls in SYM theory were determined. It has long been known, see \cite{Hori:2000ck}, that there are $N \choose k$ BPS walls between vacua $k$ units apart. Ref. \cite{Cox:2019aji} showed that the BPS $k$-walls fall into two classes, labeled by $I$ and $II$ below, which have monodromies of $\vec\sigma$, or electric fluxes $\bm\Phi^{k, I}_{i_1,...,i_k}$ and $\bm\Phi^{k, II}_{i_1,...,i_{k-1}}$,  given by:  \begin{eqnarray}
\bm\Phi^{k, I}_{i_1,...,i_k}&=& 2 \pi {k \over N} \bm\rho -  2 \pi (\bm{w}_{i_1} + \ldots + \bm{w}_{i_k}),\; {\rm all} \; k \; \bm{w}_{i} \; {\rm different};   \;{  \left( \begin{array}{c}N-1\cr k\end{array} \right)} \; {\rm   walls }, \label{fluxes1}\\
\bm\Phi^{k, II}_{i_1,...,i_{k-1}}&=& 2 \pi {k \over N} \bm\rho -   2 \pi (\bm{w}_{i_1} + \ldots + \bm{w}_{i_{k-1}}),\; {\rm all} \; k-1 \; \bm{w}_i \; {\rm different};  \;{ \left( \begin{array}{c}N-1\cr k-1\end{array} \right)} \; {\rm walls}.\nonumber \end{eqnarray} 
The total number of BPS $k$-walls is, as already stated, ${N \choose k} = {N-1 \choose k} + {N-1 \choose k-1}$. We stress that this is because
 the weights  $(\vec w_{i_1}, ..., \vec w_{i_k})$ are to be taken all different,  with indices  ranging from $1$ to $N-1$; likewise all weights in the second set, $(\vec w_{i_1}, ..., \vec w_{i_{k-1}})$, are to be taken different.\footnote{For $k=1$, the   second set in  (\ref{fluxes1}), $\bm \Phi^{1,II}$,  consists of a single wall carrying flux $ \frac{2 \pi \bm \rho}{N}$, hence the total number of $k=1$ walls is $N$.}  The above spectrum of BPS $k$-wall fluxes is invariant under $k \rightarrow N-k$ up to reversal of the overall sign of the electric flux (a parity transformation, as in \cite{Bashmakov:2018ghn}). 

\subsubsection{Quarks in the Highest Weight of the $\mathbf{k}$-index Antisymmetric Representation} \label{sec:3.2.1}

We shall now use the results for the BPS wall tensions to propose double-string configurations confining quarks of various weights.
We begin by first concentrating on quark sources of $N$-ality $k$ and of weights $\vec w_k$---the highest weight of the $k$-index antisymmetric representation. One result (ignoring DW interactions) that we can immediately obtain is that quarks of weights $\vec w_k$ can be confined by double-string configurations made out of BPS $1$-walls only, for all $k$.  This is because the fluxes of BPS $1$-walls are ${2 \pi \over N} \vec\rho - 2 \pi \vec w_p$, $p=1,...,N$, with $\vec w_N \equiv 0$. We can then take a configuration   where DW1 is a $1$-wall of flux $\vec\Phi^{1, I}_k \equiv {2 \pi \over N} \vec\rho - 2 \pi \vec w_k$ and DW2 is of flux $\vec\Phi^{1, II} \equiv {2 \pi \over N} \vec\rho - 2 \pi \vec w_N = {2 \pi \over N} \vec \rho$, as per (\ref{fluxes1}). As shown on the figure, the total monodromy is $\vec\Phi^{1, II} - \vec\Phi^{1, I}_k = 2 \pi \vec w_k$, as required for a string confining quarks of weight $\vec w_k$. Since the BPS $1$-walls are the ones of lowest tension (recall (\ref{dwtension})), we expect that the minimum energy confining configuration for quarks in an $N$-ality $k$ representation will be due to the double-string configuration made of two BPS $1$-walls. To determine the actual $N$-ality dependence of the string tensions, one requires a numerical evaluation  of the energy of the double-string configuration at large quark separation, which will incorporate the domain wall interactions. 
 
For $N$-ality $k$ quarks of weights different from $\vec w_k$, we shall argue below that
it is not possible (for every weight) to engineer a double-string configuration made out of BPS $1$-walls. In fact, as we shall see below it is not possible to construct a double-string configuration made of BPS $p$-walls for all  weights of a given $N$-ality. Thus,  non-BPS DWs become important for some weights. However, since any $N$-ality $k$ weight is related to the highest weight $\vec w_k$ by the addition of root vectors, it is clear that at sufficiently large quark separation, any such double-string configuration  made of non-BPS walls (or of BPS $p$-walls with $p>1$) will decay to the BPS $1$-wall double-string configuration by the creation of $W$-boson and superpartner pairs. The screening of any $N$-ality $k$ weight down to $\vec w_k$ is possible because $W$-boson charges take values in the root lattice.

The fact that the ``long distance" EFT (\ref{lagrangian1}) describing an abelian confining theory breaks down when distances between probe quarks are taken sufficiently large has been noted before: see \cite{Greensite:2011zz} and \cite{Greensite:2014gra} for a more recent discussion and references. The  screening of static sources by the pair creation of massive ``$W$-bosons" is responsible for ``restoring justice" and ensuring that asymptotic string tensions are only $N$-ality dependent. The importance of this screening, and the associated EFT breakdown, is not new and holds for other models with abelian confinement, as discussed for Seiberg-Witten (SW) theory \cite{Douglas:1995nw} and deformed Yang-Mills (dYM) theory \cite{Poppitz:2017ivi}.

{\flushleft Below, we  consider in some detail how quarks of $N$-ality $k$,  but of general weights, i.e. weights different from the fundamental weight $\vec w_k$ are confined. }

\subsubsection{Quarks of General Weights of N-ality ${\mathbf{k=1}}$}
\label{sec:3.2.2}

Fundamental quarks of any weight $\vec \nu_A$, $A = 1...,N$, of the fundamental representation (with $\vec\nu_1 = \vec w_1$) are all confined by BPS $1$-walls only and have the same tension. As we stress in (\ref{zeroform2}) below, this is due to the unbroken 0-form center symmetry (which holds also in dYM but not in SW theory). It can also be seen from the fact that $\vec\nu_A = \vec w_A - \vec w_{A-1}$, with $\vec w_{0} = \vec w_N = 0$, that the difference between two appropriately chosen BPS $1$-wall fluxes from (\ref{fluxes1}) is always a weight of the fundamental representation. Explicitly,  with $\vec\Phi^{1, I}_A$ taken from (\ref{fluxes1}), we have that
\begin{eqnarray}\label{Nality1}
2 \pi \vec \nu_A =2\pi \vec w_A - 2\pi \vec w_{A-1} =\vec\Phi^{1,I}_{A-1} - \vec \Phi^{1, I}_{A} ,
\end{eqnarray}
showing that for all weights of $N$-ality 1, quarks are confined by double strings of lowest tension made of two BPS 1-walls (when $A=1$ or $N$, one of the weights must be taken $\vec\Phi^{1,II}$ from (\ref{fluxes1})). Note that we could have also used BPS 2-walls of fluxes $\vec\Phi^{2,II}_A$ (from the second line of from (\ref{fluxes1})) instead of  $\vec\Phi^{1, I}_A$  to engineer a double-string configuration confining fundamental quarks;  however, these are expected to have higher tension due to the higher tension of the BPS 2-walls.

Finally, we stress that the fact that all string tensions of $N$-ality 1 are the same is because the unbroken  $\Z_N^{(1),\S^1_L}$ zero-form center symmetry (\ref{zeroform1}) permutes all the weights of the fundamental representation 
\begin{eqnarray} 
 \Z_N^{(1),\S^1_L}: ~ \vec{\nu}_A &\rightarrow& {\cal P} \vec{\nu}_A = \vec{\nu}_{A+1 ({\rm mod} N)}, \label{zeroform2} \end{eqnarray}
 and relates the strings confining the different weights of fundamental quarks to each other.
This $\Z_N$ symmetry action (\ref{zeroform2}) will be important in our discussions of higher $N$-alities below.

\subsubsection{Quarks of General Weights of N-ality $\mathbf{k=2}$}
\label{sec:3.2.3}

 We next consider the two-index representation obtained by taking two fundamental quarks close to each other. The  weights of the resulting two-index representations are $\vec\nu_A + \vec\nu_B$. There are a total of $N^2$ such weights split between the two-index symmetric and antisymmetric representations (which share  many weights). Let us arrange the $N^2$ weights into an $N \times N$ matrix with entries $\vec\nu_{A,B} \equiv \vec\nu_A + \vec\nu_B$ (for use below, we take the indices to cyclically ``wrap around", i.e. we take $\vec\nu_{N+1,l}\equiv \vec\nu_{1,l}$ and $\vec\nu_{l,N+1}\equiv \vec\nu_{l,1}$).
Further, because of the $\Z_N$ action (\ref{zeroform2}), it is sufficient to restrict to $A=1$, since all strings confining quarks of weights $\vec\nu_{A, A+p ({\rm mod} N)}$, $p=1,...N$, are related by a symmetry and so should share the same properties, including equal tensions. Thus, we shall say that the weights $\vec\nu_{A,A+p}$, with $A=1,...,N$, form a $\Z_N$ orbit and shall focus on its representative $\vec\nu_{1,1+p}$. Clearly, as $p=1,...N$, there are $N$ such orbits for the $k=2$ two-index representations.

Consider first  the $p=1$ nearest off-diagonal $\Z_N$ orbit of
$\vec\nu_{1, 1+1}$. Quarks with these weights are all confined by double strings made of BPS $1$-walls; one can see this using an expression essentially identical to (\ref{Nality1}). This is because $\vec\nu_{1,2} = \vec w_{2} - \vec w_{0} = \vec w_2$ (recall we defined $\vec w_0=0$)  is proportional to the difference between two BPS $1$-wall fluxes from (\ref{fluxes1}), $\vec\Phi^{1,II} - \vec\Phi^{1, I}_{2}$. 

Further, we take $p>1$ and consider the $p$-nearest off-diagonal $\Z_N$ orbit of $\vec\nu_{1,1+p}$ with $p \ne N$. Now, we argue that the lowest tension double strings for $p>1$, $p \ne N$, can be constructed of two BPS $2$-walls. To see this, we note that 
\begin{equation} \label{porbits}
2 \pi\vec\nu_{1,1+p} = 2 \pi(\vec w_1 - \vec w_{0} + \vec w_{1+p} - \vec w_{p})=\vec\Phi^{2,II}_{p}- \vec\Phi^{2,I}_{1,1+p},
\end{equation}  the difference between two BPS $2$-wall fluxes from (\ref{fluxes1}).
The only two-index $N$-ality 2 case left to consider is the $p=N$ diagonal weights in the $\Z_N$ orbit of  $\vec\nu_{1,1} = 2 \vec w_{1}$. Because the weights determining the fluxes of BPS  walls on the right-hand side of (\ref{fluxes1}) are all to be taken different, one cannot arrange this weight to be a difference of two BPS wall fluxes (our numerical simulations show that quarks of the diagonal weights are confined by non-BPS double-string-like configurations).  The  $N$-ality $k=2$ strings lying in different $\Z_N$ orbits are studied in more detail in Section \ref{sec:6.4}.

 As already stated,  we expect that for $1<p\le N$ all $N$-ality two strings of higher tension are metastable and decay to the lowest tension double strings  made of  BPS 1-walls, the ones  in the $\Z_N$ orbit of $\vec w_2$, by the production of $W$-boson pairs.\footnote{We also recall that adjoint quarks are also confined in the abelian regime. Consider a $W$-boson whose  weight is a simple root $\vec\alpha_a = \vec\nu_a - \vec\nu_{a+1} = 2 \vec w_a - \vec w_{a+1} - \vec w_{a-1}$. This cannot be written as the difference of two BPS $k$-wall fluxes, for any $k$, hence one expects that other configurations will be responsible for confinement of non-Cartan gluons in the abelian regime. See Section \ref{sec:6.5}, where such a triple-string configuration confining a $W$-boson of weight $\vec\alpha_1$ is found.}

\subsubsection{Quarks of General Weights of N-ality $\mathbf{2 < k \le  \lfloor {N\over 2} \rfloor}$}
 \label{sec:3.2.4}

We can now similarly consider $k$-index representations of $N$-ality $k$
 made by putting together $k$ fundamental quarks.\footnote{While focusing on the lowest string tension of a given $N$-ality, it suffices to consider $N$-alities $k \le  \lfloor {N\over 2} \rfloor$. Strings of $N$-ality ${N\over 2} < k < N$ have tensions identical to those of strings of $N$-ality $N-k$, since the BPS wall tensions (\ref{dwtension}) obey $T_k = T_{N-k}$.} Clearly, now there are $N^k$ weights. Similarly to our discussion of $k=2$ above, we label them as
\begin{eqnarray}
\label{nalitykweights}
\vec\nu_{a_1, a_2,...,a_k} = \vec\nu_{a_1} + \ldots + \vec\nu_{a_{k}}~, ~{\rm all} ~ a_i \in 
 \{ 1, \ldots N \}.
\end{eqnarray}
The highest weight of the $k$-index antisymmetric representation is $\vec w_k$ and it lies in the $\Z_N$ orbit of $\vec\nu_{a, a+1, a+2, \ldots, a+k-1}$, where the $a=1$ element is $ \vec w_k = \vec\nu_{1,2, 3,...,k}$, as implied from the above definitions. 

The  $N^k$ weights (\ref{nalitykweights}) of the $k$-index $N$-ality $k$ representations  split into $\Z_N$ orbits, labeled by $k-1$ integers $\bm{n} \equiv (n_1,...,n_{k-1})$ all from $1$ to $N$, 
\begin{eqnarray}
\label{nalitykweights1}
\vec\nu_{a, a + n_1, a+ n_2,...,a + n_{k-1}}, \; a = 1,...,N,
\end{eqnarray}
where $a=1$ and $\bm n = (1,2,3,...,k-1)$ corresponds to $\vec w_{k}$.
Again, without loss of generality, we can focus on the $a=1$ element of a given $\Z_N$ orbit, for which we can write, from the definition (\ref{nalitykweights}) and recalling (\ref{Nality1}):
\begin{eqnarray}\label{nalitykweights2}
\vec\nu_{1, 1 + n_1, 1+ n_2,...,1 + n_{k-1}} = (\vec w_1 + \vec w_{1 + n_1} + \ldots + \vec w_{1 + n_{k-1}}) - (\vec w_{n_1} + \ldots + \vec w_{n_{k-1}})~. \end{eqnarray} 
For generic\footnote{Such that the numbers $\bm n$ are all distinct, non-neighboring, and not $1$, $N-1$, or $N$.}  weights, the above equation implies that 
\begin{eqnarray}\label{nalitykweights3}
2 \pi \vec\nu_{1, 1 + n_1, 1+ n_2,...,1 + n_{k-1}} 
=   \Phi^{k, II}_{n_1, ..., n_{k-1}} - \vec \Phi^{k, I}_{1, 1+n_1,...,1+ n_{k-1}} ~, 
\end{eqnarray} 
meaning that for such weights of an $N$-ality $k$ $k$-index representation, quarks  are confined
by double strings made of two BPS $k$-walls whose fluxes $\Phi^{k, II}_{n_1, ..., n_{k-1}}$ and $\vec \Phi^{k, I}_{1, 1+n_1,...,1+ n_{k-1}}$, as per (\ref{fluxes1}), combine to provide the  monodromy required by (\ref{nalitykweights2}). 

An extreme non-generic case was  already seen: for $\bm n = (1,2,3,...,k-1)$, quarks are confined by double strings made of BPS 1-walls, instead of BPS $k$-walls. More generally, if some of the numbers $n_i$ are neighboring, there are cancellations between the two groups of weights on the right-hand side of (\ref{nalitykweights2}) reducing the ``rank" of the BPS walls required to construct a  double string of the right monodromy. Like in the $k=2$ case, there are also quarks (e.g. $\bm n = (N,N,...,N)$)  that cannot be confined by double-string configurations made of BPS-walls.

\section{Numerical Simulation of  Double-String Configurations}
\label{sec:4}

To study double-string confinement, we numerically solve the static equation of motion (\ref{eq:eom}), reproduced here with the minor modification of symmetrizing the locations of the two quarks:
\begin{equation} \label{eq:eom2}
    \nabla^2 x^a = \frac{1}{4} K^{ab} K^{cd}\; \frac{\partial W}{\partial x^c}   \frac{\partial^2 \bar W}{\partial \bar x^{b} \partial \bar x^{ d}} + 2 \pi i   (\vec{\lambda})^a \partial_z \delta(z) \int_{-R/2}^{+R/2} dy' \delta(y-y')~.
\end{equation}

It is often of interest to also solve for the associated one-dimensional domain walls, whose static eqation of motion can be thought of as discarding the source term ($\vec{\lambda}  \to 0$) and eliminating one of the spatial dimensions ($\nabla^2 \to \frac{d^2}{dz^2}$) in (\ref{eq:eom2}):
\begin{equation} \label{eq:1deom}
    \frac{d^2  x^a}{dz^2} = \frac{1}{4} K^{ab} K^{cd}\; \frac{\partial W}{\partial x^c}   \frac{\partial^2 \bar W}{\partial \bar x^{b} \partial \bar x^{ d}}~.
\end{equation}
Note that both equations should also satisfy the boundary condition that $\bm x$ is equal to one of the vacua at each connected piece of the boundary at infinity.

In what follows, we will describe the numerical methods used to solve these two equations\footnote{From this point on, we will refer to (\ref{eq:eom2}) and (\ref{eq:1deom}) as the 2D EOM and 1D EOM, respectively.} efficiently. The methods are mostly identical to those described in Sections 4 and 5 of \cite{Cox:2019aji}, with some extensions and modifications. Whenever there is overlap, we summarize the main points needed to reproduce the work here, but refer the interested reader to the aforementioned paper for detailed description.

\subsection{Gauss-Seidel Finite-Difference Relaxation}
\label{sec:4.1}

The core method used is the Gauss-Seidel finite-difference method \cite{Numerical_Method}. For the 2D EOM, we first discretize our space into a rectangular grid\footnote{The dimension of the grid in the direction of the quark separation must obviously be larger than $R$. In addition to this constraint, care should be taken to ensure that the grid size is large enough to accommodate the non-constant part of the solution, usually by trial-and-error. However, the edge effect is in general not a big concern in a confining theory such as ours, in contrast with a non-confining theory; see Appendix \ref{appendix:edge}.} with ``pixels" being squares of size $h \times h$. In all cases, we have used $h=0.1$ in units of the dual photon mass $m$.

Next, the Laplacian is converted to a finite difference using the centered-difference approximation:
\begin{equation}
\nabla^2 x^a(z_i,y_i) = \frac{1}{h^2} \left[ x^a(z_{i+1},y_i) + x^a(z_{i-1},y_i) + x^a(z_i,y_{i+1}) + x^a(z_i,y_{i-1})  - 4x^a(z_i,y_i)  \right] + O(h^2).
\end{equation}
Solving for $x^a(z_i,y_i) $, we have
\begin{equation} \label{eq:GSeq}
x^a(z_i,y_i)  = \frac{1}{4} \left[ x^a(z_{i+1},y_i) + x^a(z_{i-1},y_i) + x^a(z_i,y_{i+1}) + x^a(z_i,y_{i-1}) - h^2 \nabla^2 x^a(z_i,y_i)   \right] + O(h^2).
\end{equation}
We solve this large system of coupled linear equations using the Gauss-Seidel method: start from some initial field configuration, which can be any random\footnote{Literally, as we will soon see.} guess, with the only constraint being that the boundary of the grid, which we approximate to be spatial infinity, must equal one of the vacua  (\ref{vacua}) of the theory. Then, we iteratively update the value of the field at each point of the grid (with the exception of the boundary points) by the average of its four neighboring points, with a Laplacian term subtracted:
\begin{equation} \label{eq:GSupdate}
\xnew^a(z_i,y_i)  = \frac{1}{4} \left[ \xold^a(z_{i+1},y_i) + \xold^a(z_{i-1},y_i) + \xold^a(z_i,y_{i+1}) + \xold^a(z_i,y_{i-1}) - h^2 \nabla^2 \xold^a(z_i,y_i)   \right].
\end{equation}
If the process converges, then $\xnew^a \approx \xold^a $, and so (\ref{eq:GSupdate}) becomes the same as (\ref{eq:GSeq}). Since the boundary points were not touched, the boundary value problem is solved. Once the solution is obtained, we can easily numerically compute various quantities of interest, such as numerically integrating\footnote{While integrating, care must be taken to avoid including the discontinuous jump inside the double string, introduced by the monodromy (see Figure \ref{fig:DSexample}). For example, one can use one-sided derivatives.} to get the static dimensionless energy (\ref{dimlessenergy}).

Convergence can be measured by an error function\footnote{Note that this is different from the error function used in \cite{Cox:2019aji}.}, which we have chosen to be
\begin{equation}
e(\bm x_\text{new}, \bm x_\text{old}) = \frac{\max(|\bm x_\text{new} - \bm x_\text{old}|)}{\max(|\bm x_\text{new}|)}~.
\end{equation}
The notation here refers to the maximum value among all points and all vector components. We have experimentally found an error tolerance of $e < 10^{-9}$ to be a good convergence criterion, which was used in producing all the results in this paper. However, it is computationally costly for the standard Gauss-Seidel method to comply with such a small tolerance, especially when the computation must be repeated many times as in the present case. Two methods for speeding up the Gauss-Seidel process that are specific to the present problem are provided in Appendix \ref{appendix:speedup}.

Finally, to implement the right-hand side of (\ref{eq:eom2}), one needs to use the numerical derivative of Dirac delta function, given by
\begin{equation}
\partial_z \delta(z_k - z_i) = \frac{1}{h^2}(\delta_{k,i-1} - \delta_{k,i} ).
\end{equation}
For a derivation, see Section 5 of \cite{Cox:2019aji}.

As for the 1D EOM, the exact same method suitably adapted to one dimension also works. We will only make two comments about the 1D case and refer the interested reader to Section 4 of \cite{Cox:2019aji} for more details. First, unlike the 2D case, the 1D EOM has two disconnected pieces of boundary (namely, $z=\pm \infty$), and the nontrivial solutions are the DWs interpolating between two \textit{different} vacua. Secondly, upon obtaining a solution to the 1D EOM, one can check whether it is BPS (i.e. whether it saturates the lower bound on energy) by either comparing its numerical energy to the analytic expression (\ref{dwtension}) or checking that a plot of its first derivative matches the BPS equation \cite{Hori:2000ck}:
\begin{equation}
    \frac{d x^a}{dz} = \frac{\alpha}{2}  K^{ab} \;  \frac{\partial \bar W}{\partial \bar x^{b}}; \qquad \alpha=\frac{W(\bm x(\infty)) - W(\bm x(-\infty))}{|W(\bm x(\infty)) - W(\bm x(-\infty))|} ~.
\end{equation}

\subsection{Initial Configurations}
\label{sec:4.2}

Although the initial configuration can be arbitrary, a good choice that is close to the real solution can speed up the convergence process. So we use the ``BPS initial configuration," which we now illustrate with the most frequently-used example. Recall from the discussion in Section \ref{taxonomy} that quarks of weights \twk\ are expected to be confined by double-string configurations made of two BPS 1-walls, for any $k$. We will use the same fluxes as the previous discussion to achieve the monodromy, except that for ease of numerical implementation we shift the overall field by $i \frac{2\pi}{N} \bm \rho$.

\begin{figure}[h]
	\begin{subfigure}[h]{.47  \textwidth}
		  \includegraphics[width= 1 \textwidth]{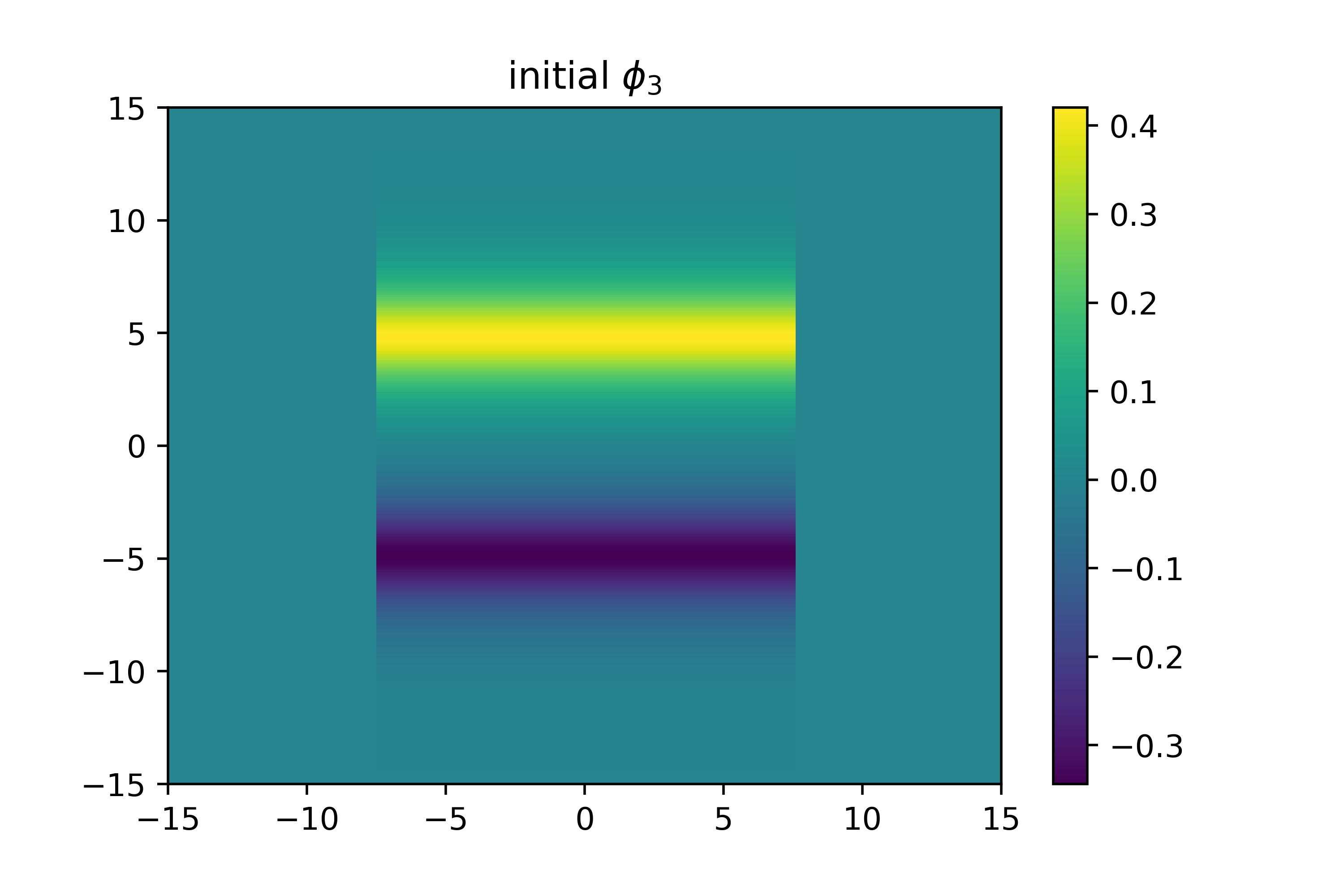}
		  \caption{}
		  \label{fig:BPSinit a}
	\end{subfigure}
	\begin{subfigure}[h]{.47  \textwidth}
		  \includegraphics[width= 1 \textwidth]{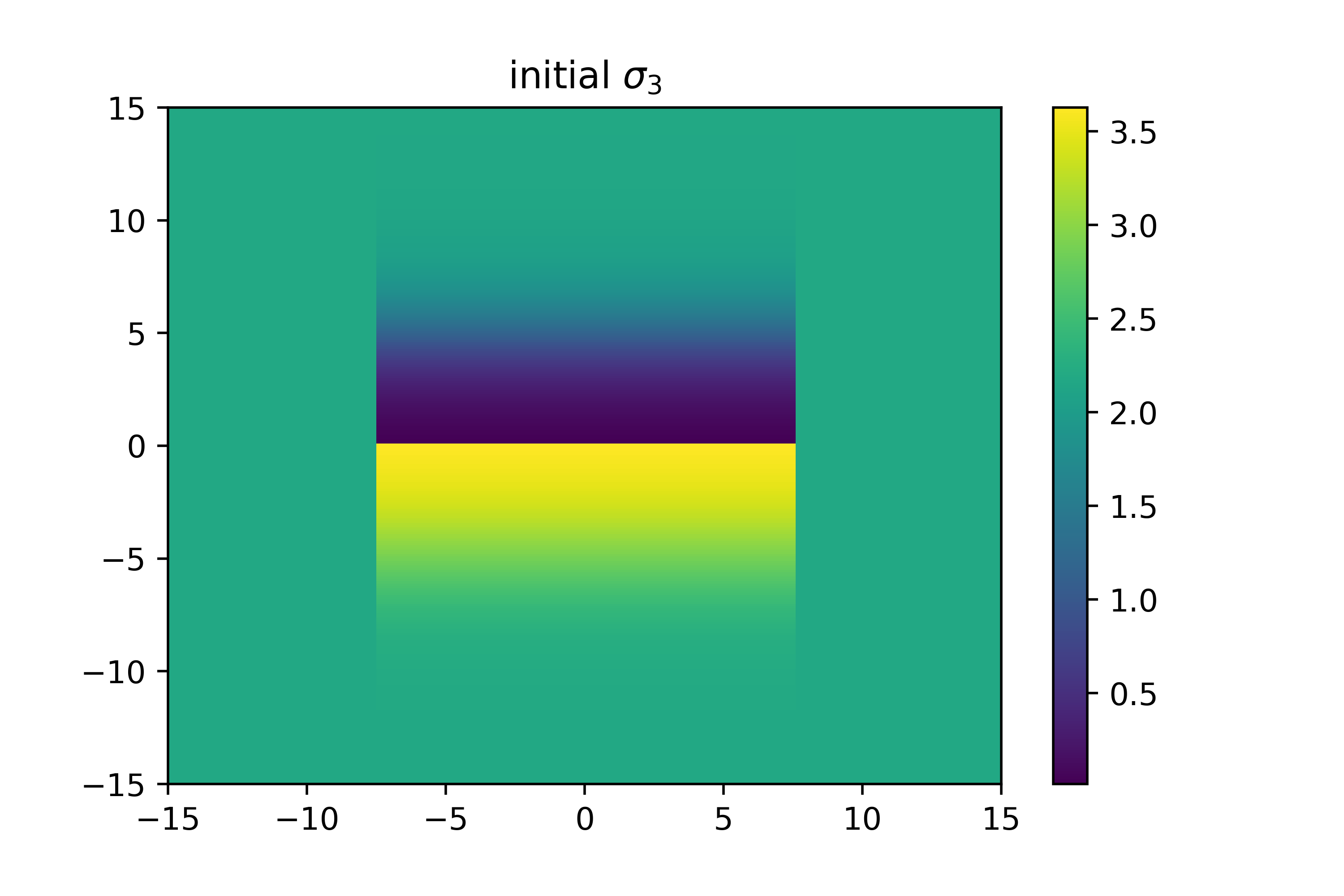}
		  \caption{}
		  \label{fig:BPSinit b}
	\end{subfigure}
\caption{A typical example of the BPS initial configuration. The parameters are: $N=5$, weight $\bm w_2$, and quark separation $R=15$. The grid size is 30 by 30 in units of the dual photon mass. Note that there are $N-1=4$ field components, each with real and imaginary parts. Only the (a) real and (b) imaginary parts of the third field are shown, though the rest are similar. Outside of the quarks, the field is set equal to the boundary value, $i \frac{2\pi}{N} \bm \rho$. In between the quarks are many copies of pairs of BPS DWs, interpolating from  the boundary value to $\bm 0$ and from $i 2 \pi \bm w_2$ back to the boundary value. The field discontinuity has a value that matches the monodromy and its location and shape match the Dirac delta function in the 2D EOM (\ref{eq:eom2}). We have used K\"ahler metric without quantum correction. Using the Gauss-Seidel iterations, this initial configuration relaxes to Figure \ref{fig:DSexample}.} 
\label{fig:BPSinit}
\end{figure}

Concretely, we impose the boundary condition $\bm x|_{\infty} = i \frac{2\pi}{N} \bm \rho$, and let the central region between the two quarks have field values $\bm w_N=\bm 0$ and \twk, separated by a field discontinuity along the $y$-axis,\footnote{In all diagrams in this paper, the $y$-axis is drawn to be horizontal, while the $z$-axis is vertical.} consistent with the location of the Dirac delta term in (\ref{eq:eom2}). We then solve for the pair of BPS DWs\footnote{Note that there is a freedom to choose the positions of the two kinks, which corresponds to a guess of the string separation. There is no way to know this value apriori.} which interpolate between $i \frac{2\pi}{N} \bm \rho \to 0$ and  $i 2\pi \bm w_k \to i \frac{2\pi}{N} \bm \rho $. To construct the initial configuration, we cover the region between the two quarks by many pairs of BPS DWs, while leaving the external region at the constant boundary value. We show a typical example in Figure \ref{fig:BPSinit}.

\begin{figure}[h]
	\begin{subfigure}[h]{.47  \textwidth}
		  \includegraphics[width= 1 \textwidth]{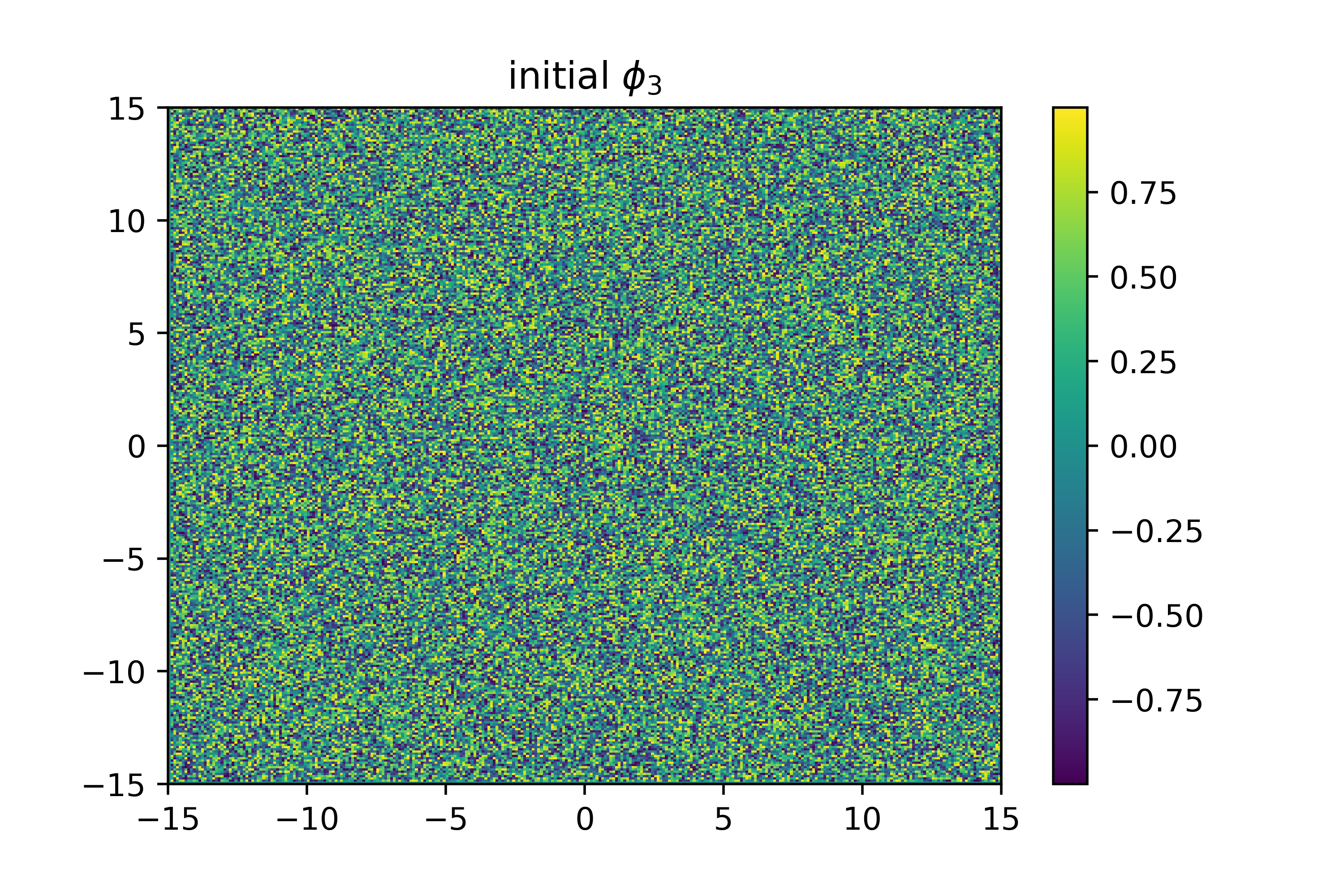}
		  \caption{}
		  \label{fig:random init a}
	\end{subfigure}
	\begin{subfigure}[h]{.47  \textwidth}
		  \includegraphics[width= 1 \textwidth]{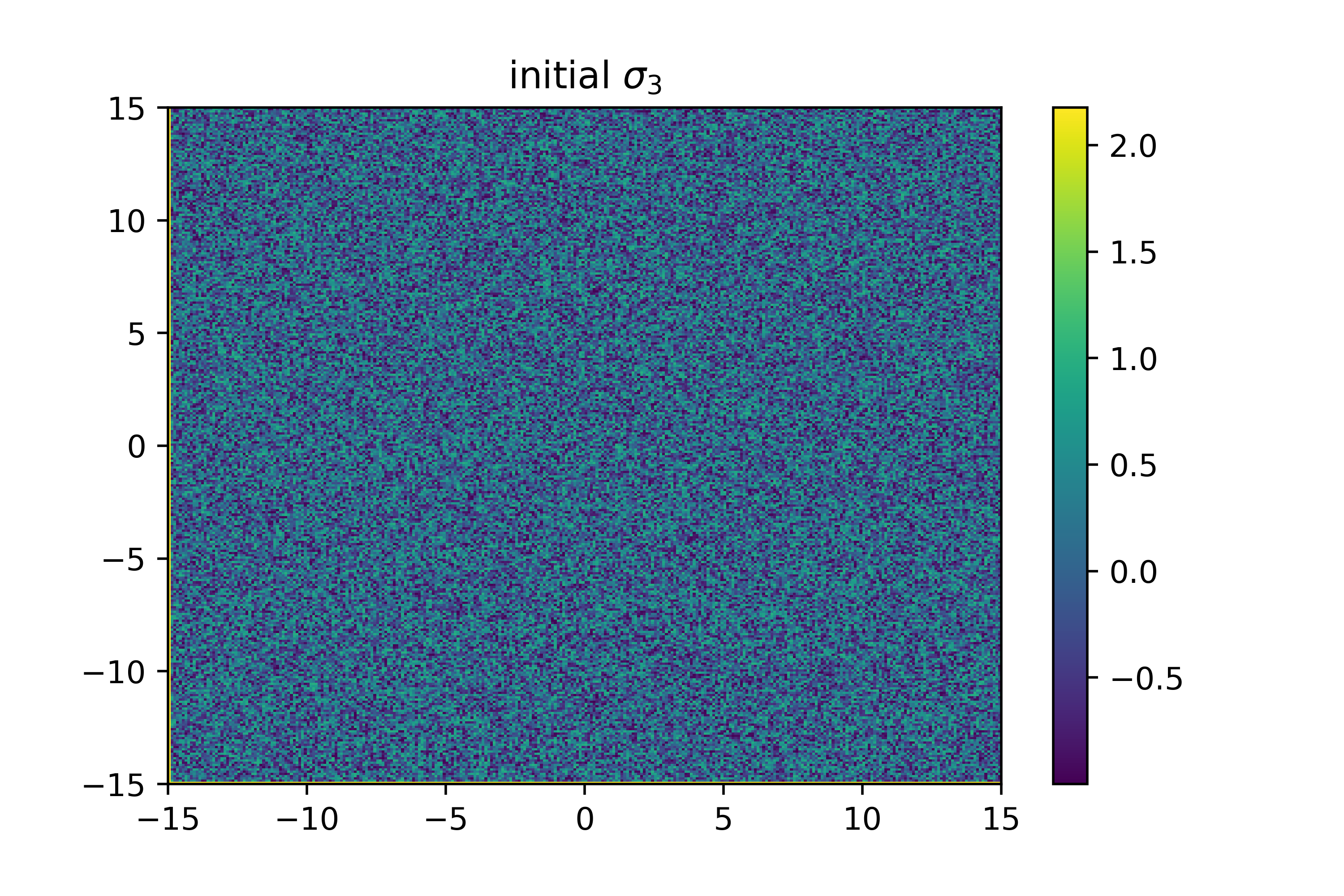}
		  \caption{}
		  \label{fig:random init b}
	\end{subfigure}
\caption{A random initial configuration. Every component at every point is a random number from -1 to 1. This relaxes to Figure \ref{fig:DSexample}. The parameters of the problem are the same as those given in the caption of Figure \ref{fig:BPSinit}. Just like there, we are only displaying the (a) real and (b) imaginary parts of the third vector component, when there are really $5-1=4$ components. Note that the maximum values (and hence colors) are different between the two pictures because they both need to satisfy the boundary condition of their respective components.} 
\label{fig:random init}
\end{figure}

There is a valid concern regarding the choice of initial configuration that needs to be addressed. In principle, there could be more than one solution that minimizes the action. Then the numerical solution might depend on the choice of initial configuration. We argue that this is not the case by testing against the most arbitrary initial condition possible: a completely random field, as shown in Figure \ref{fig:random init}. When the Gauss-Seidel method is applied with a tolerance of $10^{-9}$, both this random field\footnote{Readers who wish to reproduce this random initial configuration study should note that it takes exponentially longer time to relax it to Figure \ref{fig:DSexample} than a more reasonable choice.} and the BPS initial configuration (Figure \ref{fig:BPSinit}) relax to Figure \ref{fig:DSexample}, with the exact same numerical energy.

\subsection{A Typical Example of a Double String}
\label{sec:4.3}

Finally, we illustrate what a typical double-string solution looks like, before diving into the various interesting results we learned from mass-producing such computations. Note that although the double-string confinement prediction \cite{Anber:2015kea} was found to hold for most quarks, the fundamental quarks break this picture, as we shall see in the next section.

\begin{figure}[h]
\begin{subfigure}[h]{.47  \textwidth}
  \includegraphics[width= 1 \textwidth]{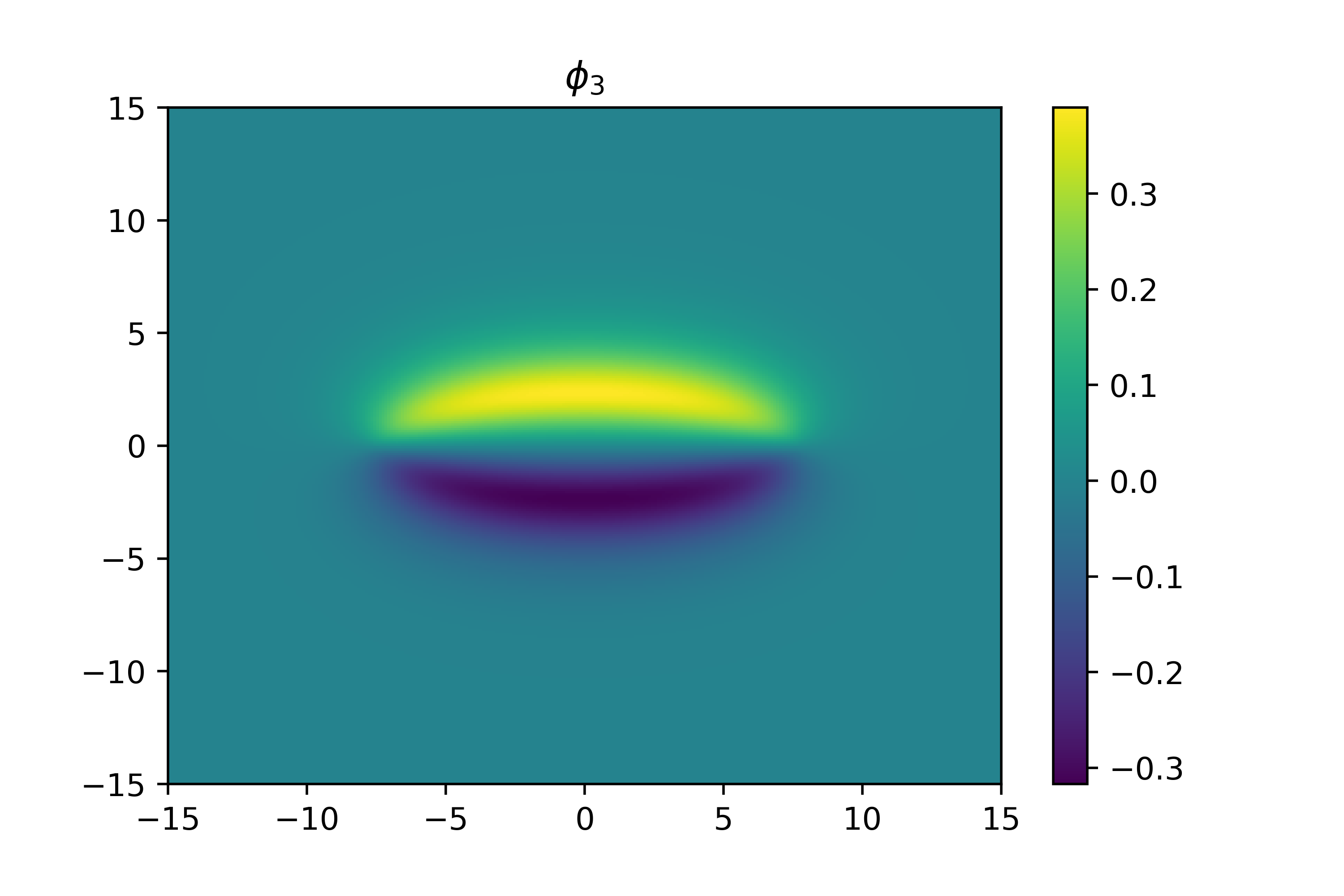}
  \caption{}
  \label{fig:DSexample(a)}
\end{subfigure}
\begin{subfigure}[h]{.47  \textwidth}
  \includegraphics[width= 1 \textwidth]{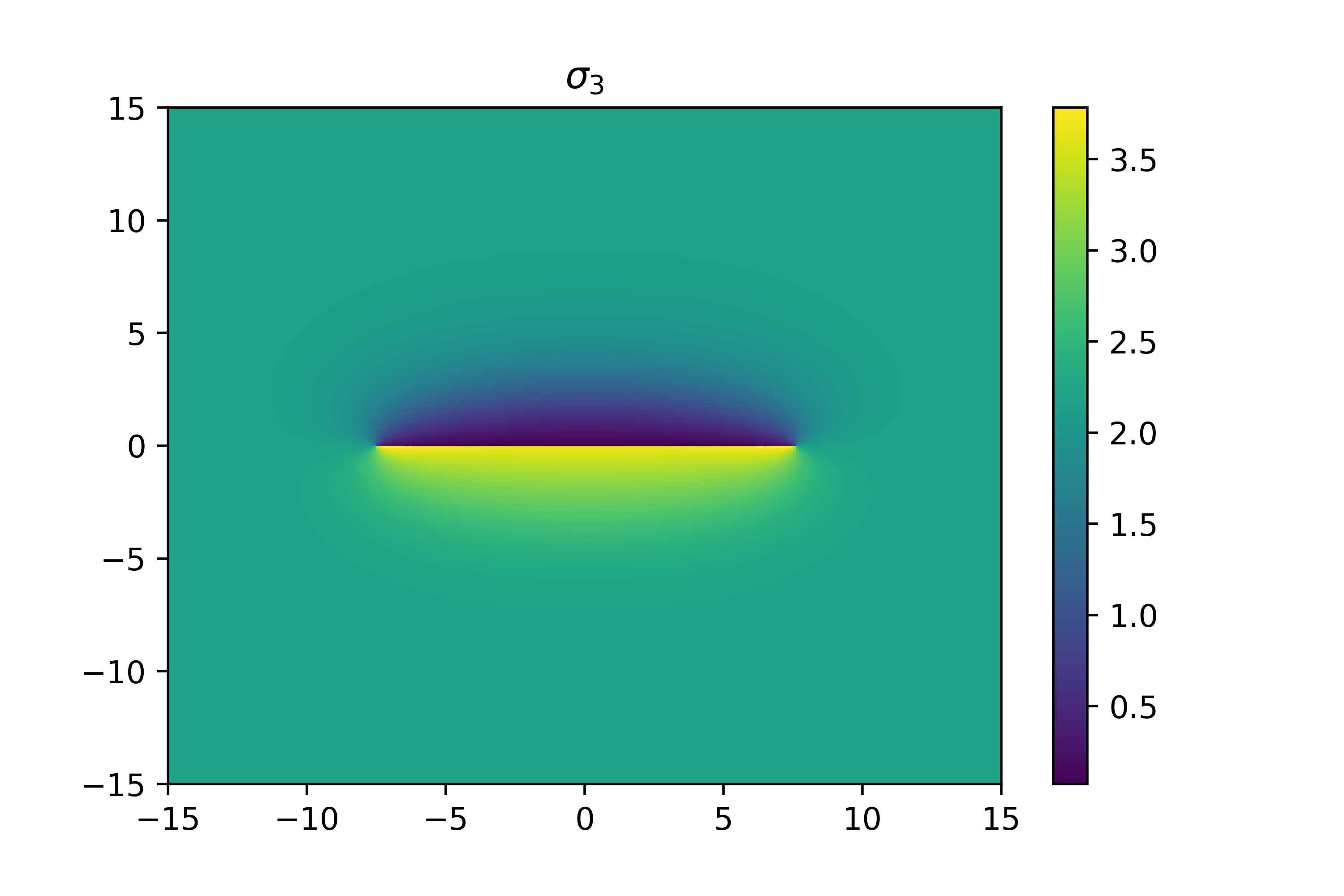}
  \caption{}
  \label{fig:DSexample(b)}
\end{subfigure}
\begin{subfigure}[h]{.47  \textwidth}
  \includegraphics[width= 1 \textwidth]{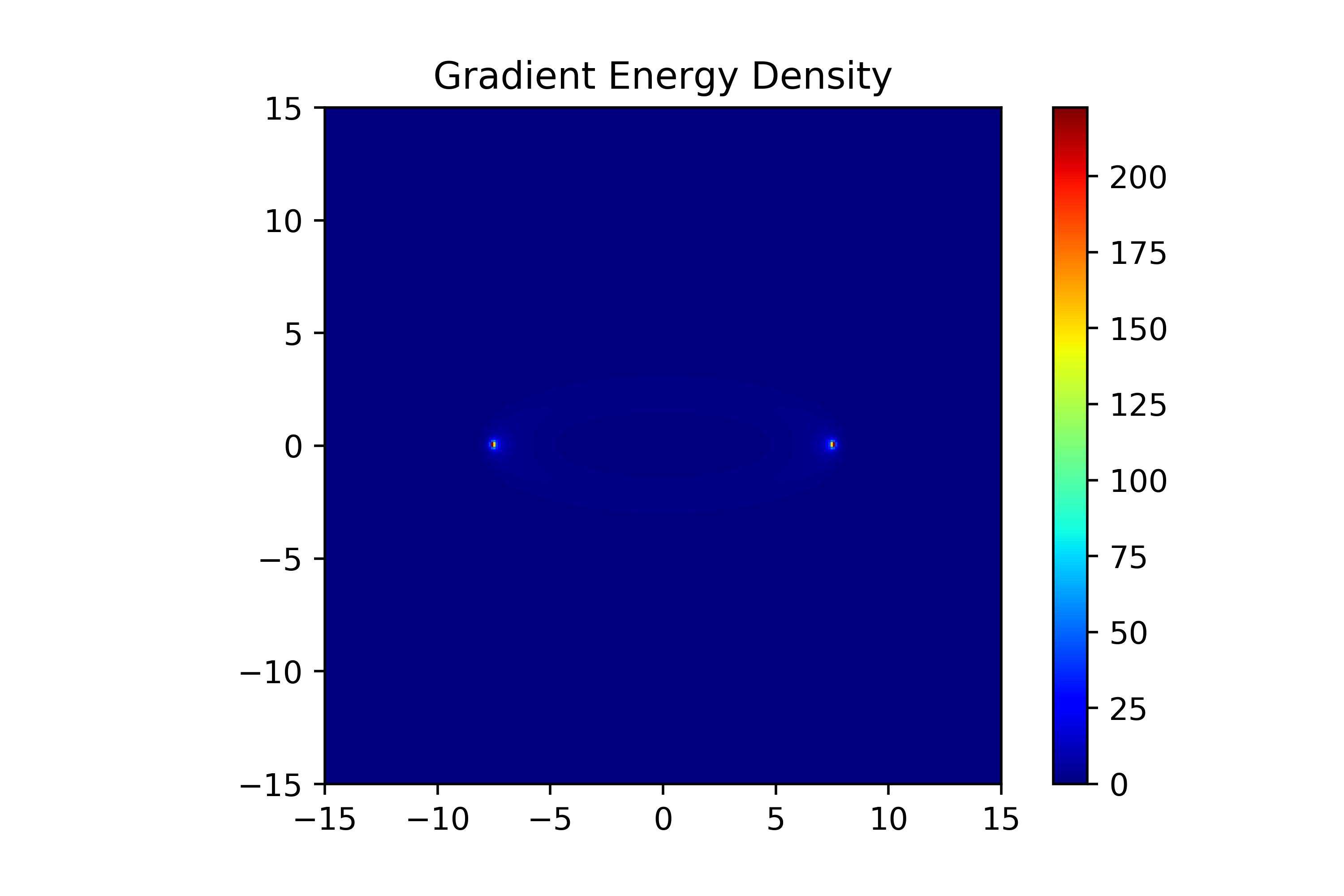}
  \caption{}
  \label{fig:DSexample(c)}
\end{subfigure}
\begin{subfigure}[h]{.47  \textwidth}
  \includegraphics[width= 1 \textwidth]{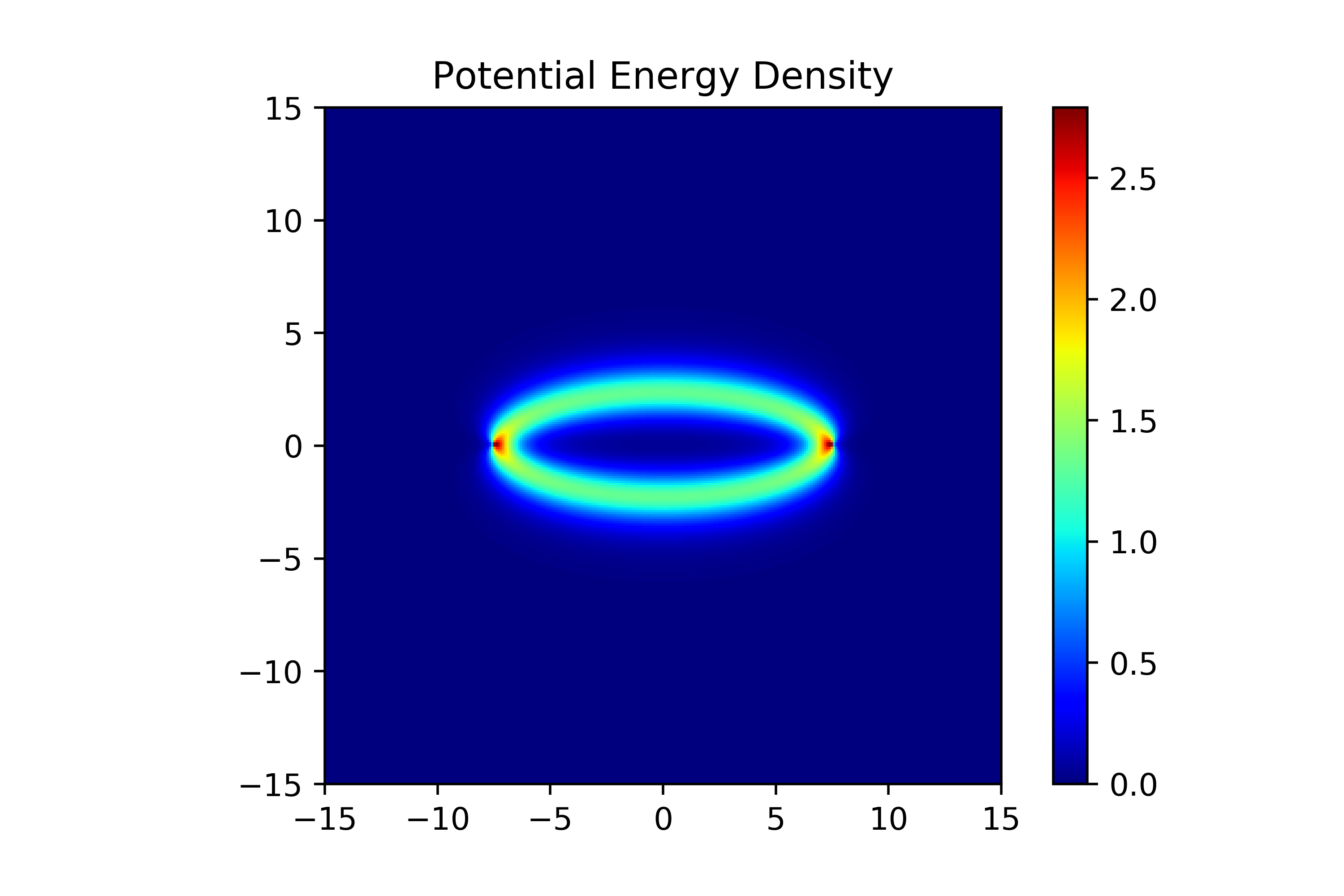}
  \caption{}
  \label{fig:DSexample(d)}
\end{subfigure}
\caption{A typical example of a double-string solution for $N=5$. All the parameters are the same as those given in the caption of Figure \ref{fig:BPSinit}.  Note that there are $N-1=4$ field components, each with real and imaginary parts. Only the (a) real and (b) imaginary parts of the third field are shown, though the rest are similar. Figure (c) and (d) are the first and second terms of the energy density in (\ref{dimlessenergy}).}
\label{fig:DSexample}
\end{figure}

We present the result for the gauge group $N=5$, with a quark-antiquark pair of weight $\pm \bm \lambda = \pm \bm w_2$, separated by a distance $R=15$, all while ignoring the one-loop correction in the K\"ahler metric (\ref{kahler}). Figure \ref{fig:DSexample(a)} and \ref{fig:DSexample(b)} show the real and imaginary component of the third field, respectively. Compare this with the schematic prediction of Figure \ref{fig:01}. We emphasize again that the field discontinuity at the $y$-axis is completely consistent with the location of the Dirac delta term in (\ref{eq:eom2}) and matches the monodromy of the quarks. Figure \ref{fig:DSexample(c)} and \ref{fig:DSexample(d)} show the potential and gradient energy density, which are the first and second term in the integrand of (\ref{dimlessenergy}), respectively. The total dimensionless energy is just the integral of the sum of these two pictures. As is clear from the scale, the total energy density is dominated by the gradient term, but a plot of the potential energy density most clearly reflects the structure of a double string, while that of the gradient energy is concentrated at the two quarks.

\begin{figure}[h]
  \centering
  \includegraphics[width= 1 \textwidth]{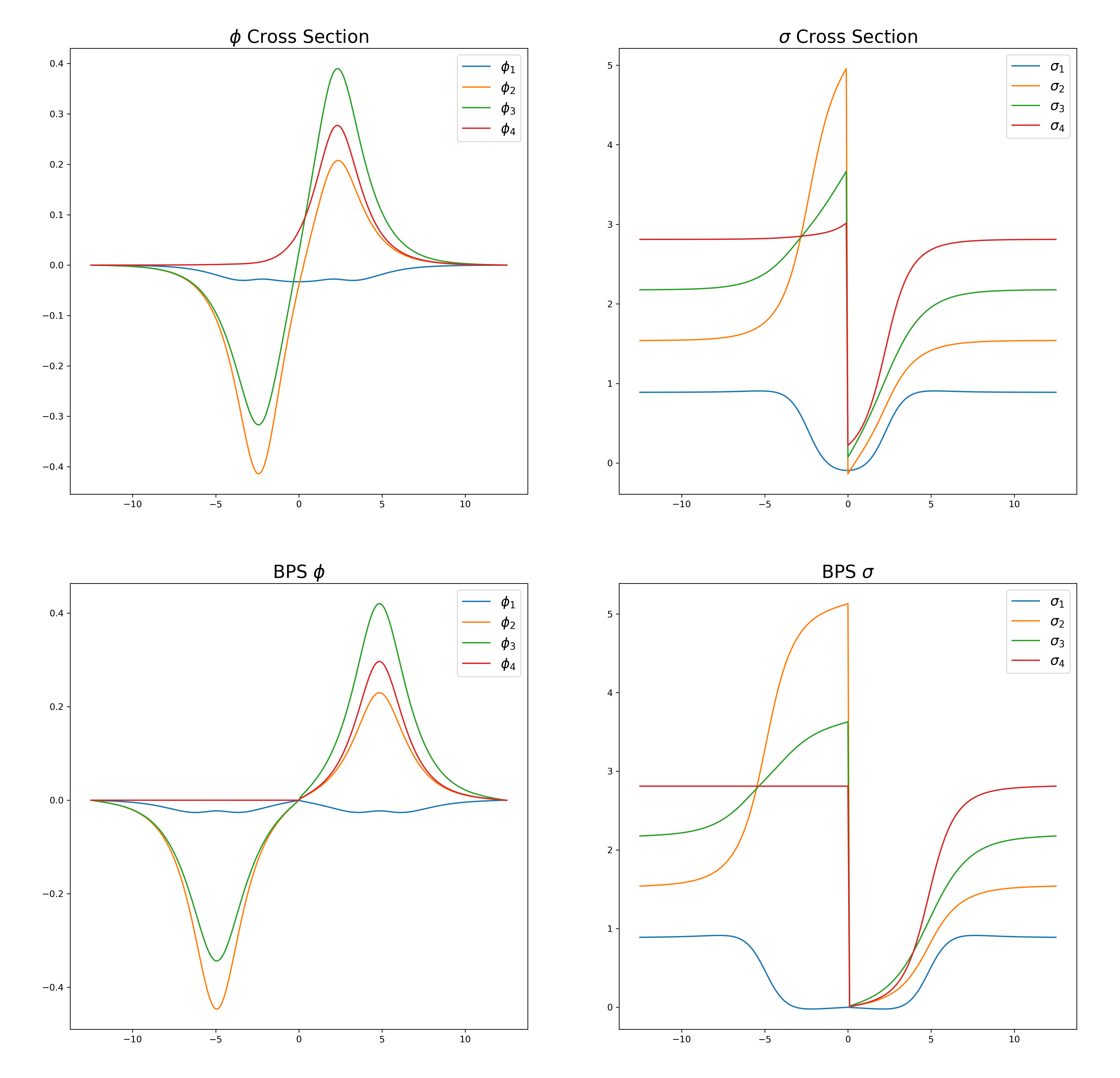}
  \caption{Comparison of the $y=0$ (the midpoint between the quarks) cross section (top row) of the double string of Figure \ref{fig:DSexample} with a corresponding pair of BPS solutions (bottom row). The fact that the quarks are far away from this location implies that the top and bottom plots ought to match. On the bottom plot, the 1D equations for  two BPS DWs are separately solved and then cut and pasted together at a distance $\Delta z \sim 10$. However, the separation of the kinks of the cross section in the 2D solution is a consequence of the physics of string separation (see Section \ref{sec:5.3}).}
  \label{fig:cross section}
\end{figure}

At the center of the double string, the two quarks are relatively far away, and so recalling the discussion of the relation between (\ref{eq:eom2}) and (\ref{eq:1deom}) at the beginning of Section \ref{sec:4}, one would expect that a cross-section corresponds to the solution of the 1D EOM. As a consistency check, we take a vertical cross section of the solution at $y=0$ and compare it to a plot of the corresponding pair of BPS solutions in Figure \ref{fig:cross section}. To a first order approximation, they do match quite well.

\section{Properties of Confining Strings}
\label{sec:5}

Upon mass production of the above computations, various interesting properties of confining strings were found. We will first present all results without the quantum correction, i.e. ignoring the second term in the K\" ahler metric (\ref{oneloopkahler}), and only afterward describe the effect of the quantum correction. Additionally, we are mainly interested in quarks of weights \twk. As explained in Section \ref{sec:3.2.1}, quarks of $N$-ality $k$ are screened down to \twk\ by the creation of $W$-bosons and superpartner pairs. Hence, we will often concentrate on these quarks without further notice.

\subsection{Collapse of the Double String for Fundamental Quarks}
\label{sec:5.1}

The picture of double-string confinement proposed in \cite{Anber:2015kea} was found to hold in most cases. However, it breaks down for fundamental quarks. To describe this, it helps to separate theories with $N \geq 5$ and $N < 5$, because it turns out the latter suffer from low-rank peculiarities.

\begin{figure}[h]
  \centering
  \includegraphics[width= 0.6 \textwidth]{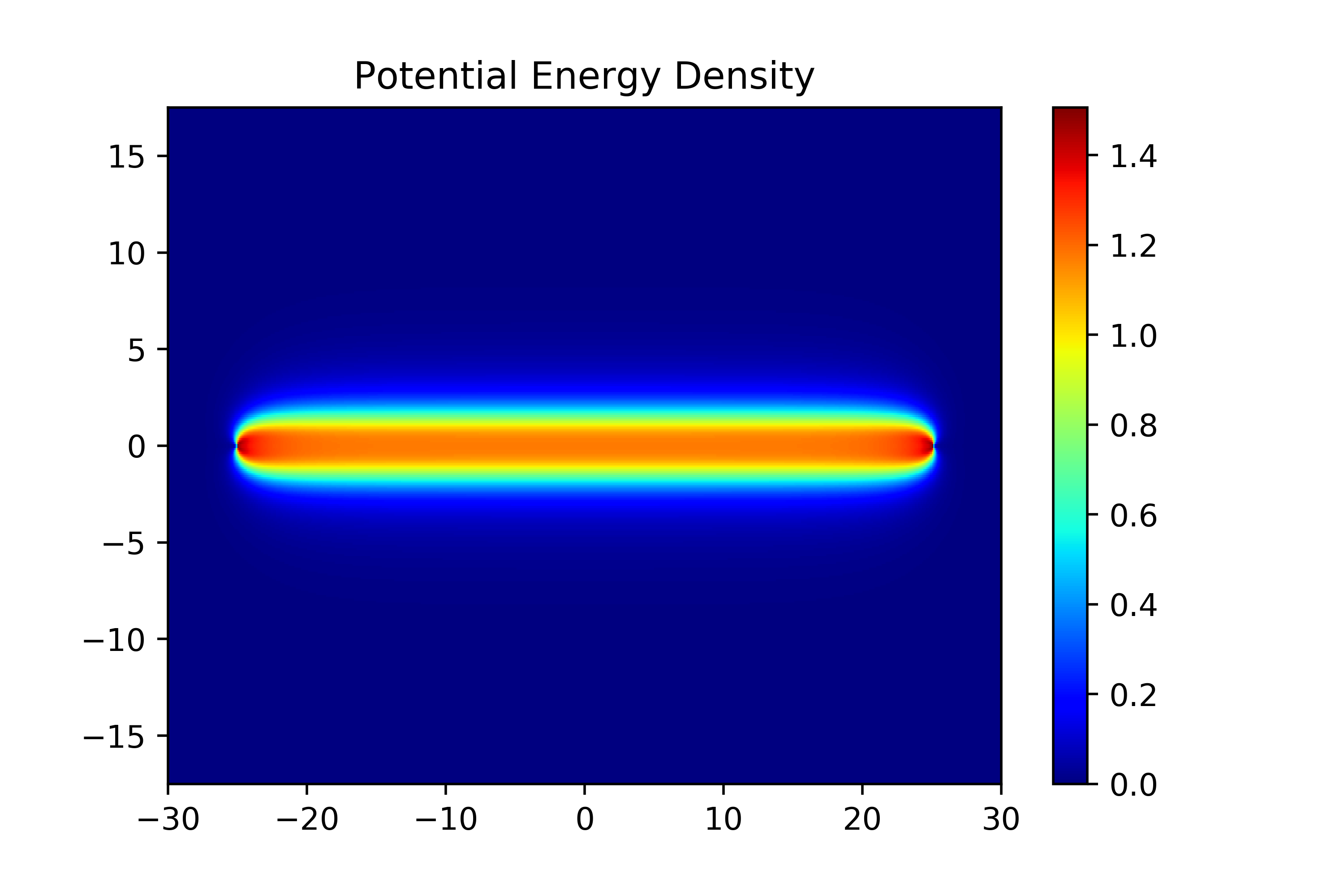}
  \caption{Single string for $N=8$ with quarks of weight $\bm w_1$ and separation of $R=50$. This can be interpreted as the collapse of two strings due to the DWs' attraction.}
  \label{fig:collapsedDS}
\end{figure}

\begin{figure}[h]
  \centering
  \includegraphics[width= 1 \textwidth]{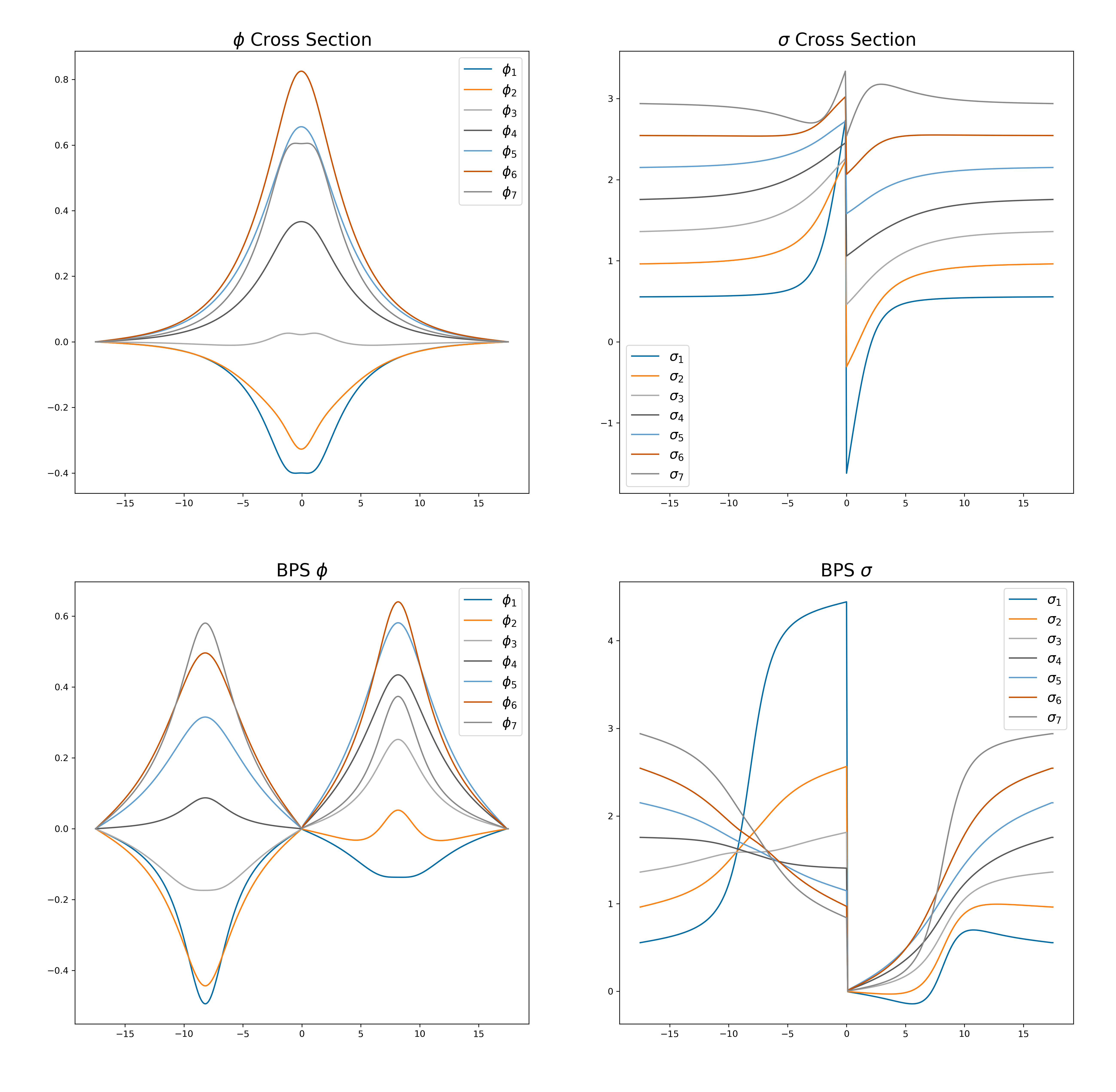}
  \caption{The cross section at the $y=0$ midpoint between the $\bm w_1$ quarks  with a collapsed string for $N=8$. The cross section does not match that of a corresponding pair of BPS solitons, in contrast with the cross section of a double-string configuration, as in Figure \ref{fig:cross section}.}
  \label{fig:collapsed cross}
\end{figure}

For $N \geq 5$, quarks of weights \twk\ with $k \neq 1$ have double-string solutions\footnote{As far as we could tell from simulations up to $N=10$.} just like that of the ``typical example" in Figure \ref{fig:DSexample}. But for $\bm w_1$, the two strings attract and form a bound state. In other words, the quarks are confined by a single string, as shown in Figure \ref{fig:collapsedDS} for $N=8$. The cross section appears to be some nonlinear merger of the corresponding pair of BPS solutions, as shown in Figure \ref{fig:collapsed cross}. This is in contrast with the double-string case for which the cross section (near the midpoint between the quarks) is almost identical to a pair of BPS solutions, as shown in Figure \ref{fig:cross section}.

\begin{figure}[h]
  \centering
  \includegraphics[width= 1 \textwidth]{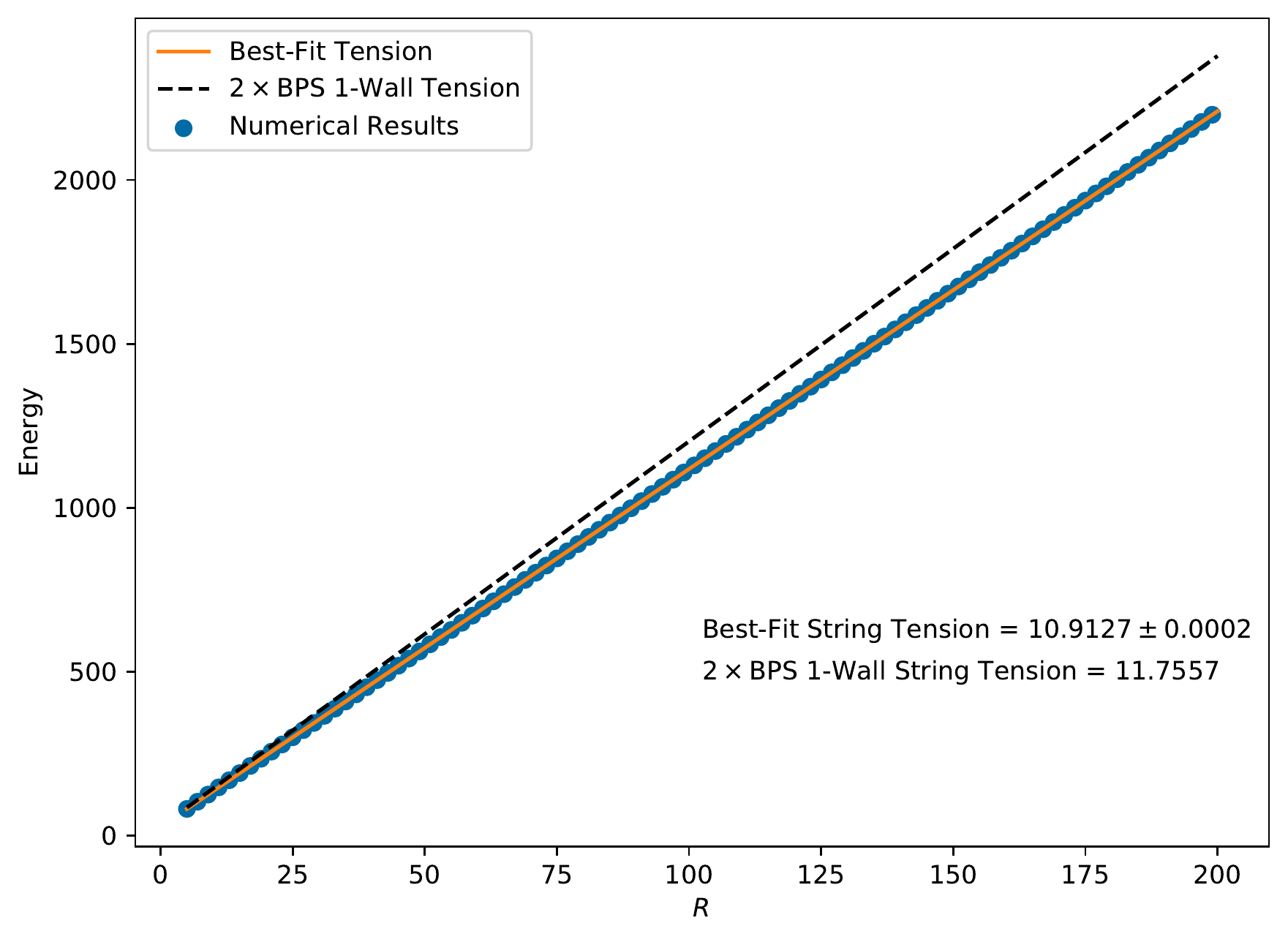}
  \caption{The energy of the fundamental confining strings for $SU(5)$, up to a quark separation of $R=200$. This shows that the binding of two BPS 1-walls into a single string persists to large distances and is not a short-distance effect.}
  \label{fig:SU5k1energy}
\end{figure}

We checked that this collapse is not just a small $R$ effect by going to as far as $R=200$, see Fig.~\ref{fig:SU5k1energy}. Stronger evidence comes from the fact that the attractive interaction between two DWs of total flux $\bm w_1$ can even be seen in 1D solutions. A comprehensive study of DW interactions in 1D is performed in Section~\ref{sec:6}.

\begin{figure}[h]
\begin{subfigure}[h]{.47  \textwidth}
  \includegraphics[width= 1 \textwidth]{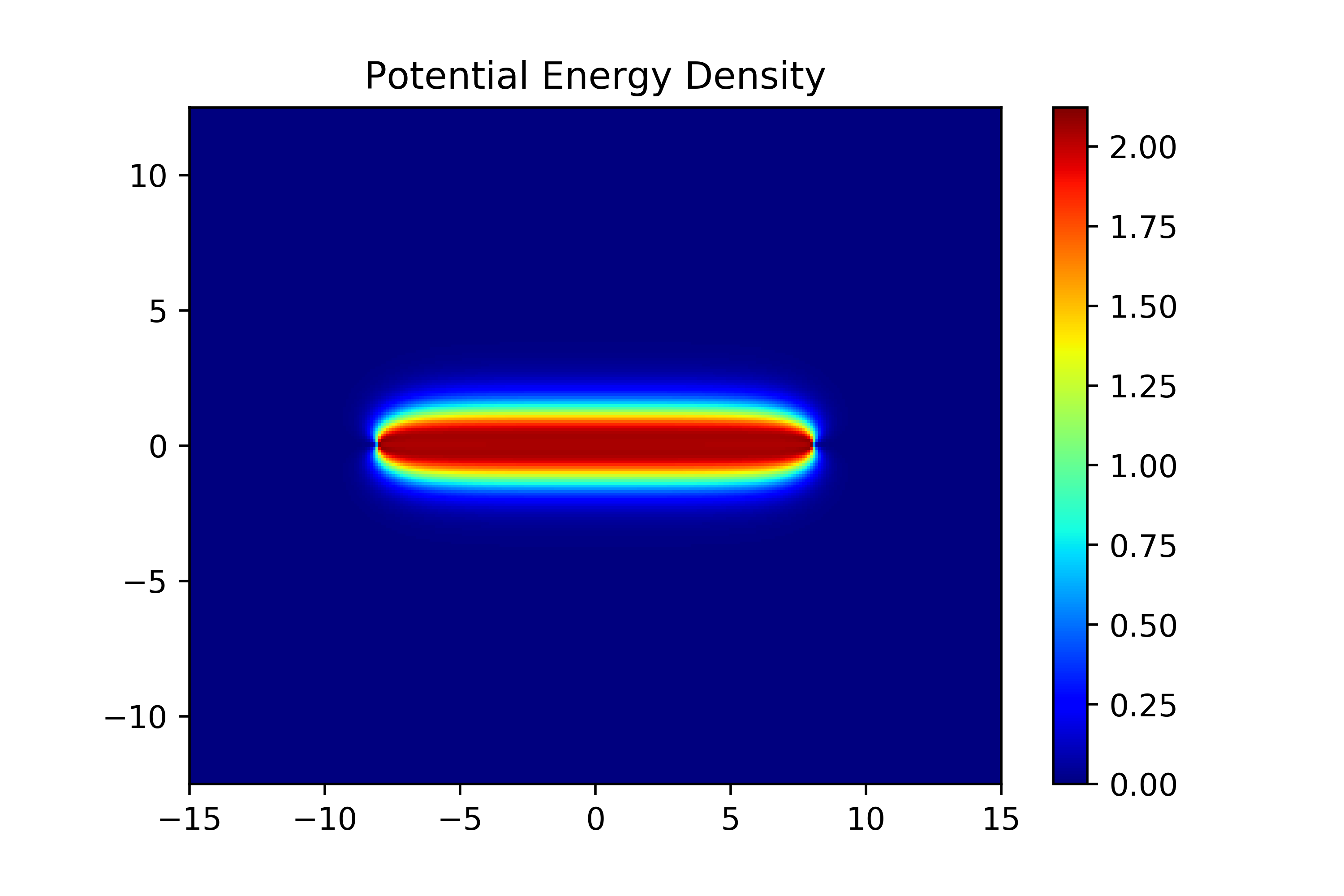}
  \caption{$R=16$}
  \label{fig:N=4 strings(a)}
\end{subfigure}
\begin{subfigure}[h]{.47  \textwidth}
  \includegraphics[width= 1 \textwidth]{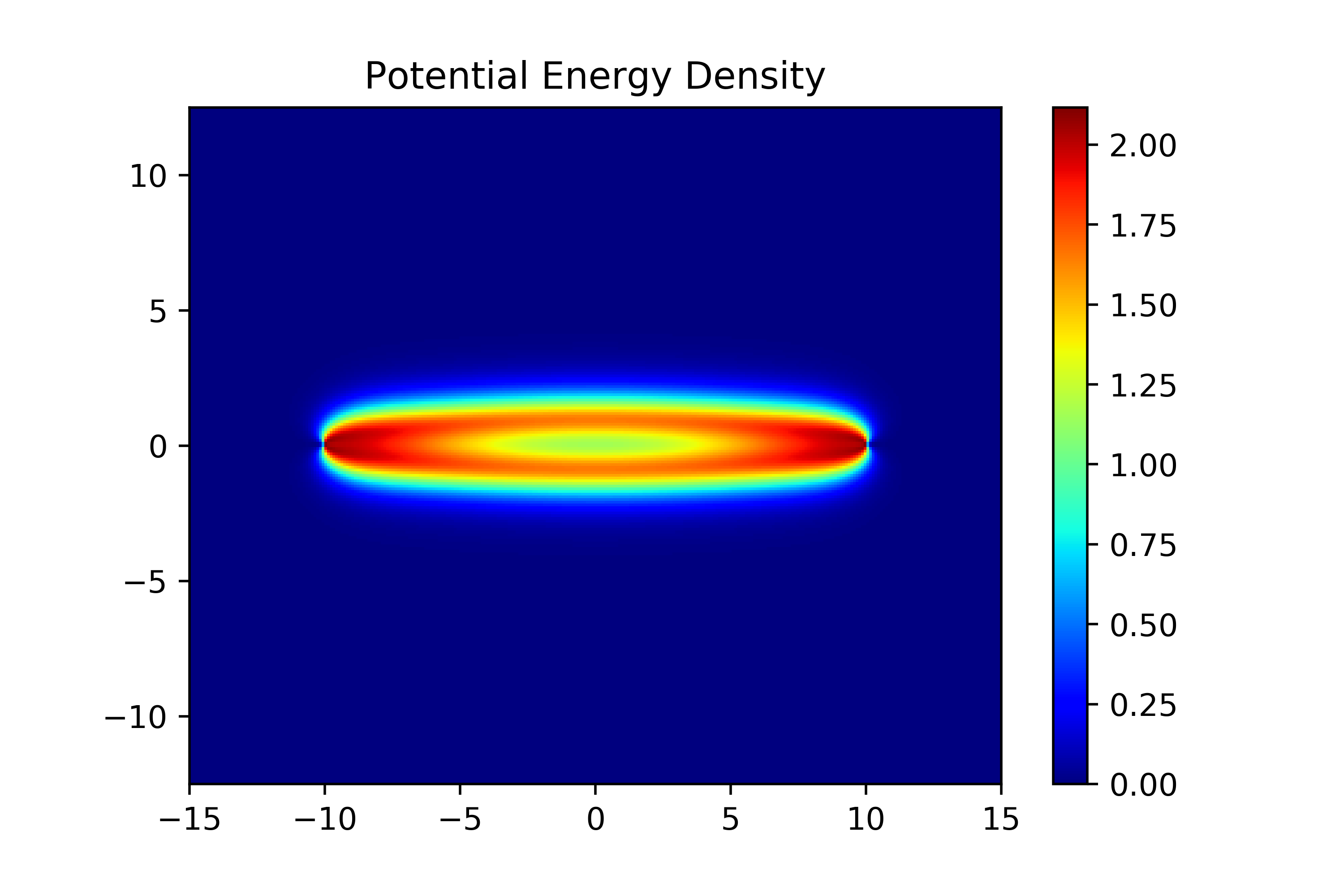}
  \caption{$R=20$}
  \label{fig:N=4 strings(b)}
\end{subfigure}
\begin{subfigure}[h]{.47  \textwidth}
  \includegraphics[width= 1 \textwidth]{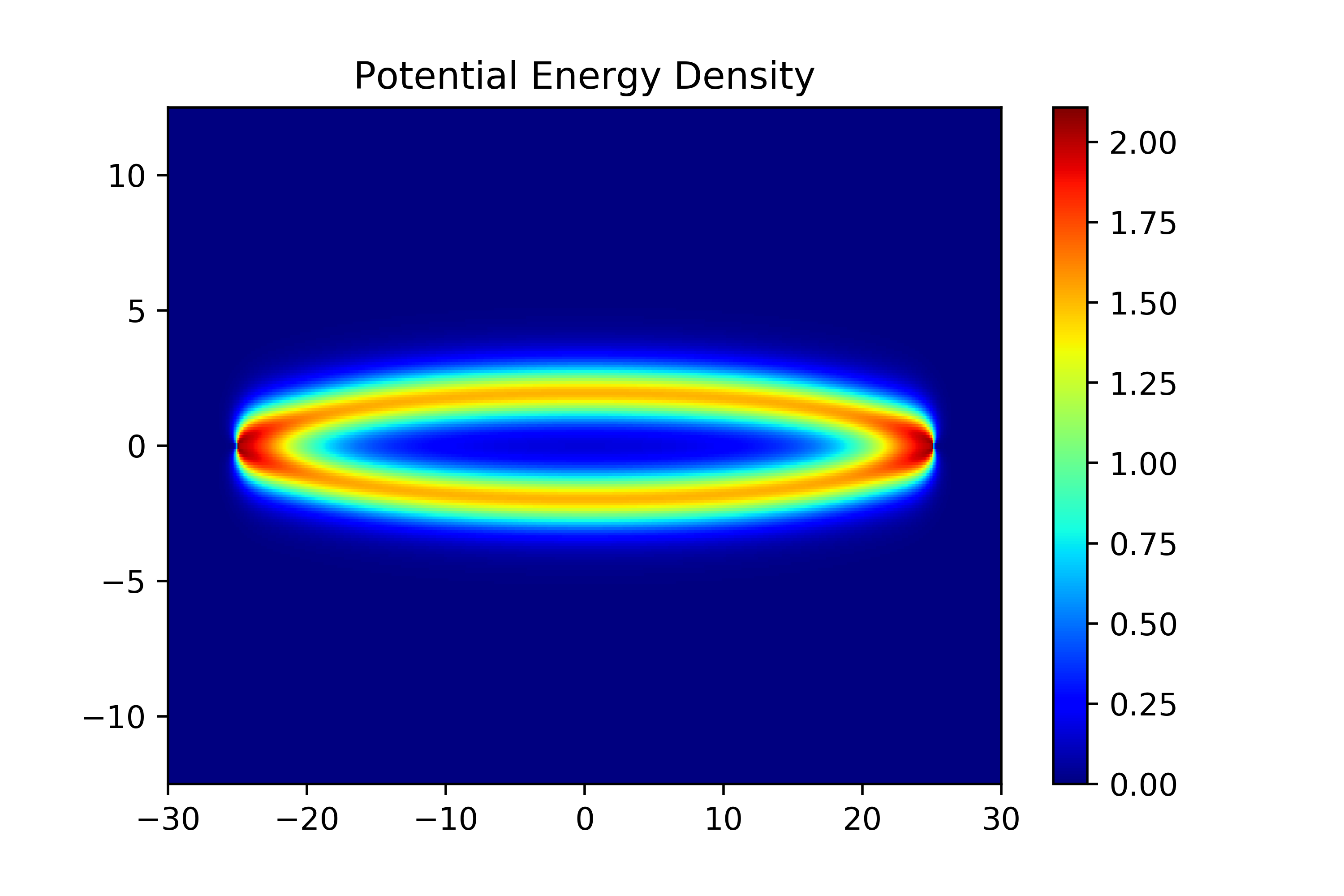}
  \caption{$R=50$}
  \label{fig:N=4 strings(c)}
\end{subfigure}
\caption{Gradual repulsion of two strings and the formation of a double-string configuration as  the quark separation $R$ increases, for $N=4$ and quarks of weight $\bm w_1$. This phenomenon is also seen in high $N$-ality cases for $N>8$.}
\label{fig:N=4 strings}
\end{figure}

For $N < 5$, this collapse does not occur and $\bm w_1$ quarks are confined by double strings. We emphasize here that the above general assertions of whether the strings collapse only pertain to large quark-separation behavior. For instance, for $N=4$ and small quark separation $R$, the DWs of $\bm w_1$ quarks do feel the attraction that would eventually become insurmountable for their higher $N$ cousins; but for large $R$, the repulsion wins over and a double-string is formed. The gradual switch starts around $R=17$, and the progression is shown in Figure \ref{fig:N=4 strings}. Similar gradual repulsion as $R$ increases also occurs in the high $N$-ality cases for $N>8$.

We currently do not have an analytical understanding of the reason for the collapse or of why unit $N$-ality is special starting at $N=5$. However, we noticed that this non-generic behavior of DWs in low dimensions and a turning point at $N=5$ has a precedent. It was found in \cite{Cox:2019aji} that BPS DWs can have certain components of their dual photons possessing a ``double-dipping"  profile; $\sigma_1$ for both of the BPS DWs in Figure \ref{fig:cross section} is a good case in point. This physically means that the corresponding double string has most of its electric flux flowing from a quark to its antiquark, but along a small portion of the DW cross section, the electric flux  flows backwards from the antiquark to the quark (of course the total flux is fixed by the quark charge). Crucially, this is a generic phenomenon for $N \geq 5$ but does not occur for $N <5$. We suspect there are some connections or common causes between the field-reversal phenomenon and the double-string collapse. Perhaps these hints can help future investigations along this direction.

\subsection{N-ality Dependence of String Tensions}
\label{sec:5.2}

It is expected that for large quark separation $R$, the static energy  $\tilde E$ (\ref{dimlessenergy}) grows linearly as a function of $R$. Furthermore, after the effect of screening is taken into account, the string tension $\tilde T$ which is the slope of an $\tilde E$ vs $R$ graph, should depend only on the rank of the gauge group $N$ and N-ality $k$ of the quarks:
\begin{equation} \label{eq:string tension}
\tilde E = \tilde T(N,k) R \quad (\text{for large} \;R).
\end{equation}
Since $N$-ality $k$ quarks are screened down to weight \twk, we can numerically test (\ref{eq:string tension}) and compute the string tensions using simulations of the confinement of \twk\ quarks.

\begin{figure}[t]
\begin{subfigure}[h]{.47  \textwidth}
  \includegraphics[width= 1 \textwidth]{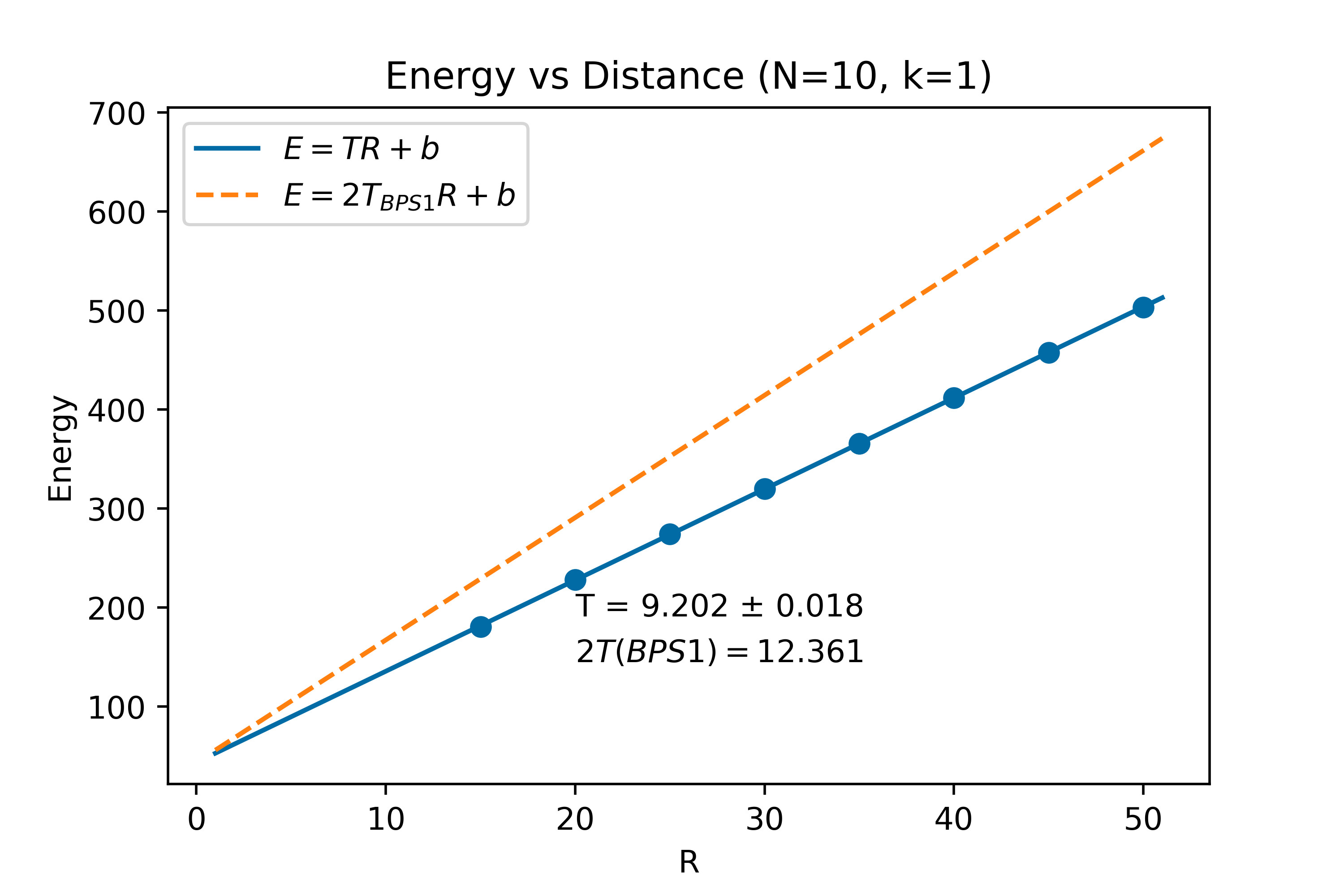}
 \caption{$k=1$}
  \label{fig:E vs R (a)}
\end{subfigure}
\begin{subfigure}[h]{.47  \textwidth}
  \includegraphics[width= 1 \textwidth]{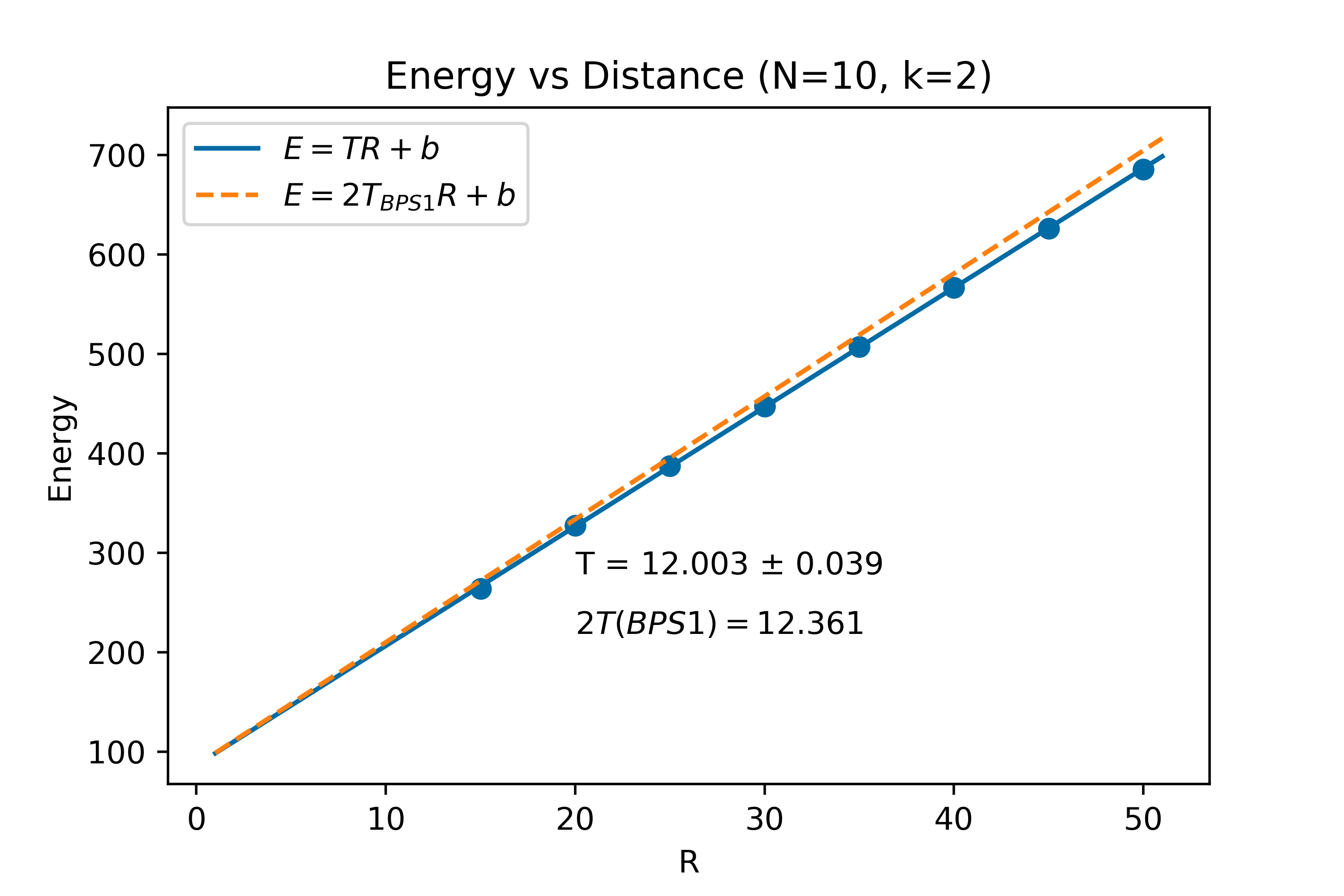}
  \caption{$k=2$}
  \label{fig:E vs R (b)}
\end{subfigure}
\caption{Linear relation between energy $\tilde E$ and quark separation $R$ for $N=10$ and (a) $k=1$ and (b) $k=2$. The string tension is the fitted slope. For comparison, twice the BPS 1-wall tension is also plotted with a dashed line.}
\label{fig:E vs R}
\end{figure}

We indeed found the asymptotic linear relation to hold for all cases tested: for $N$ up to 10 and for quarks of any $N$-ality. We also numerically extracted the tensions $\tilde T(N,k)$ using a least-squares fit. Typical plots for $k=1$ and $k \neq 1$ are shown in Figure~\ref{fig:E vs R}.

To a first order approximation, the general features of the values of the tensions can be explained using what we already know. For quarks of weight \twk\ with $k \neq 1$, we know that they are confined by double strings made of two BPS 1-walls. Of course, what we mean by this statement is that far away from the quarks, such as at the center of the double string, the one-dimensional cross section matches the shape of two BPS 1-walls (Figure \ref{fig:cross section}), whose combined monodromy produces the weight \twk.

 In fact,  from Figure 5 and our other numerical simulations, it appears that the strings look like (rotated/deformed) 1-walls for most of their length, aside from the area immediately adjacent to the quarks. If we make the approximation that the string tension per unit length at every point on the string is the same as that at the center and ignore the curvature of the string altogether, then we would expect the overall dimensionless string tension to be roughly twice the BPS 1-walls tension, $2 \tilde T_{\text{1-wall}} = 4N \sin \frac{\pi}{N} $, as in (\ref{dwtension}), for the double-string configurations that do not collapse.
  On the other hand, for fundamental strings, $\bm w_1$, we have seen that the double string collapses to form a bound state, so we expect its string tension to be less than $2 \tilde T_{\text{1-wall}}$. Indeed, just like the case for $SU(10)$ in Figure \ref{fig:E vs R}, all $k=1$ tensions are less than $2 \tilde T_{\text{1-wall}}$, reflecting the nonzero binding energy, and all $k \neq 1$ tensions are slightly above or below $2 \tilde T_{\text{1-wall}}$. 

In addition to the $k$-string tensions, of particular interest to us is their unitless ratio
\begin{equation}
f(N,k) \equiv \frac{T(N,k)}{T(N,1)}~.
\end{equation}
This quantity is considered a probe of the confinement mechanism and has been compared among different toy models of confinement as well as to lattice simulations. Some well-known scaling laws from other models are the Casimir scaling (due to one-gluon exchange, or in the strong coupling limit, see e.g.  \cite{Greensite:2011zz}),
\begin{equation}
f_{\text{Casimir}}(N,k) = \frac{k(N-k)}{N-1}~;
\end{equation}
the square root of Casimir scaling (first found in the MIT Bag Model \cite{Hasenfratz:1977dt}; it was found to also approximately hold in deformed Yang-Mills theory on $\R^3\times \S^1$, see \cite{Poppitz:2017ivi}),
\begin{equation}
f_{\text{sqrt-Casimir}}(N,k) = \sqrt{\frac{k(N-k)}{N-1}}~;
\end{equation}
and the Douglas-Shenker sine law (which holds in softly broken Seiberg-Witten theory \cite{Douglas:1995nw}, MQCD \cite{Hanany:1997hr}, and in some two-dimensional models \cite{Armoni:2011dw,Komargodski:2020mxz}),\footnote{Ref.~\cite{Anber:2017tug} found yet another different scaling law, similar to (\ref{sinelaw}),  a sine-squared one which holds  for the tension of strings wrapped around the $\S^1$, inferred from the correlator of two Polyakov loops, at small gaugino mass $m$ in softly broken SYM on $\R^3 \times \S^1$. This theory is conjectured to be continuously connected to pure Yang-Mills theory \cite{Poppitz:2012sw}.}
\begin{equation}  \label{sinelaw}
f_{\text{Douglas-Shenker}}(N,k) = \frac{\sin(\pi \frac{k}{N})}{\sin(\frac{\pi}{N})}~.
\end{equation}

With these previous results in mind, we present our numerical results of $\tilde T(N,k)$, $f(N,k)$, and the uncertainties arising from their least squares fit in Table \ref{tensiontable} in Appendix \ref{appendix:tables}. Note that $\tilde T(N,k) = \tilde T(N,N-k)$ due to charge conjugation symmetry, and so we only need to compute $N$-ality up to $\lfloor {N\over 2} \rfloor$. A plot of $f(N,k)$ is shown in Figure \ref{fig:f(N,k)}, and a comparison between the present theory and other known scaling laws for $SU(10)$ is shown in Figure \ref{fig:f(10,k)}. Among the known scaling laws, the values of $f(N,k)$ are closest to the square root of Casimir scaling, but the curve is much flatter. In fact, the curve is relatively flat apart from a large jump between the $k=1$ and $k=2$ points. This is because the tensions for $k \neq 1$ deviate only slightly from $2 \tilde T_{\text{1-wall}}$ as compared to $k=1$, which is a collapsed string. Because of the lack of string collapse, the $f(4,k)$ plot (not shown) is the only special case\footnote{$N$-ality dependence only makes sense starting at $N=4$.} where the curve almost stays constant at $1$. 

\begin{figure}[h]
\centering
  \includegraphics[width= 0.8 \textwidth]{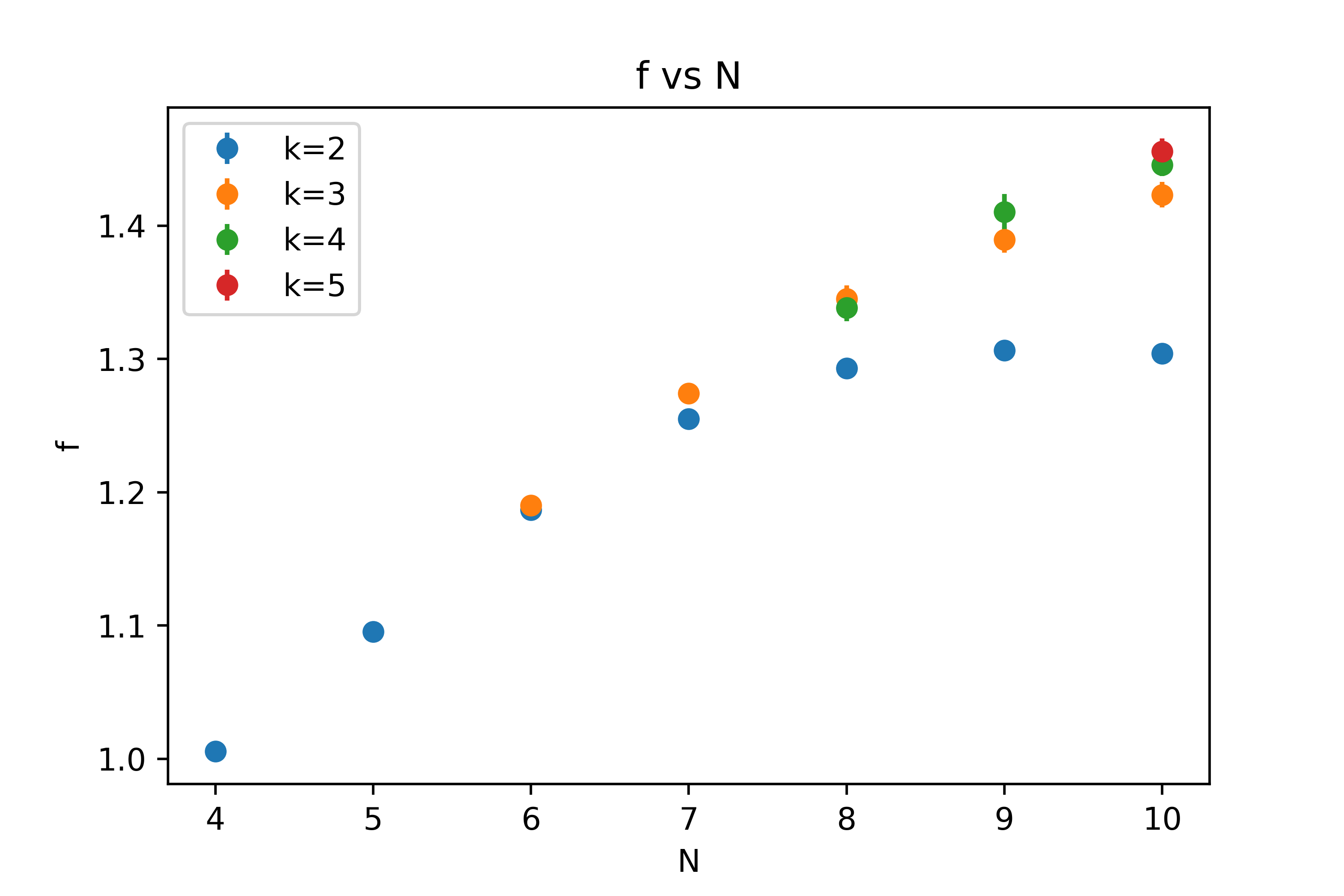}
  \caption{A plot of all computed $f(N,k)$, up to $N=10$. }
  \label{fig:f(N,k)}
\end{figure}

\subsection{Growth of the String Separation}
\label{sec:5.3}

\begin{figure}[h]
\centering
  \includegraphics[width= 0.5 \textwidth]{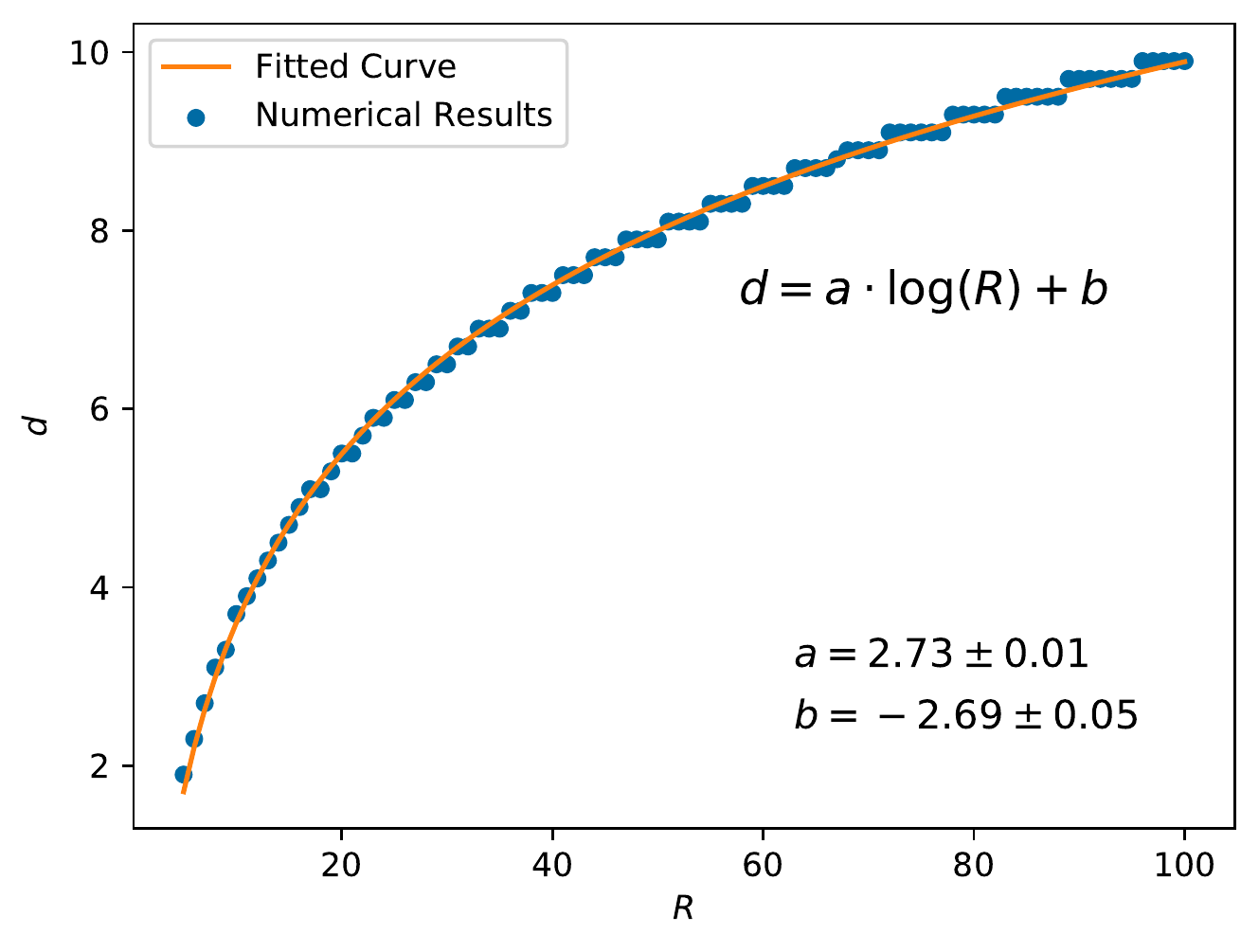}
  \caption{Logarithmic growth of string separation $d$ as a function of $R$, with fitted parameters (for $SU(5)$, $k=2$). The step-like nature of the data is entirely a numerical side-effect: the pixels of the discretization have size $h=0.1$, while $R$ is increased in intervals of $1$. When $\log(R)$ grows slower than $0.1$ upon an increase of $\Delta R=1$, the width will have a step-like growth.}
  \label{fig:log d}
\end{figure}

\begin{figure}[h]
\begin{subfigure}[h]{.47  \textwidth}
  \includegraphics[width= 1 \textwidth]{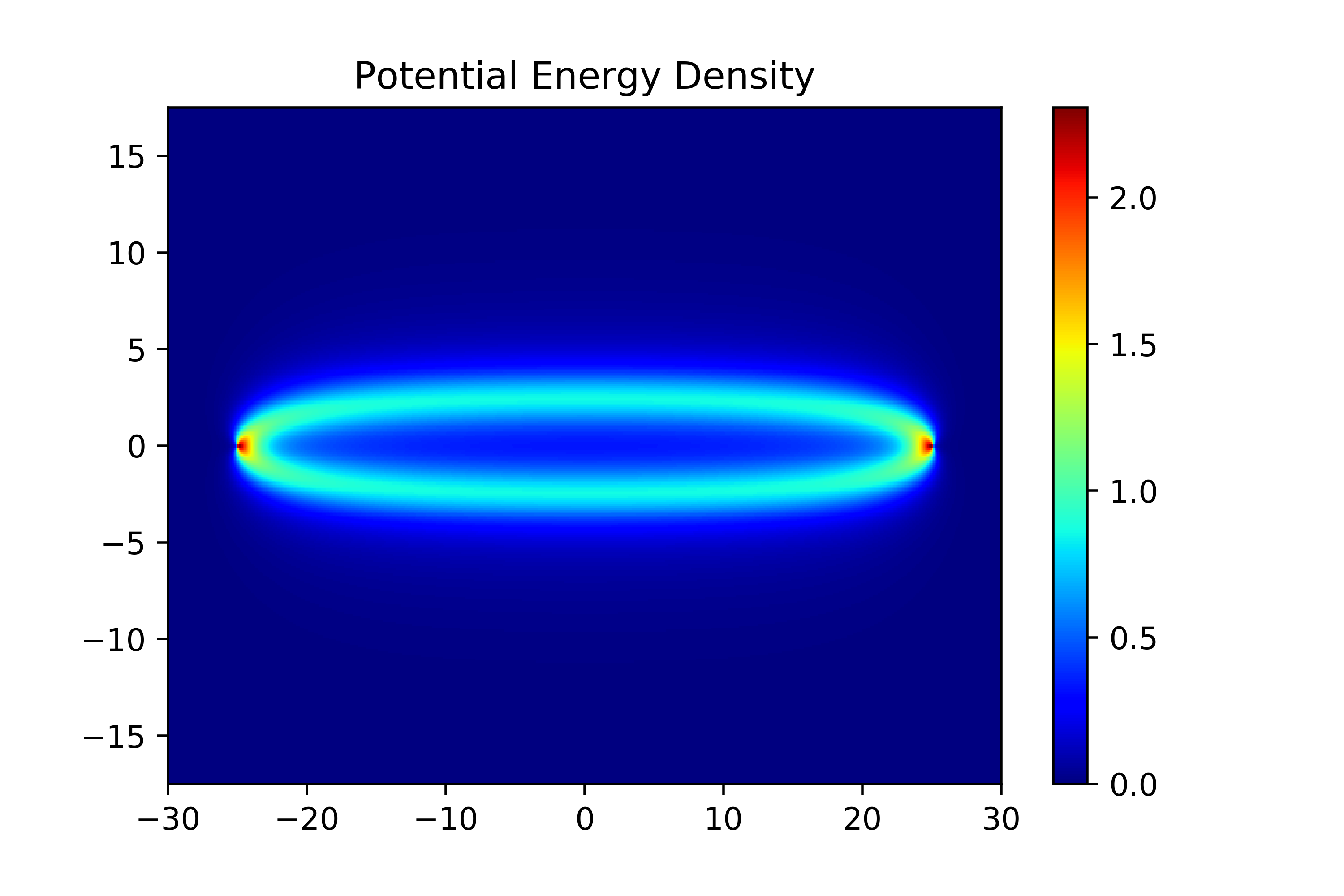}
  \caption{$k=2$}
  \label{fig:DS10(a)}
\end{subfigure}
\begin{subfigure}[h]{.47  \textwidth}
  \includegraphics[width= 1 \textwidth]{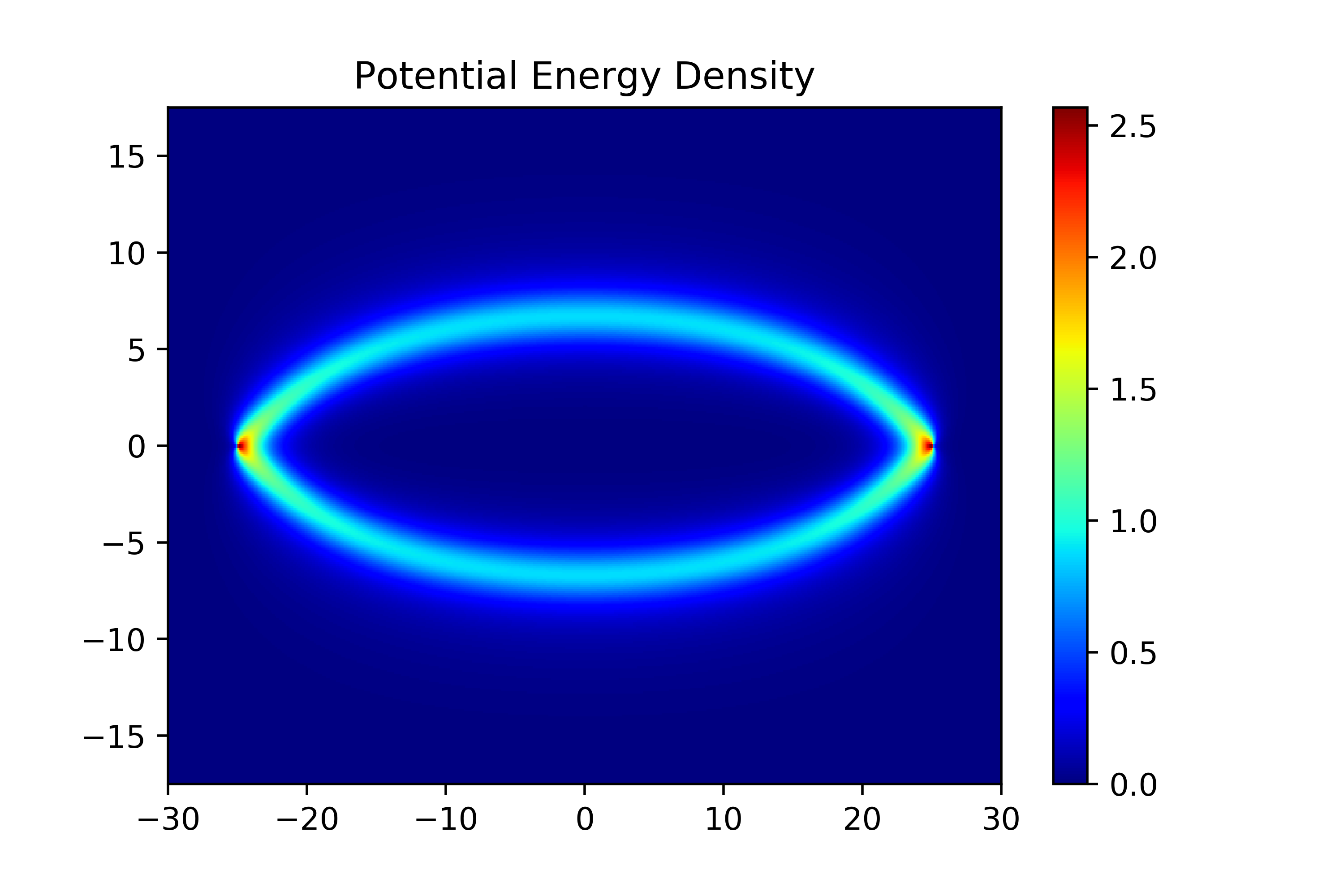}
  \caption{$k=5$}
  \label{fig:DS10(b)}
\end{subfigure}
\caption{$N$-ality dependence of string separation. In $SU(10)$, the string separation is significantly larger for $k=5$ than $k=2$ (compared at the same $R$, of course).}
\label{fig:DS10}
\end{figure}

The string separation $d$ was predicted to grow logarithmically as a function of $R$ in \cite{Anber:2015kea}. The argument goes as follows. Approximate the double string to be composed of mostly parallel DWs separated by a distance $d$, and take the double-string configuration to be a rectangle of dimensions $R$ and $d$. The energy must have a term proportional to the string tension per unit length (taken to be $Mm$, recall Section~\ref{sec:3.1}) times the length of the double string, which is  $\sim 2 Mm (R+d)$. The one-dimensional DW repulsion\footnote{Motivated by the long-distance repulsion of 1D kinks in a sine-Gordon model with a $\cos 2 \varphi$ potential (this is closest to our $SU(2)$ theory).}  per unit length of the DW can be modelled by an exponential term $\sim M m e^{-md}$, and the corresponding term in the energy must be proportional to $Mm R e^{-md}$. Thus, a larger $d$ increases the energy due to the increased length of the string,  but is compensated by a smaller repulsion energy. Up to numerical factors, the energy is
\begin{equation}
E \sim Mm (R+d) + Mm R e^{-md}~.
\end{equation}
The minimum occurs at a separation $d$ that scales as \begin{equation}
d \sim {1 \over m} \log(mR)~.
\end{equation}
From our simulations, all of the non-collapsing double strings indeed have a logarithmic growth in $d$. We define $d$ to be the distance between the two maximum points in a  $y=0$ cross-section of the potential energy density. A typical fit to a logarithm is given in Figure \ref{fig:log d}. A list of all the parameters fitted to a logarithmic description of $d$ is attached in Table \ref{separation table} in Appendix \ref{appendix:tables}.
Despite the success of the logarithmic growth prediction, this simple model does not capture the $N$-ality dependence of string separation, which turns out to be significant. The separation generally gets larger as $N$-ality increases, an effect that becomes obvious for large $N$, and most strikingly for $N=10$, as shown in Figure \ref{fig:DS10}.

\subsection{Towards Larger L: the Effect of W-Boson Loops}
\label{sec:5.4}

As discussed in Section~\ref{sec:2}, in the leading semiclassical approximation, up to now we neglected the one-loop correction to the K\" ahler metric (\ref{oneloopkahler}). In this approximation, all our results for the string tensions are expressed in dimensionless form, as given in (\ref{dwtension}), in terms of the parameters $m$ and $M$ defined in (\ref{scalemsquared}) and (\ref{scaleM}). In this section, we shall describe  the results of simulations
that take into account the one-loop correction to the K\" ahler metric. This introduces a dependence of the simulation results on the dimensionless parameter:
\begin{equation}
\label{epsilondef}
\epsilon \equiv {3 g^2 N \over 16 \pi^2}~,
\end{equation}
where $g^2$ is the 4D gauge coupling at the scale $4\pi/L$, of the order of the heaviest $W$-boson mass.
Thus, the results for the string tension are expressed as $T = M m \tilde T(\epsilon)$, where we should recall that $M$ and $m$ also depend on $\epsilon$. 
To take this into account, we use (\ref{epsilondef}), the definitions of $M$ and $m$, and the definition of $\Lambda$, (\ref{lambdascale}), now rewritten as
\begin{equation}
\label{lambdascale1}
L \Lambda = 4 \pi \epsilon^{-{1\over 3}} e^{ -{1 \over 2 \epsilon}}~,
\end{equation}
to rewrite these dimensionful parameters and the dimensionful string tension as functions of $\epsilon$, $\Lambda$, and $N$. Our goal is to keep $\Lambda$ fixed for theories with different $N$ and at different circle sizes $L$. It is clear from (\ref{lambdascale1}) that increasing $\epsilon$ is equivalent to increasing $L$ at fixed $\Lambda$. Thus, we find\footnote{The scaling of the string tension $T$ in (\ref{Mmepsilon}) in terms of parameters is usually written in a more familiar way that somewhat obscures  the $\epsilon$ dependence: $T = .675 {\Lambda^2} {\Lambda LN \over 4 \pi} \tilde T(\epsilon)$.} 
\begin{eqnarray}
\label{Mmepsilon}
M = {1 \over L} {\epsilon \over 3 N} ~,~ m =  {72 \sqrt{2} \pi \over L} {N^2 \over \epsilon^2} e^{- {3 \over 2 \epsilon}} \implies
T(\epsilon,\Lambda,N) \simeq .675 \; \Lambda^2 \; N \epsilon^{-{1\over 3}} e^{- {1 \over 2 \epsilon}} ~ \tilde T(\epsilon).
\end{eqnarray}
We  study two values of $\epsilon$: $\epsilon_1 = .09$ and $\epsilon_2 = .12$.  These are chosen to be small enough that the K\" ahler metric does not have a strong coupling singularity for the values of $N$ we study, yet large enough to produce visible deviations from the simulations with $K^{ab} = \delta^{ab}$. From (\ref{lambdascale1}), these values of $\epsilon$ correspond to $\S^1$-sizes $L_1$ and $L_2$ such that 
\begin{equation}
\label{twolengths}
{L_1 \Lambda \over 4\pi} \simeq  0.0086, \;{\rm for} \; \epsilon_1 = .09, ~ {\rm and} ~ {L_2 \Lambda \over 4 \pi} \simeq  0.0314,\;{\rm for} \; \epsilon_2 = .12,
\end{equation}
corresponding to a roughly fourfold increase of $L$, $L_2 \simeq 3.6 L_1$. We also notice that for up to $N \le 10$, the semiclassical expansion parameters $\Lambda L_1 N \over 2 \pi$ and $\Lambda L_2 N \over 2 \pi$ are smaller than unity, though hardly infinitesimal---the latter is about $.6$ for $N=10$. Thus  the validity of the semiclassical approximation is at best qualitative. As will be clear from our subsequent discussion,  the string tensions are seen to not vary wildly as a function of $\epsilon$; this small variation can be taken to qualitatively support the use of semiclassics.

The results for the string tensions and the fitted string separations for these two values of $\epsilon$ are given in Tables.~(\ref{tensiontables009}) and (\ref{separation tableeps009}), respectively. Before discussing them, let us make some comments on the simulations with $\epsilon \ne 0$. First, we note that the equation of motion (\ref{eq:eom2}) and the behavior of the K\" ahler metric (\ref{oneloopkahler}) discussed earlier imply that,  after  including $W$-boson loops, the BPS walls are expected to become wider. This is because the eigenvalues of $K^{ab}$ are less than unity and the equation of motion (\ref{eq:eom2}) implies that as the eigenvalues of the K\" ahler metric decrease, the second derivative of $\vec{x}$ is also smaller, and thus it needs to vary over a larger range to interpolate the same flux. This is indeed seen to be the case in our simulations (in addition, we have verified, as a check of their consistency, that while the 1D DWs  become wider as $\epsilon$ is increased, the BPS DW tensions (\ref{dwtension}) remain the same, as they  only depend on the boundary conditions and not on the K\" ahler metric). Second, as is seen from the fitted separation of the double strings given in (\ref{separation tableeps009}), this widening of the BPS DWs, and subsequently of the double strings, is the main qualitative effect of the inclusion of $W$-boson loops.\footnote{This transverse widening of the double string is what made the nonzero $\epsilon$ simulations challenging. The width of grid in the transverse direction had to be chosen large enough to accommodate the configuration without significant edge effects. The widening of the double string for $\epsilon$ changing from $.09$ to $.12$ is significant, especially as $N$ is increased. Perhaps, the results of (\ref{separation tableeps009}) will be useful for a future model of the double string properties. (We stress, however, that the separations in (\ref{separation tableeps009})  are based on only a few  data points for the larger values of $N$, because the strings tended to collapse at smaller quark separations. For $N=9$, $\epsilon=0.12$, for instance, the string is collapsed all the way to $R=40$.)} At the same time, we find that including nonzero $\epsilon$ does not lead to collapse of the double string for $k>1$, or to a separation of the collapsed string for $k=1$. In other words, the main conclusion from our results is that the qualitative behavior of double-string confinement is not changed by the inclusion of virtual $W$-boson effects.

Let us now discuss the results in some more detail. 
The string tensions $T(\epsilon, \Lambda, N)$ increase as $\epsilon$ (or $L$) is increased at fixed $\Lambda$: as (\ref{Mmepsilon}) shows, increasing $\epsilon$ increases the prefactor, and  numerics shows that the dimensionless part $\tilde T(\epsilon)$ also increases. 
The results for $\tilde T(\epsilon)$, shown in eqn.~(\ref{tensiontables009}) for up to $N=9$, show that $\tilde T(\epsilon)$ slightly  increases as $\epsilon$ changes from $.09$ to $.12$. Thus, the increase of the dimensionful string tension at fixed $\Lambda$ is dominated by the increase of the prefactor in (\ref{Mmepsilon}). The  effect of the quantum correction is small, as expected in the semiclassical regime. As our results show, this smallness holds even when the expansion parameter is not so small (e.g. for $\epsilon = .12$). Thus, as expected from (\ref{Mmepsilon}), the $k$-string tensions at fixed $N$ and $\Lambda$ increase with $L$, and are essentially proportional to $L$ (i.e., they increase roughly $3.6$ times), with small variations due to $W$-boson loop effects.

\begin{figure}[ht]
    \includegraphics[width=1\textwidth]{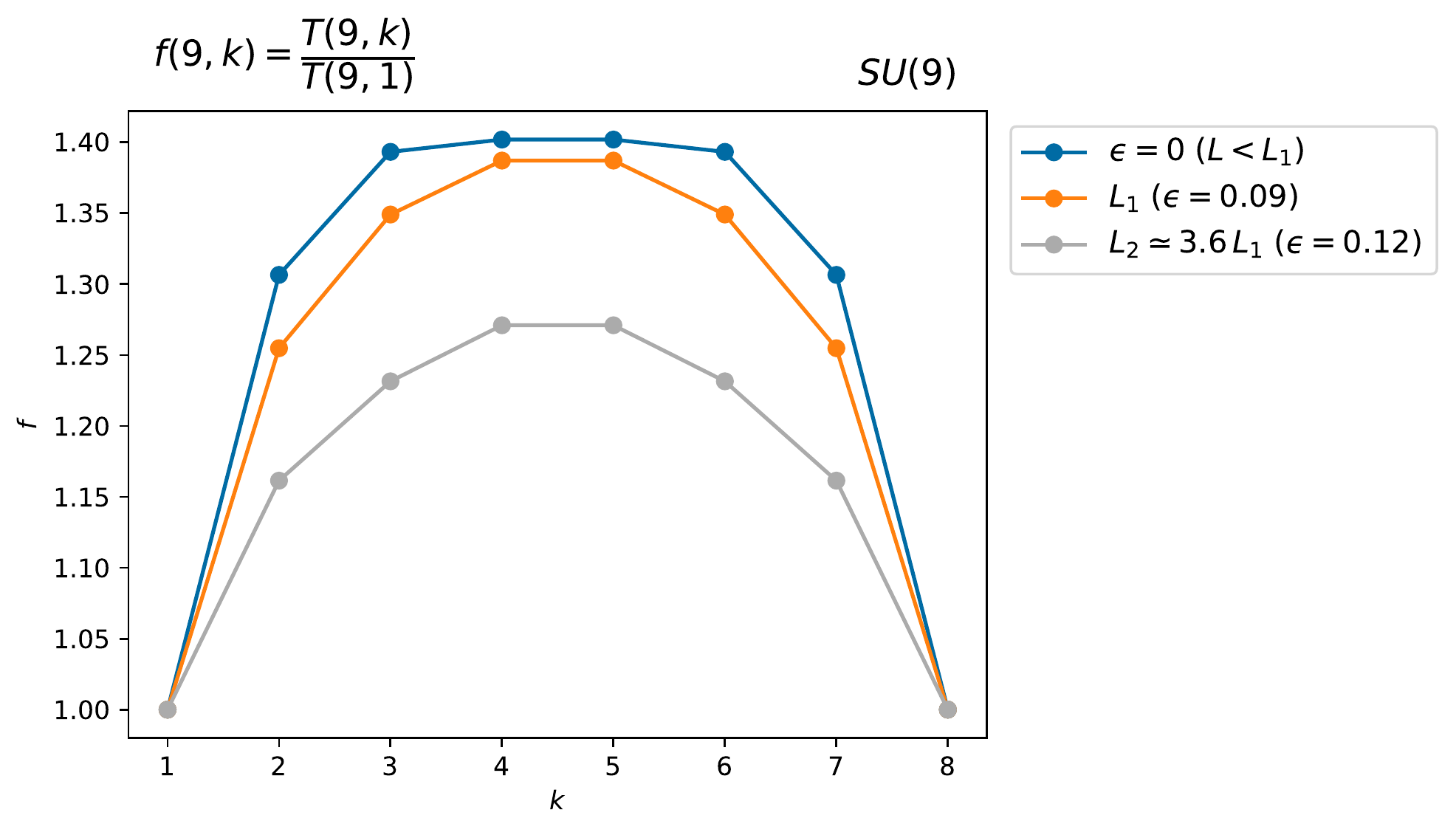}
    \caption{The $k$-string tension ratio for $SU(9)$, as a function of $L$ at fixed $\Lambda$, for three different values of $L$, increasing from the top curve towards the bottom one. Clearly, as $L$ increases, the $k$-string ratios are seen to decrease. Qualitatively, however, the curve near its maximum is not as flat as for $\epsilon=0$. This effect is entirely due to $W$-boson (and superpartner) loops. The values of the semiclassical expansion parameters corresponding to the chosen $\epsilon$ are discussed after eqn.~(\ref{twolengths}).}
    \label{fig:kratioL}
\end{figure}

However, the behavior of the $k$-string tension ratio, $f(N,k)$,  as a function of $L$ cannot be deduced without knowledge of $\tilde T(\epsilon)$ of (\ref{tensiontables009}), as the prefactor in (\ref{Mmepsilon}) cancels when taking the ratio. Even before discussing these data, we can make the following qualitative statement. If the double-string picture remains intact (as it is seen to), we expect that the string tension is approximately twice the BPS 1-wall tension. Then the K\" ahler potential-independence of the latter implies that the approximately flat behavior of $f(N,k)$ as a function of $k$ will persist. This is indeed seen from the results of (\ref{tensiontables009}). 
The result which is difficult to predict without  simulating is that $f(N,k)$ varies even less with $k$ as $L$ increases between the two values shown in (\ref{twolengths}). In Fig.~\ref{fig:kratioL}, we show the $k$-string tension ratio for the largest value $N=9$, for the two values of $L$ of (\ref{twolengths}) as well as for $\epsilon=0$.\footnote{We stress that a ratio of $L_1$ to the size of $\S^1$ that corresponds to $\epsilon=0$ cannot be determined from our results; one can only say that the $L$ which corresponds to $\epsilon=0$ is smaller than $L_1$.} One qualitative observation is that while the range of variation of $f(9,k)$ is smaller for larger $L$, the $f(9,k)$ appears to have more curvature near its maximum.

To conclude this section, we find that the qualitative double-string picture is not changed by including the leading virtual $W$-boson correction, upon increase of $L$ within the semiclassical regime. The main qualitative effect of the one-loop correction is the widening of the double string. At the same time, the  range of variation of  the   $k$-string tension ratio $f(N,k)$ decreases upon increase of $L$.

\section{Domain Wall Interactions in One Dimension} 
\label{sec:6}

In order to  to elucidate the behavior of the double-string configurations observed in two dimensions, we now
 perform a study of domain wall interactions in one spatial dimension. The morphology and behavior of individual BPS solitons in 1D was studied extensively in \cite{Cox:2019aji}. Here, we extend these results with a numerical study of the interaction of multiple DWs. In this section, we set the K\"ahler metric to $K^{ab} = \delta^{ab}$.

\subsection{Interaction Energy of Two BPS Domain Walls}
 \label{sec:6.1}

\begin{figure}[ht]
    \includegraphics[width=\textwidth]{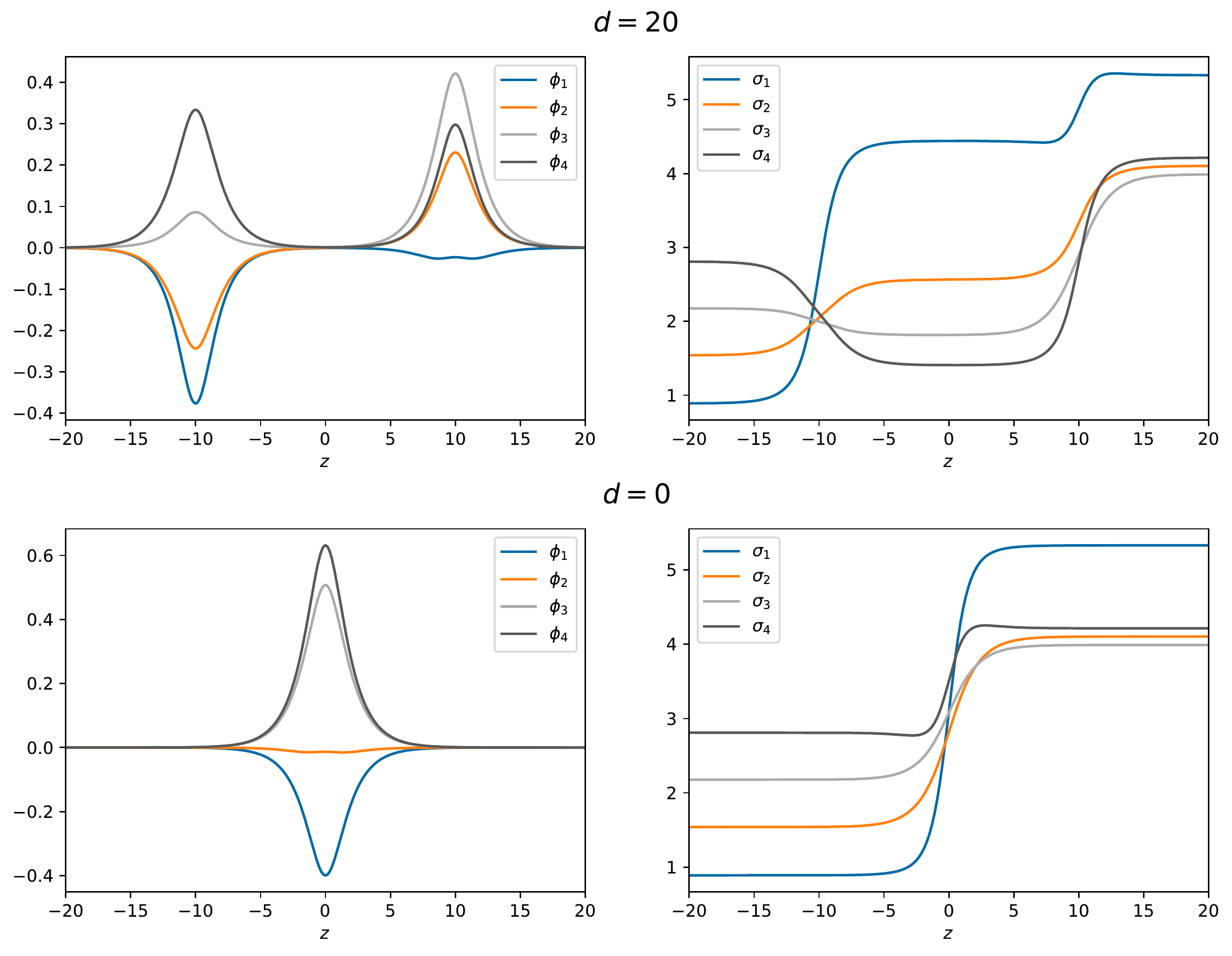}
    \caption{The sum of two BPS 1-walls for $N=5$ and $N$-ality $k=1$, taken at two different kink separations $d = 20$ and $d = 0$. The 1-wall situated to the left of the origin has flux $\bm{\Phi}_k^{1,I} =2 \pi \bm{\rho} / N - 2 \pi \bm{w}_k$, and the 1-wall to the right has flux $\bm{\Phi}^{1, II} = 2 \pi \bm{\rho} / N$. The corresponding boundary values for the $\bm{\sigma}$ field are $2 \pi \bm{\rho} / N$ at $z=-\infty$ and $2 \pi \bm{\rho} / N + 2 \pi \bm{w}_1$ at $z=\infty$.}
    \label{fig:string_separation}
\end{figure}

As observed in our 2D numerical study, quarks of $N$-ality $k = 1$ are confined by a single string for $N > 4$, with a string tension lower than that of a BPS 1-wall. We may verify that this behavior is consistent with the 1D picture by considering the interaction energy of two parallel 1-walls as a function of their spatial separation. To do so, we first numerically solve for two BPS 1-walls with electric fluxes $\bm{\Phi}_k^{1,I} =2 \pi \bm{\rho} / N - 2 \pi \bm{w}_k$ and $\bm{\Phi}^{1, II} = 2 \pi \bm{\rho} / N$ according to (\ref{eq:1deom}), and using the notation for fluxes from (\ref{fluxes1}). Note that these match the fluxes of our double-string configuration with weight $\bm{w}_k$. We then sum the two walls, separating them by a certain distance $d$ as measured from the centers of the kinks. Here, the ``center'' of a 1-wall is taken as the peak of the $\bm{\phi}$ fields.  An example of this configuration for $N=5$ and $k=1$ is shown in Figure \ref{fig:string_separation}, for two different values of $d$. The total energy of this configuration as a function of the separation $d$ can then be computed via (\ref{dimlessenergy}). This gives an idea of the long-distance  (compared to their size) interactions of these DWs, as in the ``sum ansatz" used to study  instanton-anti-instanton interactions, see e.g.~\cite{Schafer:1996wv} for a review.

\begin{figure}[ht]
    \includegraphics[width=\textwidth]{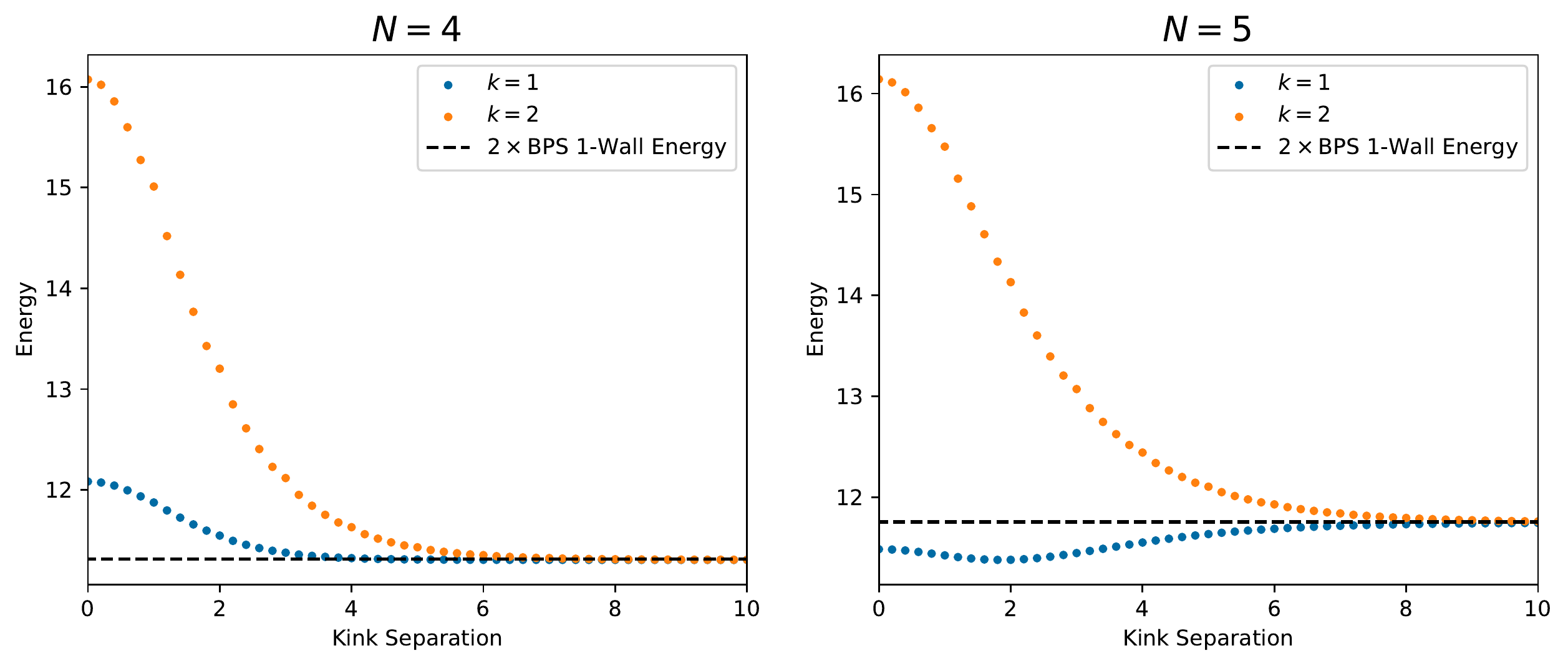}
    \caption{Total energy of the sum of two BPS 1-walls as a function of kink separation. Energies are depicted for $N = 4, 5$ and $N$-alities $k = 1, 2$. Twice the energy of a BPS 1-wall, as given by the dimensionless form of (\ref{dwtension}), is also marked by a dashed line. At large kink separations, the energy of the summed configuration approaches twice the BPS 1-wall energy asymptotically. At small kink separations, we can see from the slope of the energy curves that DWs of $N$-ality $k=2$ tend to repel, whereas DWs of $N$-ality $k=1$ are either only weakly repulsive in the case of $N=4$, or slightly attractive in the case of $N=5$. Note that these results hold for larger and smaller studied gauge groups: $N$-ality $k=1$ continues to be attractive for $N > 5$ and repulsive for $N < 4$, whereas all other $N$-alities are repulsive for all $N$ studied.}
    \label{fig:interaction_energy}
\end{figure}

For several different values of $N$ and $k$, we computed the energy of these interacting DWs for kink separations $d$ between 0 and 10. These energies are depicted for $N=4,5$ in Figure \ref{fig:interaction_energy}. Here, we observe that DWs of $N$-ality $k=2$ have a tendency to repel at small kink separations, as expected from our 2D simulations. In contrast, for $N$-ality $k=1$, DWs exhibit either weak repulsion\footnote{Here, ``weak'' is defined relative to the repulsion of the $N$-ality $k=2$ configurations.} in the case of $N=4$, or slight attraction in the case of $N=5$. For all $N$ and $k$, the energy converges asymptotically to twice the BPS 1-wall tension as the kink separation is taken to infinity. These results hold for all other gauge groups considered in this study: $N$-ality $k=1$ remains attractive at small kink separations for $N > 5$, and repulsive for $N < 4$; similarly, for all $N$-alities $k > 1$ and all $N$, the DWs are repulsive.

These results allow us to put the behavior of the 2D confining strings into greater context. As noted above, the repulsion of the DWs for $k > 1$ is consistent with the emergence of the double-string configurations observed in the 2D simulations. Moreover, the attraction of the DWs for $k = 1$ is consistent with the observed collapse into a single-string configuration for $N \geq 5$ and $k = 1$. Notably, the behavior for $N=4$, $k=1$ is more complicated. Although these DWs seem to be slightly repulsive at small kink separations, in our 2D numerical simulations we observed these walls collapse into a single string at smaller quark separations. It is possible that since these walls are merely weakly repulsive relative to the $k=2$ configuration, there exists some other property of the confining string for which a single string is energetically preferable at small quark separations. For example, morphological properties such as the additional string length and/or curvature of the double-string configuration may make a single string the preferred solution at these small quark separations.

\subsection{Attraction and Repulsion of Domain Walls} \label{sec:6.2}

We further investigated the attraction and repulsion of DWs by numerically relaxing configurations consisting of the sum of two DWs similar to those depicted in Figure \ref{fig:string_separation} (this way of studying DW interactions is motivated by the ``streamline" or ``valley" method of \cite{Balitsky:1986qn}, see also \cite{Schafer:1996wv}).

\begin{figure}[ht]
    \includegraphics[width=\textwidth]{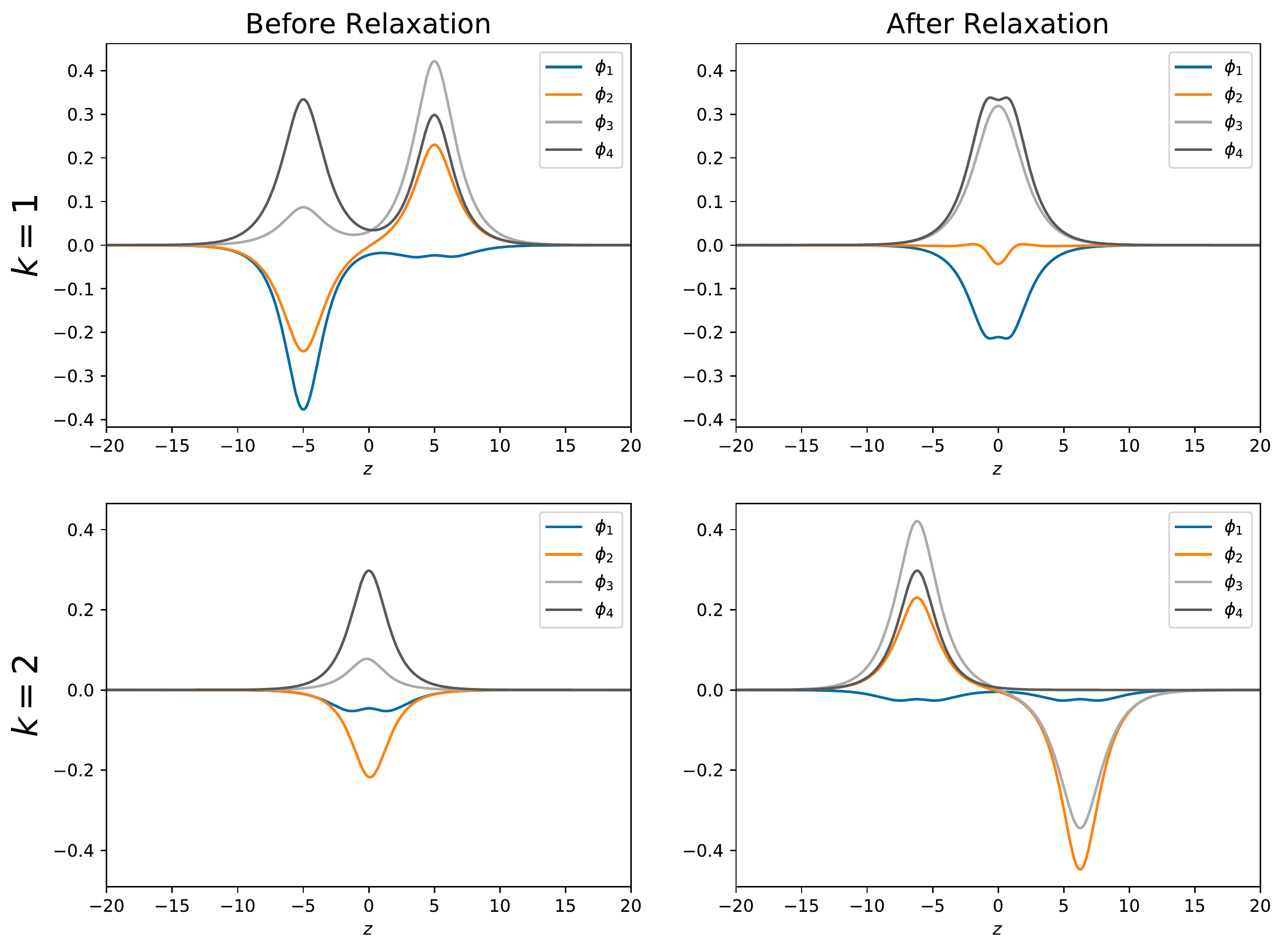}
    \caption{Attraction and repulsion of DWs for $N=5$ and $k=1,2$. The initial conditions, of which the $\bm{\phi}$ fields are depicted in the left column, consist of the sum of two BPS 1-walls similar to those depicted in Figure \ref{fig:string_separation}. These initial conditions were relaxed for up to 100,000 iterations of the relaxation algorithm, with the usual error tolerance of $10^{-9}$. The $\bm{\phi}$ fields for these solutions are depicted in the right column. In the top row, an $N$-ality $k=1$ configuration of two DWs separated by 10 units is seen to attract to form a single string. The resultant $\bm{\phi}$ field is comparable with---but not identical to---the sum of the $\bm{\phi}$ fields of the individual 1-walls, having a smaller magnitude and an additional central divot as compared to the summed field (compare with the lower-left panel of Figure \ref{fig:string_separation}). Although not depicted here, the $\bm{\sigma}$ field of the relaxed solution appears as the sum of the $\bm{\sigma}$ fields of the individual 1-walls, as depicted in the lower-right panel of Figure \ref{fig:string_separation}. In the bottom row, an $N$-ality $k=2$ configuration with an initial separation of 0 is seen to repel to form a double string. In this case, the resultant $\bm{\phi}$ and $\bm{\sigma}$ fields both appear as the simple sum of the individual 1-walls, reminiscent of the top row of Figure \ref{fig:string_separation}.}
    \label{fig:dw_attraction}
\end{figure}

 An example of this for $N = 5$ and $k = 1,2$ is shown in Figure \ref{fig:dw_attraction}. Here, a configuration of two DWs of $N$-ality $k=1$ is seen to attract to form a single string after relaxation, whereas a similar configuration of $N$-ality $k=2$ is seen to repel after a similar procedure. In general, we observe that the attraction or repulsion of DWs configured as described above is consistent with the interaction energy of the walls studied in the previous section by the ``sum ansatz." For example, configurations of $N$-ality $k=1$ are seen to repel for $N < 4$ and attract for $N \geq 5$.

For those DW configurations which are observed to repel, the solution resulting from numerical relaxation is seen to closely correspond with the simple sum of the $\bm{\phi}$ and $\bm{\sigma}$ fields of the individual 1-walls taken a distance apart. For those configurations which attract, although the $\bm{\sigma}$ fields of the 1-walls appear to simply sum together, the $\bm{\phi}$ fields behave differently. The $\bm{\phi}$ fields of these ``collapsed'' solutions are reminiscent of sum of the two BPS DWs, but have a smaller magnitude and posses an additional ``divot'' at their centers (for instance, compare the lower-left panel of Figure \ref{fig:string_separation} with the upper-right panel of Figure \ref{fig:dw_attraction}). We make no attempt to explain this phenomenon in this work, other than to point out the obvious fact that these DW interactions are highly nonlinear. We are fortunate, therefore, that the $\bm{\sigma}$ fields of these DWs appear to interact linearly or almost linearly.\footnote{Scalar theories with a complex potential equal to our superpotential $W$, the affine Toda theories studied in \cite{Hollowood:1992by}, are integrable. We do not know if there is any connection between this fact and the ``almost linear" property of the $\sigma$ fields alluded to above. }

\subsection{Comparison with Confining Strings} \label{sec:6.3}
We also compare the profile of these one-dimensional DWs with the $y = 0$ (midpoint between the quarks) cross sections of our confining string configurations. Although these cross sections are discontinuous at $z = 0$ due to the imposition of the monodromy, we can remove this discontinuity by translating the $\bm{\sigma}$ field in the region $z > 0$. In particular, for a quark-antiquark pair of weight $\pm \bm{w}_k$, translating the $\bm{\sigma}$ field by $2 \pi \bm{w}_k$ makes the field continuous. This allows us to directly compare the cross sections of the confining strings with the 1D solutions from Section \ref{sec:6.2} above.

\begin{figure}[ht]
    \includegraphics[width=\textwidth]{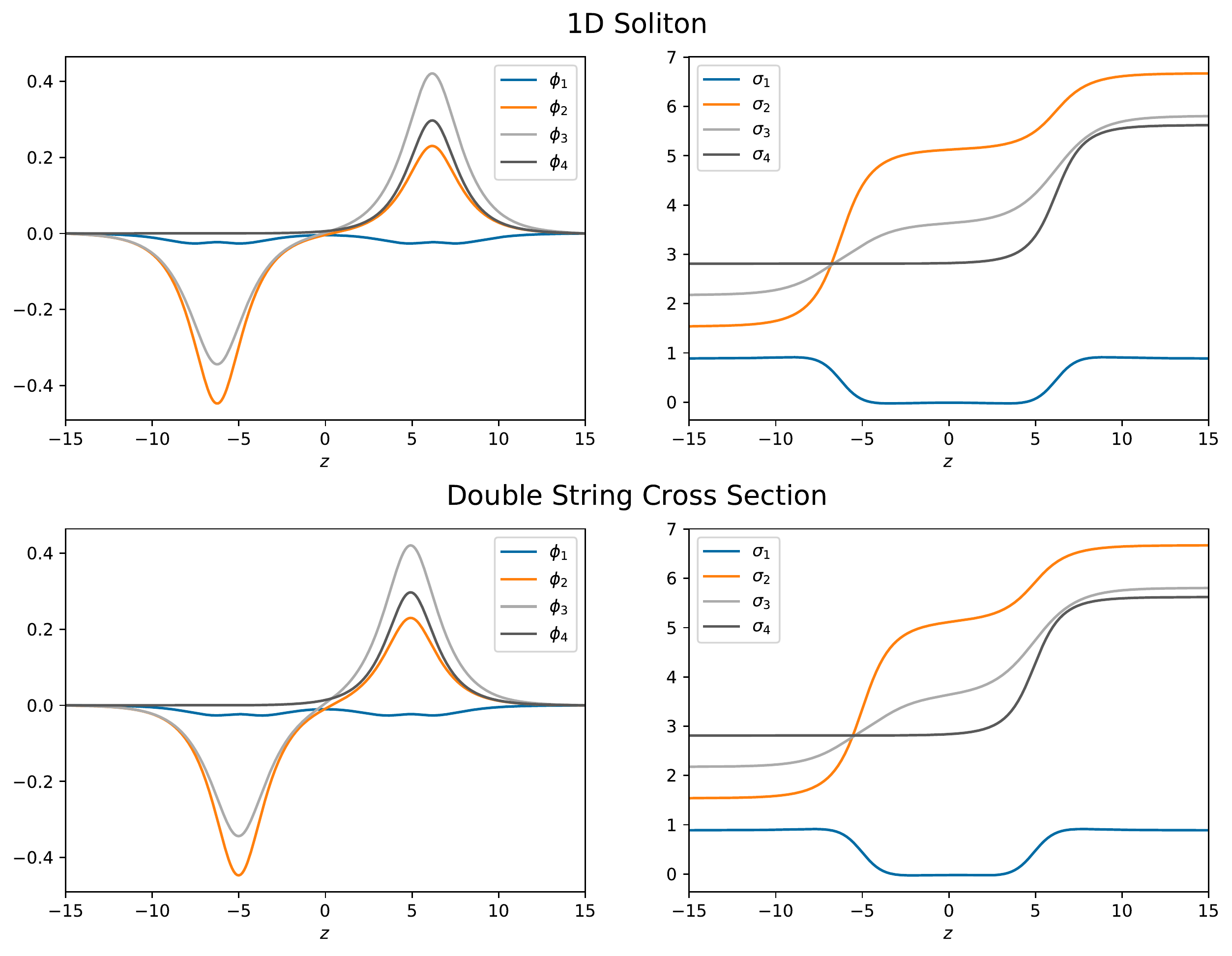}
    \caption{One-dimensional DW versus the $y=0$ cross section of a double string for $N=5$ and $N$-ality $k=2$. The top row shows the field of the 1D DW, which was produced by relaxing a configuration consisting of the sum of two 1-walls, as described in Section \ref{sec:6.2}. The bottom row shows a $y=0$ cross section similar to that depicted in Figure \ref{fig:cross section}, although here the $\bm{\sigma}$ field has been translated by $2 \pi \bm{w}_2$ in the region $z > 0$ such that it is continuous.}
    \label{fig:cross_section_vs_1d}
\end{figure}

An example of such a comparison for $N=5$ and $k=2$ is shown in Figure \ref{fig:cross_section_vs_1d}. Here, the relaxed sum of two 1D DWs is compared with a similar cross section to that shown in Figure \ref{fig:cross section}, although translated as described above to impose continuity. It is clear from this figure that the cross section closely replicates the 1D solution aside from a small difference in the string separations, for which the double string obeys the log-growth property described in Section \ref{sec:5.3}.

In general, we observed that the $y = 0$ cross sections for all of our confining strings were very similar to the corresponding one-dimensional configurations of two summed and relaxed BPS solitons. This held true for both the double-string configurations and the collapsed single-string configurations. Indeed, features of the collapsed 1D configurations---such as the nonlinear interaction of the $\bm{\phi}$ fields like that depicted in the upper-right panel of Figure \ref{fig:dw_attraction}---are also replicated in the corresponding confining-string cross sections. As such, we believe that many properties of the confining strings, such as string tension and the tendency to collapse, can be predicted from the behavior of 1D DWs. This allows us to consider properties such as the behavior of higher gauge groups, or of weights in $\ZN$ orbits different from that of $\bm w_k$, without the significant computational demands of the two-dimensional configurations.

\subsection{String Tensions of Differing $\ZN$ Orbits} \label{sec:6.4}
In this section, we study  the tensions of double-string configurations with $N$-ality $k$ weights belonging to different $\ZN$ orbits. Recall that our work up to this point has focused primarily on the string tensions of the orbits belonging to the highest weight, $\bm{w}_k$, of the $k$-index antisymmetric representation. As discussed in Section \ref{taxonomy}, there exist double-string configurations of $N$-ality $k$ belonging to differing $\ZN$ orbits, which can be confined by up-to BPS $k$-walls. However, if the string tensions of these configurations are higher than that of the highest-weight orbit, we would expect that such weights would be screened by the pair-production of $W$-bosons down to $\bm{w}_k$, i.e., such that the configuration is confined by BPS 1-walls. As discussed above, we can study the string tensions of these $\ZN$ orbits in a one-dimensional environment with the expectation that the corresponding 2D string tensions will be similar. In this section, therefore, we confirm that the orbits of $N$-ality $k=2$ are expected to decay to the highest-weight orbit.

Recall from Section \ref{sec:3.2.3} that the $\ZN$ orbits of $N$-ality $k=2$ can be labeled by representative weights $\bm{\nu}_{1,1+p} = \bm{\nu}_1 + \bm{\nu}_{1 + p}$, where $p = 0, \ldots, N-1$. We may estimate the string tension of these weights by numerically relaxing a 1D configuration with total flux $\bm{\Phi} = 2 \pi \bm{\nu}_{1,1+p}$. In particular, we set up our field with boundary values $\bm{\sigma}(-\infty) = 0$ and $\bm{\sigma}(\infty) = 2 \pi \bm{\nu}_{1,1+p}$, and with initial conditions linearly interpolating between these boundary values. When relaxing such a configuration, we found that the algorithm converged to the expected DW configuration of the given flux, and in particular replicated the $y=0$ cross section of the corresponding 2D confining string whenever the 2D solution was available. Therefore, the tensions of such configurations can be used to approximate the string tensions of the 2D confining strings. We stress that these 1D tensions are only approximations of the 2D tensions, and are subject to some variation depending on the initial conditions used. In particular, these solutions may relax to some local minimum of energy instead of the global minimum. The string tensions reported herein represent the minimum energy configurations observed in our tests.

\begin{figure}[ht]
    \includegraphics[width=\textwidth]{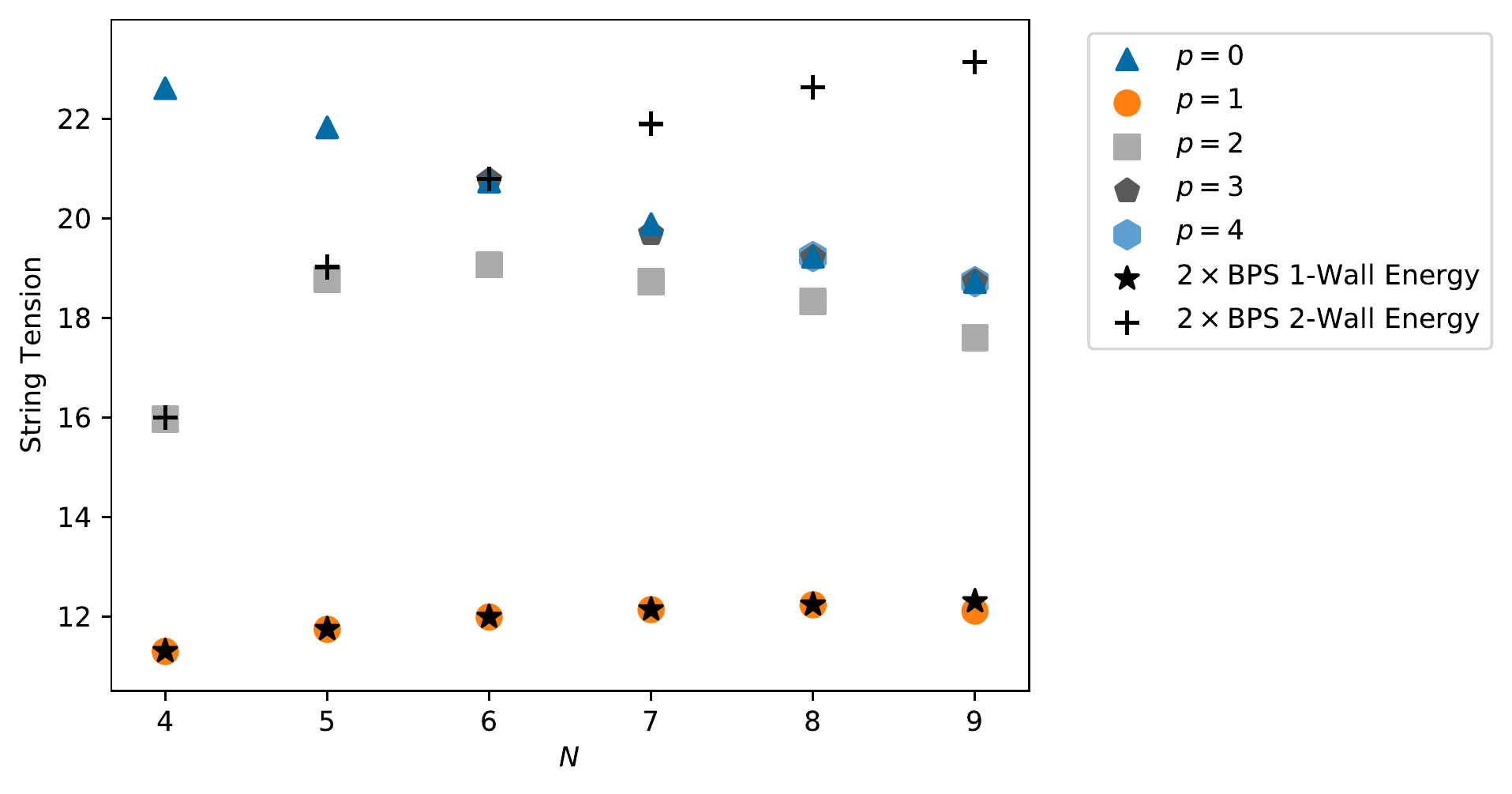}
    \caption{Approximate string tensions for weights in differing $\ZN$ orbits of $N$-ality $k=2$. The weight of each point is $\bm{\nu}_{1,1+p} = \bm{\nu}_1 + \bm{\nu}_{1 + p}$. Also marked are twice the BPS 1-wall and 2-wall tensions for each $N$. The lowest tension orbit belongs to the highest weight, $\bm{\nu}_{1,2} = \bm{w}_2$, and aligns closely with the tension of two BPS 1-walls. Notably, orbits with $p > 1$ and $N > 5$ appear to form a bound state with string tension lower than twice the BPS 2-wall tension, despite the expected double-string configuration for these weights being composed of 2-walls. This may be related to the $N$-ality $k=1$ collapsing effect which led to the single-string configurations.}
    \label{fig:orbits}
\end{figure}

Figure \ref{fig:orbits} plots the approximate string tensions of each $N$-ality $k$$=$$2$ orbit for $N=4,...,9$, computed from the one-dimensional setup described above. Also plotted are twice the BPS 1-wall and 2-wall tensions for comparison. As expected, the $p=1$ orbit belonging to the highest weight consistently exhibits the lowest string tension, and closely follows twice the BPS 1-wall string tension for all $N$. As such, the assumption that strings will decay to the highest weight appears to be valid for $N$-ality $k=2$. Note that the string tensions of the $p=1$ orbit differ slightly from twice the BPS 1-wall tensions due to the effect of the binding energy of the two walls, although this effect is too small to be visible on the plot for $N<9$. It is also notable that the $\ZN$ orbits with $p > 1$ do not strictly match the tension of two BPS 2-walls, despite the expectation that the double-string configurations for these orbits would consist of 2-walls. Instead, for $N > 5$ some of these orbits appear to form a lower-energy bound state. This may be related to the collapse of the confining string for $N$-ality $k=1$ studied earlier.

\begin{figure}[ht]
    \includegraphics[width=\textwidth]{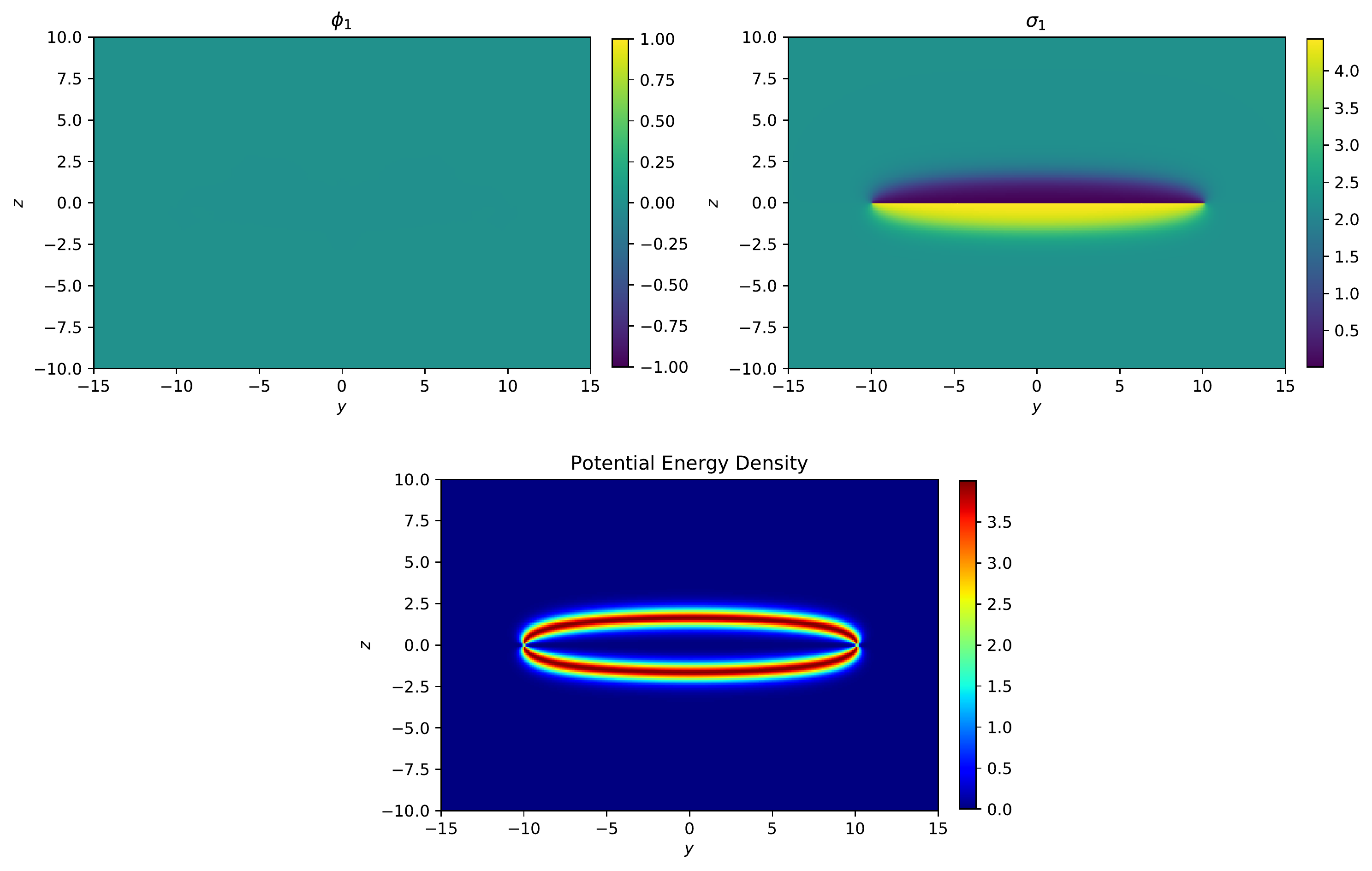}
    \caption{$N=4$ double-string configuration with weight $\bm{\nu}_{1,3}$. As a member of the $p=2$ orbit, this configuration is confined by two BPS 2-walls, as expected from eqn.~(\ref{porbits}). Note that these walls are ``magnetless,'' i.e., the $\bm{\phi}$ field vanishes everywhere (see \cite{Cox:2019aji}). It is expected that, for a large enough separation, pair production of $W$-bosons will cause this double-string configuration to decay to the energetically more favorable one made of two BPS 1-walls.}
    \label{fig:2-wall}
\end{figure}

We also confirmed that the $p=2$ orbit is confined by 2-walls for $N=4,5$ by considering the two-dimensional double-string configurations with weight $\nu_{1,3}$. Figure \ref{fig:2-wall} gives an example of this for $N=4$. Here, the configuration is confined by a ``magnetless'' 2-wall, i.e., a 2-wall where the $\bm{\phi}$ field vanishes identically. Indeed, it was shown in \cite{Cox:2019aji} that such magnetless $k$-walls can exist only for even $N$ and for $k = N/2$. Thus, the vanishing of the $\bm{\phi}$ field in Figure \ref{fig:2-wall} is consistent with confinement by a BPS 2-wall. 

\subsection{``Gluon Confinement'' and Triple Strings}
\label{sec:6.5}

\begin{figure}[ht]
    \includegraphics[width=\textwidth]{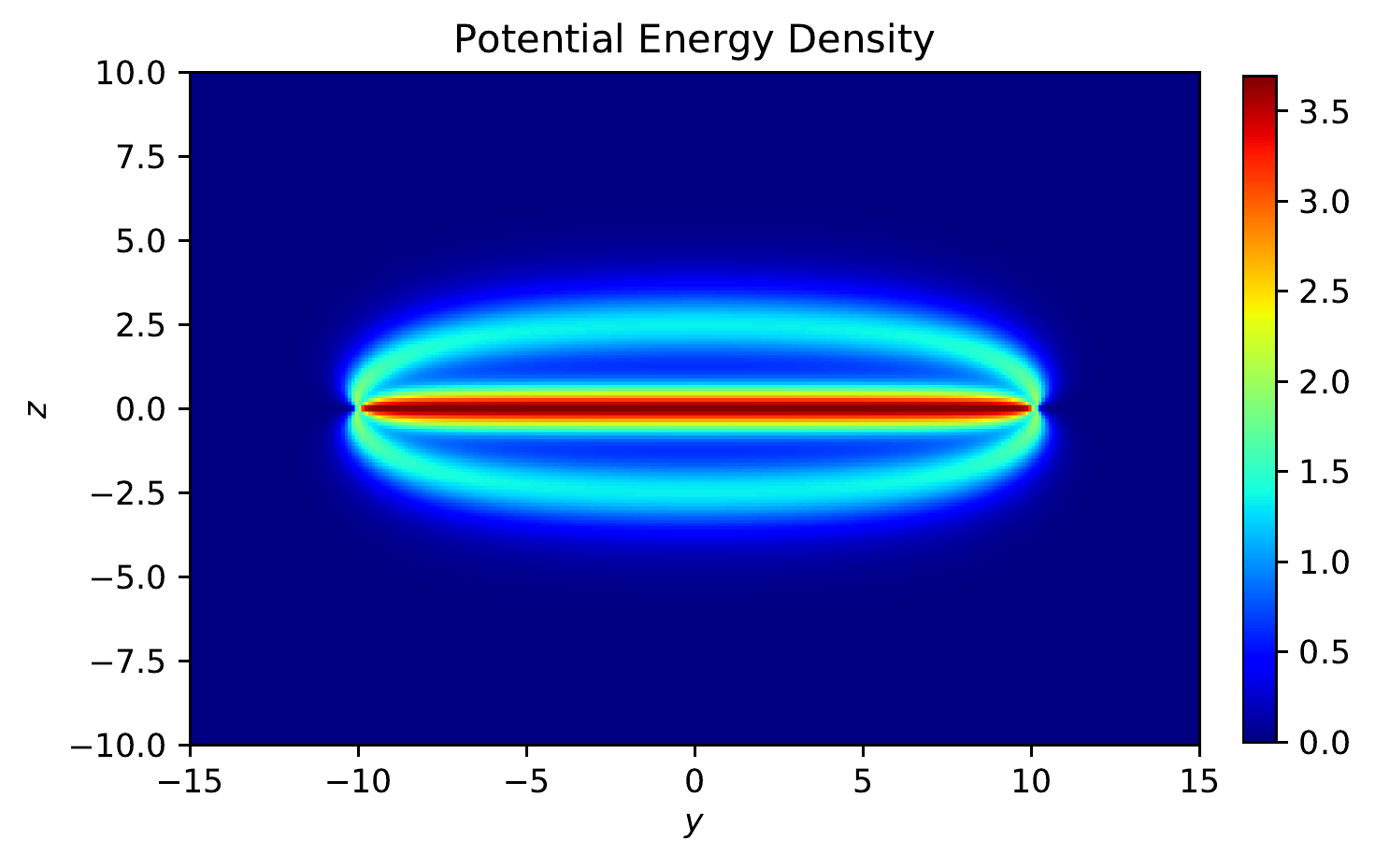}
    \caption{``Gluon confinement" by a triple string. Plotted is the potential energy density of an $N=5$ confining string with a root-lattice weight $\bm{\alpha}_1$. Uniquely, this configuration possesses three distinct flux tubes: two BPS 1-walls and a single BPS 3-wall (the two distinct vacua inside the triple-string configuration are now the $k=1$ and $k=4$ vacua of $SU(5)$ SYM, while the string is embedded in the $k=0$ vacuum). At large enough separation, this triple string should break due to the pair production of $W$-bosons, i.e., the adjoint Wilson loop should show perimeter law.}
    \label{fig:root}
\end{figure}

As a final note, we also considered the behavior of a confining string with a weight belonging to the root lattice, corresponding to replacing the quark static sources by gluons ($W$-bosons). Such a configuration has $N$-ality 0, and accordingly does not belong to any of the $\ZN$ orbits described earlier. The potential energy density of such confining strings for $N=5$ and with weight $\bm{\alpha}_1$ is shown in Figure \ref{fig:root}. Unlike all previous confining strings, this configuration possesses three distinct flux tubes: a single tube with a strong flux forming a straight line between the gluons, as well as two weaker flux tubes above and below this line. In fact, the uppermost and lowermost flux tubes are BPS 1-walls with fluxes $\bm\Phi^{1,I}_1 = {2 \pi \bm\rho\over 5} - 2 \pi \bm w_1$, while the thick wall is a BPS 3-wall with flux $\bm\Phi^{3,I}_{1,3,4} = 3 {2 \pi  \bm\rho\over 5} - 2 \pi (\bm w_1 + \bm w_3 + \bm w_4)$, using the notation from (\ref{fluxes1}).  The two vacua inside the triple strings are the $k=1$ and $k=4$ vacua (\ref{vacua}) and the non-Cartan gluon sources are embedded in the $k=0$ vacuum. We took the signs of the fluxes such that the total flux is the sum of the fluxes of the three DWs: $2 \bm\Phi^{1,I}_1  + \bm\Phi^{3,I}_{1,3,4} = - 2 \pi \bm\alpha_1$.
This effect is consistent with the one-dimensional configuration of equivalent flux when compared with the $y=0$ cross section of the confining string. As explained earlier, at large enough separation between the root-lattice sources our abelian EFT breaks down and pair production of $W$-bosons is expected to break this triple flux tube.

\bigskip

{\flushleft{\bf Acknowledgements:}} We thank Andrew Cox for collaboration in the early stage of the project. We further thank Mohamed Anber and Aleksey Cherman for discussions and useful suggestions. This work is supported by an NSERC Discovery Grant. MB also acknowledges support from an undergraduate University of Toronto Excellence Award. Computations were performed on the Niagara supercomputer at the SciNet HPC Consortium \cite{niagara, scinet}. SciNet is funded by: the Canada Foundation for Innovation; the Government of Ontario; Ontario Research Fund - Research Excellence; and the University of Toronto.

\appendix
\section{A Note on Edge Effects} \label{appendix:edge}
In our simulations of the double string, we approximate the edge of the rectangular grid to be spatial infinity, sometimes even if the quarks are located relatively close to the boundary. Why is this allowed? After all, in standard electrodynamics, the classic image-charge problem of solving the Poisson's equation in a grounded box is very different from that of just solving it in empty space.

The answer is that unlike standard electromagnetism, our theory is confining, that is, the field lines do not naturally spread to infinity and so the edge effect is weak. In fact, it is easy to show that if we turn off the superpotential $W$ in our problem, the desired logarithmic growth of the  electrostatic energy as a function of quark separation $R$ quickly gets distorted by the edge effect.

\begin{figure}[h]
    \centering
    \includegraphics[width=0.8 \textwidth]{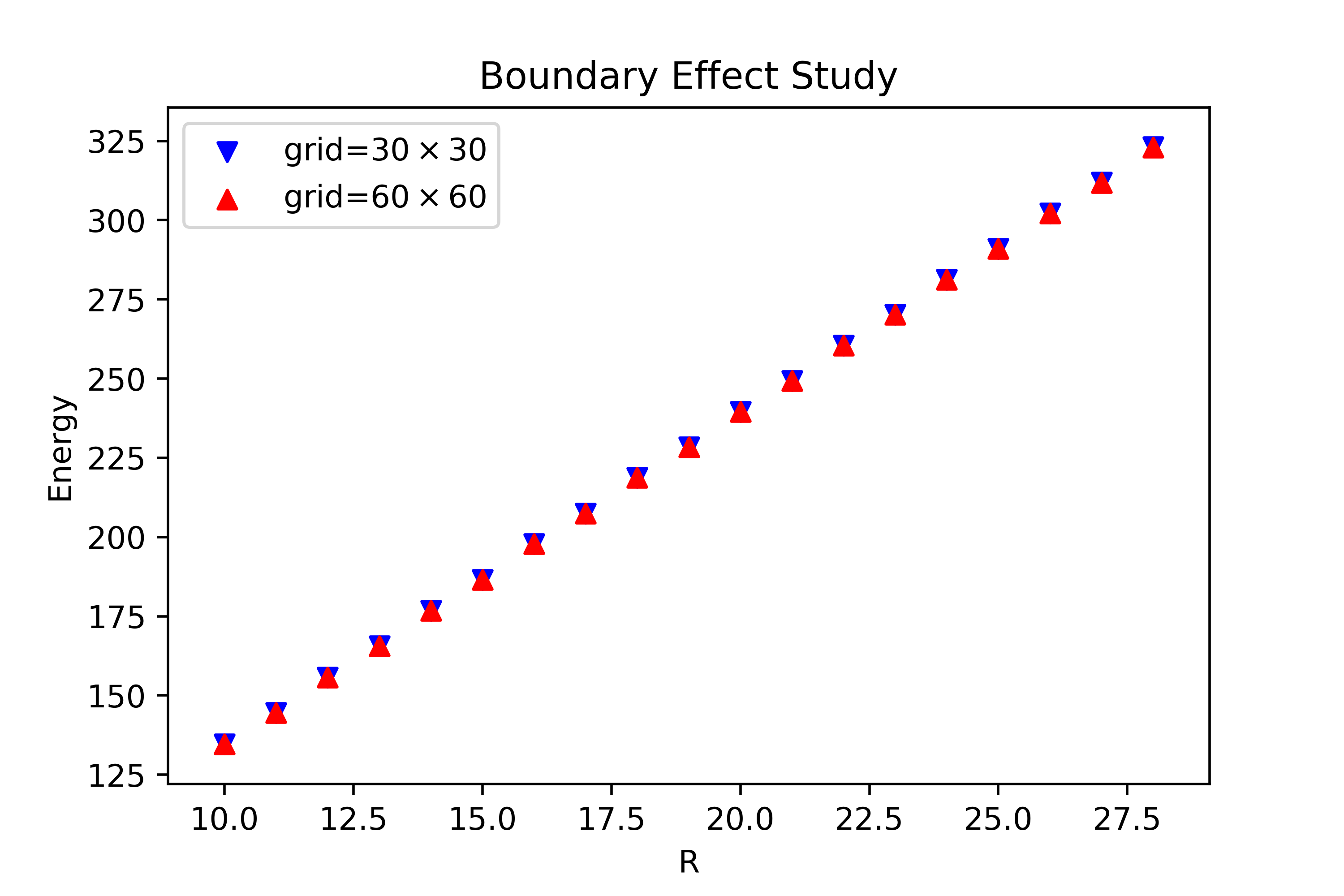}
    \caption{The energy dependence on quark separation $R$ in $SU(3)$ and N-ality $k=1$ is tested separately in a 30 by 30 grid and a 60 by 60 grid. The fact that they agree even when the quarks are right at the edge of the smaller grid ($R=28$) shows that the boundary effect in this confining theory is minimal, and that taking the finite distance boundary to approximate spatial infinity is justified.}
    \label{fig:boundary_effect}
\end{figure}

To dispel all doubts that the edge effect is minimal here, we tested the energy dependence on $R$ separately in a grid of size 30 by 30 and another of size 60 by 60.  We push the quarks right up to the edge in the size 30 grid (with $R=28$) and show that the energy still reproduces the ``true" value, that of the larger grid; see Figure \ref{fig:boundary_effect}.

\section{Methods for Speed Up} \label{appendix:speedup}
There are various standard methods to speed up the Gauss-Seidel convergence process. We will not describe them in any detail, except for mentioning that while using the successive overrelaxation method \cite{Numerical_Method}, we found that the overrelaxation parameter can be set to as high as $1.96$ in most cases tested, while a value any larger might prevent convergence.

We now briefly describe two speed-up methods specific to the present problem (a third one is the BPS initial condition, as described in Section \ref{sec:4.2}.

{\flushleft \bf Half-Grid Method:} We can exploit the fact that (\ref{eq:eom2}) and its boundary condition are symmetric under $y \to -y$ to reduce the computational time by a factor of 2. We first cut the standard grid in half along the $y$-direction. Along the three original sides of the resultant half-rectangle, the boundary condition remains the same, that is, $\bm x$ is still equal to one of the vacua there. Those three sides continue to be untouched by the Gauss-Seidel iterations, i.e. they respect a Dirichlet boundary condition. The new boundary that comes from slicing the grid now respects a Neumann boundary condition: it must have a vanishing derivative in the $y$-direction. We implement this by setting the values along this boundary to be equal to its neighboring value in the $y$-direction, at the end of every iteration. At the end, we reflect the solution to restore the full grid. Unfortunately, this trick cannot be repeated in the $z$-direction because $\partial_z \delta(z)$ is odd.

{\flushleft \bf Simplifying the EOM:} We are often interested in the regime where the one-loop correction can be ignored, in which case the K\" ahler metric is just the Kronecker delta: $K^{ab} = \delta^{ab}$. Then, (\ref{eq:eom2}) becomes
\begin{equation}
    \nabla^2 x^a = \frac{1}{4}\dwddw{a}{b} + 2 \pi i   (\vec{\lambda})^a \partial_z \delta(z) \int_{-R/2}^{+R/2} dy' \delta(y-y').
\end{equation}
Explicitly writing out the first term and using the identity $ \bm \alpha_i \cdot \bm \alpha_j = 2 \delta_{ij} - \delta_{i,j+1} - \delta_{i,j-1} $, where the index is taken modulo $N$, we have
\begin{eqnarray}
\dwddw{a}{b} &=& \sum_{i=1}^N \sum_{j=1}^N e^{\bm \alpha_i \cdot \bm x + \bm \alpha_j \cdot  \bar{\bm x}}  (\bm \alpha_j)^a \bm \alpha_i \cdot \bm \alpha_j \\
&=& \sum_{i=1}^N \left[ 2e^{\bm \alpha_i \cdot \bm x + \bm \alpha_i \cdot \bar{\bm x}} (\bm \alpha_i)^a - e^{\bm \alpha_i \cdot \bm x + \bm \alpha_{i-1} \cdot \bar{\bm x}} (\bm \alpha_{i-1})^a  - e^{\bm \alpha_i \cdot \bm x + \bm \alpha_{i+1} \cdot \bar{\bm x}} (\bm \alpha_{i+1})^a  \right] \\
&=& \sum_{i=1}^N (\bm \alpha_i)^a e^{\bm \alpha_i \cdot \bar{\bm x}} \left[ 2e^{\bm \alpha_i \cdot \bm x} - e^{\bm \alpha_{i+1} \cdot \bm x} - e^{\bm \alpha_{i-1} \cdot \bm x} \right].
\end{eqnarray}
This reduces a double nested for-loop to a single for-loop, which is a significant improvement in speed as this term must be evaluated at every point for every iteration. In addition, the 1D EOM (\ref{eq:1deom}) admits the same simplification.

\section{Tables of Numerical Results}
 \label{appendix:tables}

\subsubsection{String Tensions $T(N,k)$ for $N \le 10$}
In eqn.~(\ref{tensiontable}), we summarize the numerical results for string tensions $T(N,k)$ and their ratios $f(N,k) = T(N,k) / T(N,1)$, as well as the uncertainties arising from the least squares fit.

The string tension ratios $f(N,k)$ for the largest number of colors we studied, $N=10$, are plotted in Fig.~\ref{fig:f(10,k)} alongside other scaling laws; see also the discussion in Section~\ref{sec:5.2}.
\begin{eqnarray}\label{tensiontable}
\begin{array}{ |c|c|c|c| }
\hline
N & k & \tilde T & f \\
 \hline
 \hline
 2    &  1    &  7.9905 \pm  0.0009 &  1  \\
 \hline
 \hline
 3    &  1    & 10.401 \pm  0.002 &  1  \\
 \hline
 \hline
  4    &  1    & 11.321 \pm  0.001 &  1  \\
 \hline
    &  2    & 11.351 \pm  0.004 &  1.0026 \pm  0.0004 \\
 \hline
 \hline
 5    &  1    & 10.9127 \pm  0.0002 &  1   \\
 \hline
    &  2    & 11.838 \pm  0.007 &  1.0848 \pm  0.0007 \\
 \hline
 \hline
 6    &  1    & 10.3685 \pm  0.0006 &  1   \\
 \hline
    &  2    & 12.14 \pm  0.01 &  1.171 \pm  0.001 \\
 \hline
     &  3    & 12.14 \pm  0.01 &  1.171 \pm  0.001 \\
 \hline
 \hline
 7    &  1    &  9.9428 \pm  0.0003 &  1  \\
 \hline
     &  2    & 12.51 \pm  0.03 &  1.258 \pm  0.003 \\
 \hline
    &  3    & 12.66 \pm  0.05 &  1.274 \pm  0.005 \\
 \hline
 \hline
  8    &  1    &  9.6158 \pm  0.0004 &  1   \\
 \hline
     &  2    & 12.44 \pm  0.02 &  1.294 \pm 0.002 \\
 \hline
     &  3    & 12.94 \pm  0.06 &  1.346 \pm  0.006 \\
 \hline
     &  4    & 12.95 \pm  0.06 &  1.347 \pm  0.007 \\
 \hline
 \hline
 9    &  1    &  9.3627 \pm  0.0006 &  1   \\
 \hline
     &  2    & 12.23 \pm  0.02 &  1.307 \pm  0.002 \\
 \hline
     &  3    & 13.04 \pm  0.05 &  1.393 \pm  0.005 \\
 \hline
     &  4    & 13.13 \pm  0.06 &  1.402 \pm  0.006 \\
 \hline
 \hline
  10    &  1    &  9.20 \pm  0.02 &  1  \\
 \hline
    &  2    & 12.00 \pm  0.04 &  1.304 \pm  0.005  \\
 \hline
    &  3    & 13.10 \pm  0.09 &  1.42 \pm  0.01 \\
 \hline
   &  4    & 13.31 \pm  0.07 &  1.446 \pm  0.009 \\
 \hline
    &  5    & 13.40 \pm  0.09 &  1.46 \pm  0.01 \\
 \hline
\end{array}
\end{eqnarray}

\subsubsection{Double-String Separations}
The numerical results for fitted parameters of the logarithmic growth of string separation are summarized in (\ref{separation table}). We fitted the separation with the formula $d(R) = a \log(R) + b$. The two parameters $a$ and $b$ with their uncertainties from the least square fits are shown. For $N \geq 5$, note that because of the double-string collapse for $k=1$, string separation only makes sense when $k \neq 1$.
\begin{eqnarray}\label{separation table}
\begin{array}{ |c|c|c|c| }
        \hline
N & k & a & b \\
       \hline
       \hline
  2 &    1 &    0.98  \pm  0.01 &   0.33 \pm    0.05 \\
       \hline
       \hline
  3 &    1 &    1.37 \pm    0.01 &   -0.53 \pm    0.05  \\
       \hline
       \hline 
  4 &   1 &    1.87  \pm  0.02 &   -3.48 \pm    0.09 \\
       \hline
  4 &    2 &    1.97  \pm  0.01 &   -0.99 \pm    0.05 \\
       \hline
       \hline
  5 &    2 &    2.73  \pm    0.01 &   -2.69 \pm    0.05  \\
       \hline
       \hline
  6 &    2 &    3.46 \pm    0.01 &   -4.96 \pm    0.05 \\
       \hline
   &    3 &    3.73 \pm    0.01 &   -4.79 \pm    0.06 \\
       \hline
       \hline
  7 &    2 &    3.11 \pm    0.4 &   -4.6 \pm    0.1 \\
       \hline
   &    3 &    4.72 \pm    0.04 &   -7.6 \pm    0.1 \\
       \hline
       \hline
  8 &    2 &    2.31 \pm    0.04 &   -2.9 \pm    0.1 \\
       \hline
   &    3 &    5.62  \pm    0.04 &  -10.7 \pm    0.1 \\
       \hline
   &    4 &    6.26 \pm   0.05 &  -12.0 \pm    0.2 \\
       \hline
       \hline
  9 &    2 &    1.81  \pm    0.04 &   -1.7 \pm    0.1 \\
       \hline
   &    3 &    5.52 \pm    0.05 &  -10.8 \pm    0.2 \\
       \hline
   &    4 &    7.44 \pm    0.06 &  -16.0 \pm    0.2 \\
       \hline
       \hline
 10 &    2 &    1.31 \pm    0.06  &   -0.2 \pm    0.2 \\
       \hline
  &    3 &    4.9 \pm    0.2  &   -9.7 \pm    0.6 \\
       \hline
  &    4 &    8.0 \pm    0.1 &  -18.5 \pm    0.3 \\
       \hline
 &    5 &    8.8 \pm    0.2 &  -20.9 \pm    0.7 \\
\hline
\end{array}
\end{eqnarray}

\subsubsection{String Tensions $T(N,k)$ for $N \le 9$, with $W$-boson Loops: $\epsilon = 0.09$ vs. $\epsilon = 0.12$.}\label{appx:C.0.3}

In eqn.~(\ref{tensiontables009}), we summarize the numerical results for string tensions $T(N,k)$ and their ratios $f(N,k) = T(N,k) / T(N,1)$, as well as the uncertainties arising from the least squares fit, for $\epsilon = 0.09$ and $\epsilon = 0.12$.
\begin{eqnarray}\label{tensiontables009}
\begin{array}{ |c|c|c|c|||c|c|c|c| }
\hline
\epsilon=.09 & & & &\epsilon=.12 & & &   \\
\hline
N & k & \tilde T & f & N & k & \tilde T & f\\
 \hline
 \hline
 2    &  1    &  8.012 \pm  0.002 &  1  &2&1&8.08\pm 0.01&1\\
 \hline
 \hline
 3    &  1    & 10.445 \pm  0.005 &  1 &3&1&10.67\pm 0.04&1 \\
 \hline
 \hline
  4    &  1    & 11.358 \pm  0.003 &  1  &4&1&11.50\pm 0.01&1\\
 \hline
    &  2    & 11.431 \pm  0.009 &  1.0065 \pm  0.0009 &&2&11.77\pm 0.05&1.024\pm 0.004\\
 \hline
 \hline
 5    &  1    & 11.2785 \pm  0.0006 &  1 &5&1&11.628\pm 0.004&1  \\
 \hline
    &  2    & 11.96 \pm  0.01 &  1.060 \pm  0.001&&2&12.46\pm 0.06&1.071\pm 0.005 \\
 \hline
 \hline
 6    &  1    & 10.8364 \pm  0.0008 &  1 &6&1&11.37\pm 0.02& 1 \\
 \hline
    &  2    & 12.28 \pm  0.01 &  1.133\pm  0.001 &&2&12.84\pm 0.05&1.129\pm 0.005\\
 \hline
     &  3    & 12.32 \pm  0.02 &  1.137 \pm  0.002&&3&13.00\pm 0.06&1.143\pm 0.006 \\
 \hline
 \hline
 7    &  1    &  10.486 \pm  0.004 &  1 &7&1&11.19\pm 0.03&1 \\
 \hline
     &  2    & 12.77 \pm  0.04 &  1.218 \pm  0.004&&2&13.01\pm 0.04&1.163\pm 0.004 \\
 \hline
    &  3    & 13.10 \pm  0.06 &  1.250\pm  0.006&&3&13.43\pm 0.06&1.201\pm 0.006 \\
 \hline
 \hline
  8    &  1    &  10.218\pm  0.007&  1&8&1&11.19\pm 0.06& 1  \\
 \hline
     &  2    & 12.73 \pm  0.03 &  1.245 \pm 0.003&&2&12.99\pm 0.03&1.160\pm 0.006 \\
 \hline
     &  3    & 13.43 \pm  0.06 &  1.314 \pm  0.006 &&3&13.70\pm 0.05&1.224\pm 0.008\\
 \hline
     &  4    & 13.54 \pm  0.07 &  1.325 \pm  0.007 &&4&13.90\pm 0.06&1.242\pm 0.008\\
 \hline
 \hline
 9    &  1    &  10.02 \pm  0.01 &  1 &9&1&11.25\pm 0.07&1  \\
 \hline
     &  2    & 12.58 \pm  0.02 &  1.255 \pm  0.003&&2&13.061\pm 0.005&1.161\pm 0.008 \\
 \hline
     &  3    & 13.52 \pm  0.05 &  1.349 \pm  0.005&&3&13.85\pm 0.05&1.231\pm 0.009 \\
 \hline
     &  4    & 13.90 \pm  0.07 &  1.387 \pm  0.008&&4&14.29\pm 0.04&1.271\pm 0.009 \\
 \hline
\end{array}
\end{eqnarray}

\subsubsection{Double-String Separations, with $W$-boson Loops: $\epsilon = 0.09$ vs. $\epsilon=0.12$.}
\label{appx:C.0.4}

The numerical results for fitted parameters of the logarithmic growth of string separation for $\epsilon = 0.09$ and $\epsilon = 0.12$ are summarized in (\ref{separation tableeps009}). We fitted the separation with the formula $d(R) = a \log(R) + b$. The two parameters $a$ and $b$ with their uncertainties from the least square fits are shown. For $N \geq 5$, note that because of the double-string collapse for $k=1$, string separation only makes sense when $k \neq 1$. 
 \begin{eqnarray}\label{separation tableeps009}
\begin{array}{ |c|c|c|c|||c|c|c|c|}
\hline
\epsilon=.09 & & & &\epsilon=.12 & & &   \\
\hline
N & k & a & b&N &k &a &b \\
\hline
\hline  
  2 &    1 &    1.49  \pm  0.01 &   -0.04 \pm    0.05 &2&1&1.76 \pm 0.03&-0.2 \pm 0.1 \\
       \hline
       \hline
  3 &    1 &    2.36\pm    0.01 &   -2.31 \pm    0.05 &3&1&2.93	\pm 0.03& -3.3\pm	0.1 \\
       \hline
       \hline 
  4 &   1 &    3.42  \pm  0.02 &   -7.5\pm    0.1 &4&1&5.5\pm	0.1&	-13.9\pm	0.4 \\
       \hline
  &    2 &    3.44  \pm  0.01 &   -3.78\pm    0.06&&2&4.80\pm	0.04&	-7.3\pm	0.1 \\
       \hline
       \hline
  5 &    2 &    4.88 \pm    0.02 &   -7.86\pm    0.06  &5&2&7.23\pm	0.05&	-15.0\pm	0.2\\
       \hline
       \hline
  6 &    2 &    6.16 \pm    0.02 &   -12.47 \pm    0.07&6&2&9.47\pm	0.08&	-23.7\pm	0.3 \\
       \hline
   &    3 &    6.84 \pm    0.02 &   -13.34 \pm    0.07 &&3&10.20\pm	0.08&	-24.6\pm	0.3\\
       \hline
       \hline
  7 &    2 &    6.18 \pm    0.06 &   -13.5 \pm    0.2&7&2&13.5\pm	0.1&	-40.0\pm	0.4 \\
       \hline
   &    3 &    8.91 \pm    0.07 &   -20.4 \pm    0.2 &&3&14.3\pm	0.1&	-40.2\pm	0.4\\
       \hline
       \hline
  8 &    2 &    5.97 \pm    0.07 &   -13.7 \pm    0.2&8&2&16.9\pm	0.3&	-56\pm	1 \\
       \hline
   &    3 &    10.93 \pm    0.08 &  -28.4 \pm    0.3&&3&18.7\pm	0.2&	-58.8\pm	0.8 \\
       \hline
   &    4 &    12.58 \pm   0.09 &  -33.5 \pm    0.3&&4&21.1\pm	0.2&	-66.9\pm	0.8 \\
       \hline
       \hline
  9 &    2 &    5.89 \pm    0.08 &   -13.9 \pm    0.3&9&2&31\pm	2&	-116\pm	7 \\
       \hline
   &    3 &    11.3\pm    0.1 &  -30.9 \pm    0.4&&3&30.3\pm	0.5&	-106\pm	2 \\
       \hline
   &    4 &   17.0 \pm    0.1 &  -50.7 \pm    0.4&&4&37.8\pm	0.5&	-135\pm	2 \\
       \hline
\end{array}
\end{eqnarray}

  \bibliography{SYMNality.bib}

\providecommand{\href}[2]{#2}\begingroup\raggedright\begin{thebibliography}{10}

\bibitem{Greensite:2011zz}
J.~Greensite, {\it {An introduction to the confinement problem}},  {\em Lect.
  Notes Phys.} {\bf 821} (2011) 1--211.

\bibitem{Seiberg:1996nz}
N.~Seiberg and E.~Witten, {\it {Gauge dynamics and compactification to
  three-dimensions}},  in {\em {The mathematical beauty of physics: A memorial
  volume for Claude Itzykson. Proceedings, Conference, Saclay, France, June
  5-7, 1996}}, pp.~333--366, 1996.
\newblock \href{http://arxiv.org/abs/hep-th/9607163}{{\tt hep-th/9607163}}.

\bibitem{Aharony:1997bx}
O.~Aharony, A.~Hanany, K.~A. Intriligator, N.~Seiberg, and M.~J. Strassler,
  {\it {Aspects of N=2 supersymmetric gauge theories in three-dimensions}},
  {\em Nucl. Phys.} {\bf B499} (1997) 67--99,
  [\href{http://arxiv.org/abs/hep-th/9703110}{{\tt hep-th/9703110}}].

\bibitem{Davies:1999uw}
N.~M. Davies, T.~J. Hollowood, V.~V. Khoze, and M.~P. Mattis, {\it {Gluino
  condensate and magnetic monopoles in supersymmetric gluodynamics}},  {\em
  Nucl. Phys.} {\bf B559} (1999) 123--142,
  [\href{http://arxiv.org/abs/hep-th/9905015}{{\tt hep-th/9905015}}].

\bibitem{Davies:2000nw}
N.~M. Davies, T.~J. Hollowood, and V.~V. Khoze, {\it {Monopoles, affine
  algebras and the gluino condensate}},  {\em J. Math. Phys.} {\bf 44} (2003)
  3640--3656, [\href{http://arxiv.org/abs/hep-th/0006011}{{\tt
  hep-th/0006011}}].

\bibitem{Unsal:2007jx}
M.~Unsal, {\it {Magnetic bion condensation: A New mechanism of confinement and
  mass gap in four dimensions}},  {\em Phys. Rev.} {\bf D80} (2009) 065001,
  [\href{http://arxiv.org/abs/0709.3269}{{\tt arXiv:0709.3269}}].

\bibitem{Unsal:2007vu}
M.~Unsal, {\it {Abelian duality, confinement, and chiral symmetry breaking in
  QCD(adj)}},  {\em Phys. Rev. Lett.} {\bf 100} (2008) 032005,
  [\href{http://arxiv.org/abs/0708.1772}{{\tt arXiv:0708.1772}}].

\bibitem{Poppitz:2011wy}
E.~Poppitz and M.~Unsal, {\it {Seiberg-Witten and 'Polyakov-like' magnetic bion
  confinements are continuously connected}},  {\em JHEP} {\bf 07} (2011) 082,
  [\href{http://arxiv.org/abs/1105.3969}{{\tt arXiv:1105.3969}}].

\bibitem{Poppitz:2012sw}
E.~Poppitz, T.~Schafer, and M.~Unsal, {\it {Continuity, Deconfinement, and
  (Super) Yang-Mills Theory}},  {\em JHEP} {\bf 10} (2012) 115,
  [\href{http://arxiv.org/abs/1205.0290}{{\tt arXiv:1205.0290}}].

\bibitem{Argyres:2012ka}
P.~C. Argyres and M.~Unsal, {\it {The semi-classical expansion and resurgence
  in gauge theories: new perturbative, instanton, bion, and renormalon
  effects}},  {\em JHEP} {\bf 08} (2012) 063,
  [\href{http://arxiv.org/abs/1206.1890}{{\tt arXiv:1206.1890}}].

\bibitem{Argyres:2012vv}
P.~Argyres and M.~Unsal, {\it {A semiclassical realization of infrared
  renormalons}},  {\em Phys. Rev. Lett.} {\bf 109} (2012) 121601,
  [\href{http://arxiv.org/abs/1204.1661}{{\tt arXiv:1204.1661}}].

\bibitem{Poppitz:2012nz}
E.~Poppitz, T.~Schafer, and M.~Unsal, {\it {Universal mechanism of
  (semi-classical) deconfinement and theta-dependence for all simple groups}},
  {\em JHEP} {\bf 03} (2013) 087, [\href{http://arxiv.org/abs/1212.1238}{{\tt
  arXiv:1212.1238}}].

\bibitem{Lee:1997vp}
K.-M. Lee and P.~Yi, {\it {Monopoles and instantons on partially compactified
  D-branes}},  {\em Phys. Rev.} {\bf D56} (1997) 3711--3717,
  [\href{http://arxiv.org/abs/hep-th/9702107}{{\tt hep-th/9702107}}].

\bibitem{Kraan:1998pm}
T.~C. Kraan and P.~van Baal, {\it {Periodic instantons with nontrivial
  holonomy}},  {\em Nucl. Phys.} {\bf B533} (1998) 627--659,
  [\href{http://arxiv.org/abs/hep-th/9805168}{{\tt hep-th/9805168}}].

\bibitem{Polyakov:1976fu}
A.~M. Polyakov, {\it {Quark Confinement and Topology of Gauge Groups}},  {\em
  Nucl. Phys.} {\bf B120} (1977) 429--458.

\bibitem{Bergner:2015cqa}
G.~Bergner, P.~Giudice, G.~M�nster, and S.~Piemonte, {\it {Witten index and
  phase diagram of compactified $\mathcal N=1$ supersymmetric Yang-Mills theory
  on the lattice}},  {\em PoS} {\bf LATTICE2015} (2016) 239,
  [\href{http://arxiv.org/abs/1510.05926}{{\tt arXiv:1510.05926}}].

\bibitem{Bergner:2018unx}
G.~Bergner, S.~Piemonte, and M.~Unsal, {\it {Adiabatic continuity and
  confinement in supersymmetric Yang-Mills theory on the lattice}},  {\em JHEP}
  {\bf 11} (2018) 092, [\href{http://arxiv.org/abs/1806.10894}{{\tt
  arXiv:1806.10894}}].

\bibitem{Bergner:2019dim}
G.~Bergner, C.~Lopez, and S.~Piemonte, {\it {A study of center and chiral
  symmetry realization in thermal $\mathcal{N}=1$ super Yang-Mills theory using
  the gradient flow}},  \href{http://arxiv.org/abs/1902.08469}{{\tt
  arXiv:1902.08469}}.

\bibitem{Dunne:2016nmc}
G.~V. Dunne and M.~Unsal, {\it {New Nonperturbative Methods in Quantum Field
  Theory: From Large-N Orbifold Equivalence to Bions and Resurgence}},  {\em
  Ann. Rev. Nucl. Part. Sci.} {\bf 66} (2016) 245--272,
  [\href{http://arxiv.org/abs/1601.03414}{{\tt arXiv:1601.03414}}].

\bibitem{Anber:2015kea}
M.~M. Anber, E.~Poppitz, and T.~Sulejmanpasic, {\it {Strings from domain walls
  in supersymmetric Yang-Mills theory and adjoint QCD}},  {\em Phys. Rev.} {\bf
  D92} (2015), no.~2 021701, [\href{http://arxiv.org/abs/1501.06773}{{\tt
  arXiv:1501.06773}}].

\bibitem{Gaiotto:2014kfa}
D.~Gaiotto, A.~Kapustin, N.~Seiberg, and B.~Willett, {\it {Generalized Global
  Symmetries}},  {\em JHEP} {\bf 02} (2015) 172,
  [\href{http://arxiv.org/abs/1412.5148}{{\tt arXiv:1412.5148}}].

\bibitem{Gaiotto:2017yup}
D.~Gaiotto, A.~Kapustin, Z.~Komargodski, and N.~Seiberg, {\it {Theta, Time
  Reversal, and Temperature}},  {\em JHEP} {\bf 05} (2017) 091,
  [\href{http://arxiv.org/abs/1703.00501}{{\tt arXiv:1703.00501}}].

\bibitem{Gaiotto:2017tne}
D.~Gaiotto, Z.~Komargodski, and N.~Seiberg, {\it {Time-reversal breaking in
  QCD$_{4}$, walls, and dualities in 2 + 1 dimensions}},  {\em JHEP} {\bf 01}
  (2018) 110, [\href{http://arxiv.org/abs/1708.06806}{{\tt arXiv:1708.06806}}].

\bibitem{Sulejmanpasic:2016uwq}
T.~Sulejmanpasic, H.~Shao, A.~Sandvik, and M.~Unsal, {\it {Confinement in the
  bulk, deconfinement on the wall: infrared equivalence between compactified
  QCD and quantum magnets}},  {\em Phys. Rev. Lett.} {\bf 119} (2017), no.~9
  091601, [\href{http://arxiv.org/abs/1608.09011}{{\tt arXiv:1608.09011}}].

\bibitem{Komargodski:2017smk}
Z.~Komargodski, T.~Sulejmanpasic, and M.~Unsal, {\it {Walls, anomalies, and
  deconfinement in quantum antiferromagnets}},  {\em Phys. Rev.} {\bf B97}
  (2018), no.~5 054418, [\href{http://arxiv.org/abs/1706.05731}{{\tt
  arXiv:1706.05731}}].

\bibitem{Hsin:2018vcg}
P.-S. Hsin, H.~T. Lam, and N.~Seiberg, {\it {Comments on One-Form Global
  Symmetries and Their Gauging in 3d and 4d}},  {\em SciPost Phys.} {\bf 6}
  (2019), no.~3 039, [\href{http://arxiv.org/abs/1812.04716}{{\tt
  arXiv:1812.04716}}].

\bibitem{Witten:1997ep}
E.~Witten, {\it {Branes and the dynamics of QCD}},  {\em Nucl. Phys.} {\bf
  B507} (1997) 658--690, [\href{http://arxiv.org/abs/hep-th/9706109}{{\tt
  hep-th/9706109}}].

\bibitem{Cox:2019aji}
A.~A. Cox, E.~Poppitz, and S.~S.~Y. Wong, {\it {Domain walls and deconfinement:
  a semiclassical picture of discrete anomaly inflow}},  {\em JHEP} {\bf 12}
  (2019) 011, [\href{http://arxiv.org/abs/1909.10979}{{\tt arXiv:1909.10979}}].

\bibitem{niagara}
M.~Ponce, R.~van Zon, S.~Northrup, D.~Gruner, J.~Chen, F.~Ertinaz, A.~Fedoseev,
  L.~Groer, F.~Mao, B.~C. Mundim, M.~Nolta, J.~Pinto, M.~Saldarriaga,
  V.~Slavnic, E.~Spence, C.-H. Yu, and W.~R. Peltier, {\it {Deploying a Top-100
  Supercomputer for Large Parallel Workloads: The Niagara Supercomputer}},  in
  {\em PEARC '19 Proceedings}, Association for Computing Machinery, 2019.
\newblock \href{http://arxiv.org/abs/1907.13600}{{\tt arXiv:1907.13600}}.

\bibitem{scinet}
C.~Loken, D.~Gruner, L.~Groer, R.~Peltier, N.~Bunn, M.~Craig, T.~Henriques,
  J.~Dempsey, C.-H. Yu, J.~Chen, L.~J. Dursi, J.~Chong, S.~Northrup, J.~Pinto,
  N.~Knecht, and R.~V. Zon, {\it {{SciNet}: Lessons Learned from Building a
  Power-efficient Top-20 System and Data Centre}},  {\em J. Phys.: Conf. Ser.}
  {\bf 256} (2010) 012026.

\bibitem{Douglas:1995nw}
M.~R. Douglas and S.~H. Shenker, {\it {Dynamics of SU(N) supersymmetric gauge
  theory}},  {\em Nucl. Phys.} {\bf B447} (1995) 271--296,
  [\href{http://arxiv.org/abs/hep-th/9503163}{{\tt hep-th/9503163}}].

\bibitem{Cherman:2016jtu}
A.~Cherman and E.~Poppitz, {\it {Emergent dimensions and branes from large-$N$
  confinement}},  {\em Phys. Rev.} {\bf D94} (2016), no.~12 125008,
  [\href{http://arxiv.org/abs/1606.01902}{{\tt arXiv:1606.01902}}].

\bibitem{Poppitz:2017ivi}
E.~Poppitz and M.~E. Shalchian~T., {\it {String tensions in deformed Yang-Mills
  theory}},  {\em JHEP} {\bf 01} (2018) 029,
  [\href{http://arxiv.org/abs/1708.08821}{{\tt arXiv:1708.08821}}].

\bibitem{Anber:2017pak}
M.~M. Anber and L.~Vincent-Genod, {\it {Classification of compactified
  $su(N_c)$ gauge theories with fermions in all representations}},  {\em JHEP}
  {\bf 12} (2017) 028, [\href{http://arxiv.org/abs/1704.08277}{{\tt
  arXiv:1704.08277}}].

\bibitem{Anber:2019nfu}
M.~M. Anber, {\it {Self-conjugate QCD}},  {\em JHEP} {\bf 10} (2019) 042,
  [\href{http://arxiv.org/abs/1906.10315}{{\tt arXiv:1906.10315}}].

\bibitem{Anber:2013xfa}
M.~M. Anber, {\it {The abelian confinement mechanism revisited: new aspects of
  the Georgi-Glashow model}},  {\em Annals Phys.} {\bf 341} (2014) 21--55,
  [\href{http://arxiv.org/abs/1308.0027}{{\tt arXiv:1308.0027}}].

\bibitem{Unsal:2020yeh}
M.~Unsal, {\it {Strongly coupled QFT dynamics via TQFT coupling}},
  \href{http://arxiv.org/abs/2007.03880}{{\tt arXiv:2007.03880}}.

\bibitem{Anber:2014lba}
M.~M. Anber, E.~Poppitz, and B.~Teeple, {\it {Deconfinement and continuity
  between thermal and (super) Yang-Mills theory for all gauge groups}},  {\em
  JHEP} {\bf 09} (2014) 040, [\href{http://arxiv.org/abs/1406.1199}{{\tt
  arXiv:1406.1199}}].

\bibitem{Anber:2014sda}
M.~M. Anber and T.~Sulejmanpasic, {\it {The renormalon diagram in gauge
  theories on $ {\mathrm{\mathbb{R}}}^3 \times {\mathbb{S}}^1 $}},  {\em JHEP}
  {\bf 01} (2015) 139, [\href{http://arxiv.org/abs/1410.0121}{{\tt
  arXiv:1410.0121}}].

\bibitem{Aitken:2017ayq}
K.~Aitken, A.~Cherman, E.~Poppitz, and L.~G. Yaffe, {\it {QCD on a small
  circle}},  {\em Phys. Rev.} {\bf D96} (2017), no.~9 096022,
  [\href{http://arxiv.org/abs/1707.08971}{{\tt arXiv:1707.08971}}].

\bibitem{Anber:2017ezt}
M.~M. Anber and E.~Poppitz, {\it {New nonperturbative scales and glueballs in
  confining supersymmetric gauge theories}},  {\em JHEP} {\bf 03} (2018) 052,
  [\href{http://arxiv.org/abs/1711.00027}{{\tt arXiv:1711.00027}}].

\bibitem{Anber:2015wha}
M.~M. Anber and E.~Poppitz, {\it {On the global structure of deformed
  Yang-Mills theory and QCD(adj) on $ {\mathrm{\mathbb{R}}}^3\times
  {\mathbb{S}}^1 $}},  {\em JHEP} {\bf 10} (2015) 051,
  [\href{http://arxiv.org/abs/1508.00910}{{\tt arXiv:1508.00910}}].

\bibitem{Argurio:2018uup}
R.~Argurio, M.~Bertolini, F.~Bigazzi, A.~L. Cotrone, and P.~Niro, {\it {QCD
  domain walls, Chern-Simons theories and holography}},  {\em JHEP} {\bf 09}
  (2018) 090, [\href{http://arxiv.org/abs/1806.08292}{{\tt arXiv:1806.08292}}].

\bibitem{Bashmakov:2018ghn}
V.~Bashmakov, F.~Benini, S.~Benvenuti, and M.~Bertolini, {\it {Living on the
  walls of super-QCD}},  {\em SciPost Phys.} {\bf 6} (2019), no.~4 044,
  [\href{http://arxiv.org/abs/1812.04645}{{\tt arXiv:1812.04645}}].

\bibitem{Anber:2018jdf}
M.~M. Anber and E.~Poppitz, {\it {Anomaly matching, (axial) Schwinger models,
  and high-T super Yang-Mills domain walls}},  {\em JHEP} {\bf 09} (2018) 076,
  [\href{http://arxiv.org/abs/1807.00093}{{\tt arXiv:1807.00093}}].

\bibitem{Anber:2018xek}
M.~M. Anber and E.~Poppitz, {\it {Domain walls in high-T SU(N) super Yang-Mills
  theory and QCD(adj)}},  {\em JHEP} {\bf 05} (2019) 151,
  [\href{http://arxiv.org/abs/1811.10642}{{\tt arXiv:1811.10642}}].

\bibitem{Poppitz:2020tto}
E.~Poppitz and F.~D. Wandler, {\it {Topological terms and anomaly matching in
  effective field theories on $\mathbb{R}^3\times S^1$: I. Abelian symmetries
  and intermediate scales}},  \href{http://arxiv.org/abs/2009.14667}{{\tt
  arXiv:2009.14667}}.

\bibitem{Hori:2000ck}
K.~Hori, A.~Iqbal, and C.~Vafa, {\it {D-branes and mirror symmetry}},
  \href{http://arxiv.org/abs/hep-th/0005247}{{\tt hep-th/0005247}}.

\bibitem{Greensite:2001nx}
J.~Greensite and C.~B. Thorn, {\it {Gluon chain model of the confining force}},
   {\em JHEP} {\bf 02} (2002) 014,
  [\href{http://arxiv.org/abs/hep-ph/0112326}{{\tt hep-ph/0112326}}].

\bibitem{Ambjorn:1999ym}
J.~Ambjorn, J.~Giedt, and J.~Greensite, {\it {Vortex structure versus monopole
  dominance in Abelian projected gauge theory}},  {\em JHEP} {\bf 02} (2000)
  033, [\href{http://arxiv.org/abs/hep-lat/9907021}{{\tt hep-lat/9907021}}].

\bibitem{Polyakov:1987ez}
A.~M. Polyakov, {\em {Gauge Fields and Strings}}, vol.~3.
\newblock 1987.

\bibitem{Greensite:2014gra}
J.~Greensite and R.~Hollwieser, {\it {Double-winding Wilson loops and monopole
  confinement mechanisms}},  {\em Phys. Rev. D} {\bf 91} (2015), no.~5 054509,
  [\href{http://arxiv.org/abs/1411.5091}{{\tt arXiv:1411.5091}}].

\bibitem{Numerical_Method}
S.~Chapra and R.~Canale, {\em {Numerical Methods for Engineers}}.
\newblock McGraw-Hill, 2015.

\bibitem{Hasenfratz:1977dt}
P.~Hasenfratz and J.~Kuti, {\it {The Quark Bag Model}},  {\em Phys. Rept.} {\bf
  40} (1978) 75--179.

\bibitem{Hanany:1997hr}
A.~Hanany, M.~J. Strassler, and A.~Zaffaroni, {\it {Confinement and strings in
  MQCD}},  {\em Nucl. Phys. B} {\bf 513} (1998) 87--118,
  [\href{http://arxiv.org/abs/hep-th/9707244}{{\tt hep-th/9707244}}].

\bibitem{Armoni:2011dw}
A.~Armoni, D.~Dorigoni, and G.~Veneziano, {\it {k-String Tension from
  Eguchi-Kawai Reduction}},  {\em JHEP} {\bf 10} (2011) 086,
  [\href{http://arxiv.org/abs/1108.6196}{{\tt arXiv:1108.6196}}].

\bibitem{Komargodski:2020mxz}
Z.~Komargodski, K.~Ohmori, K.~Roumpedakis, and S.~Seifnashri, {\it {Symmetries
  and Strings of Adjoint QCD${}_2$}},
  \href{http://arxiv.org/abs/2008.07567}{{\tt arXiv:2008.07567}}.

\bibitem{Anber:2017tug}
M.~M. Anber and V.~Pellizzani, {\it {Representation dependence of k -strings in
  pure Yang-Mills theory via supersymmetry}},  {\em Phys. Rev.} {\bf D96}
  (2017), no.~11 114015, [\href{http://arxiv.org/abs/1710.06509}{{\tt
  arXiv:1710.06509}}].

\bibitem{Schafer:1996wv}
T.~Schafer and E.~V. Shuryak, {\it {Instantons in QCD}},  {\em Rev. Mod. Phys.}
  {\bf 70} (1998) 323--426, [\href{http://arxiv.org/abs/hep-ph/9610451}{{\tt
  hep-ph/9610451}}].

\bibitem{Balitsky:1986qn}
I.~Balitsky and A.~Yung, {\it {Collective - Coordinate Method for Quasizero
  Modes}},  {\em Phys. Lett. B} {\bf 168} (1986) 113--119.

\bibitem{Hollowood:1992by}
T.~J. Hollowood, {\it {Solitons in affine Toda field theories}},  {\em Nucl.
  Phys. B} {\bf 384} (1992), no.~3 523--540,
  [\href{http://arxiv.org/abs/hep-th/9110010}{{\tt hep-th/9110010}}].

\end{thebibliography}\endgroup
  
  \bibliographystyle{JHEP}

\end{document}